\documentclass[prb,twocolumn,showpacs,amsmath,amssymb,superscriptaddress,floatfix]{revtex4-1}
\usepackage[english]{babel}
\usepackage{graphicx,color}
\usepackage{bm}
\usepackage{comment}

\newcommand{\sub}[1]{{\mbox{\footnotesize #1}}}

\definecolor{darkred}{rgb}{0.90,0,0}
\definecolor{darkgreen}{rgb}{0,0.60,.2}
\definecolor{darkblue}{rgb}{0,0,1}
\definecolor{grey}{cmyk}{0,0,0,0.25}
\definecolor{orange}{cmyk}{0,0.6,0.8,0}

\begin{document}

\title{\boldmath Sudden expansion of Mott insulators in one dimension}

\author{L.\ Vidmar}
\affiliation{Department of Physics and Arnold Sommerfeld Center for Theoretical Physics,
Ludwig-Maximilians-Universit\"at M\"unchen, D-80333 M\"unchen, Germany}
\affiliation{J. Stefan Institute, SI-1000 Ljubljana, Slovenia}
\author{S. Langer}
\affiliation{Department of Physics and Astronomy, University of Pittsburgh, Pittsburgh, Pennsylvania 15260, USA}
\author{I. P. McCulloch} \affiliation{School of Physical Sciences, The University of Queensland, Brisbane, QLD 4072, Australia}
\author{U. Schneider}
\affiliation{Department of Physics,
Ludwig-Maximilians-Universit\"at M\"unchen, D-80799 M\"unchen, Germany}
\author{U. Schollw\"ock}
\affiliation{Department of Physics and Arnold Sommerfeld Center for Theoretical Physics,
Ludwig-Maximilians-Universit\"at M\"unchen, D-80333 M\"unchen, Germany}
\author{F. Heidrich-Meisner}
\affiliation{Department of Physics and Arnold Sommerfeld Center for Theoretical Physics,
Ludwig-Maximilians-Universit\"at M\"unchen, D-80333 M\"unchen, Germany}

\begin{abstract}
We investigate the expansion of bosons and fermions in a homogeneous lattice after a sudden removal of the trapping potential using exact numerical methods. 
As a main result, we show that in one dimension, both bosonic and fermionic Mott insulators expand with the same velocity, irrespective of the interaction strength, provided the expansion starts from the ground state of the trapped gas.
Furthermore, their density profiles become identical during the expansion:
The asymptotic density dynamics is identical to that of initially localized, non-interacting particles, and the asymptotic velocity distribution is flat.
The expansion velocity for initial correlated Mott insulating states is therefore independent of the interaction strength and particle statistics.
Interestingly, this non-equilibrium dynamics is sensitive to the interaction driven quantum phase transition in the Bose-Hubbard model:
While being constant in the Mott phase, the expansion velocity decreases in the superfluid phase and vanishes for large systems in the non-interacting limit.
These results are compared to the set-up of a recent experiment [Ronzheimer et al., Phys. Rev. Lett. {\bf 110}, 205301 (2013)], where the trap opening was combined with an interaction quench from infinitely strong interactions to finite values.
In the latter case, the interaction quench breaks the universal dynamics in the asymptotic regime and the expansion depends on the  interaction strength.
We carry out an analogous analysis for a two-component Fermi gas, with similar observations.
In addition, we study the effect of breaking the integrability of hard-core bosons in different ways: 
While the fast ballistic expansion from the ground state of Mott insulators in one dimension remains unchanged for finite interactions, we observe strong deviations from this behavior on a two-leg ladder even in the hard-core case. 
This change in dynamics bares similarities with  the dynamics in the dimensional crossover from one to two dimensions observed in the aformentioned experimental study.
\end{abstract}

%\pacs{71.27.+a, 75.10.Pq, 75.40.Mg, 05.60.Gg}
\maketitle

% 75.10.Pq Magnetic ordering, general theory and models of spin chain models
% 71.27.+a Strongly correlated electron systems
% 75.40.Mg Computer modeling and simulation of magnetic critical points
% 05.60.Gg Transport processes, quantum

\section{Introduction} \label{sec:intro}

The possibility of realizing various many-body Hamiltonians in the laboratory with ultra-cold atomic gases~\cite{bloch08} 
allows one to address outstanding questions from condensed matter theory in well-controlled experiments.
In the context of low-dimensional systems with strong correlations, two topics currently receive a lot of attention, namely non-equilibrium dynamics~\cite{greiner02,kinoshita06,hofferberth07,trotzky12,meinert13} and transport properties.~\cite{fertig05,brantut12,stadler12,schneider12,ronzheimer13,reinhard13,li13}
In the former case, theorists seek to understand, e.g., relaxation processes and the conditions for thermalization,~\cite{polkovnikov11,rigol08} whereas in the latter case,
qualitative questions such as whether transport is ballistic or rather diffusive in microscopic models of strongly interacting systems remain actively debated~(see Refs.~\onlinecite{zotos97,prosen11,sirker11,heidrichmeisner07} and references therein).

\begin{figure}[!t]
\includegraphics[width=0.99\columnwidth,clip]{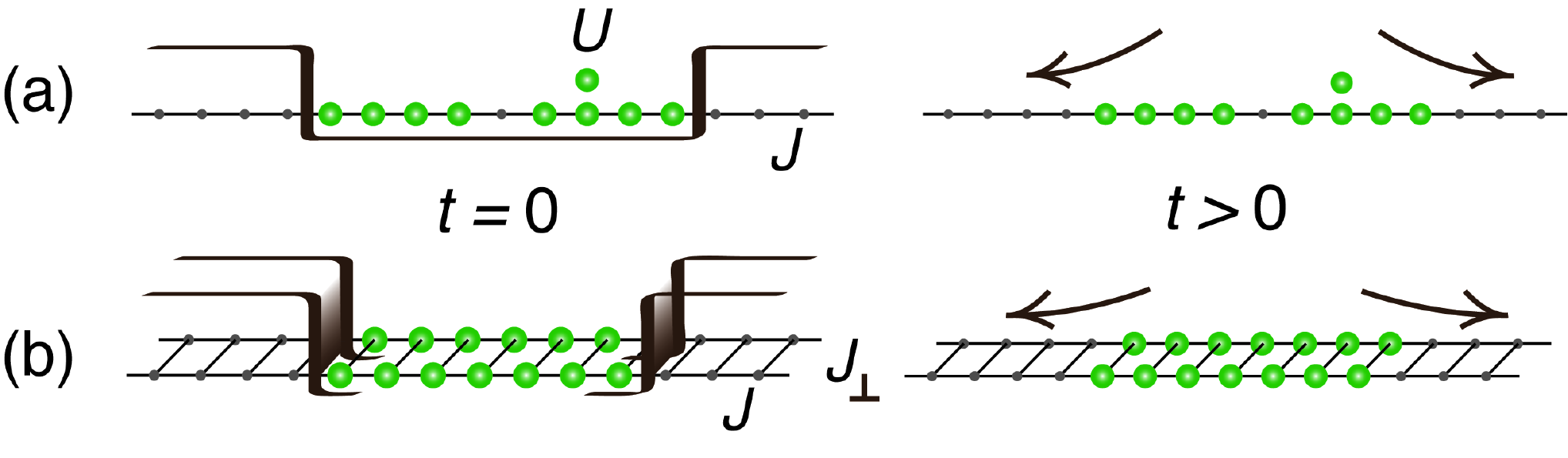}
\caption{(Color online)
{\it
Sketch of the setup.}
(a) Single chain with hopping amplitude $J$ and on-site repulsion $U$.
(b) Two-leg ladder with hopping amplitude %$J$ along the chains and
$J_{\perp}$ perpendicular to the chains.
}\label{fig1}
\end{figure} 

While in higher dimensions almost all interacting models show the same generic behavior, governed by local thermalization and leading to diffusive transport and therefore finite dc-conductivities, many one-dimensional (1D) models such as the Heisenberg model, the Fermi-Hubbard model or hard-core bosons~\cite{peredes04,kinoshita04}
are believed to exhibit anomalous behavior.
This can include, for example, non-ergodic dynamics,~\cite{caux11} non-thermal stationary states (see, e.g., Refs.~\onlinecite{kinoshita06,rigol07,manmana07,barthel08}), or ballistic transport with divergent dc-conductivities despite the presence of interactions.~\cite{zotos97,heidrichmeisner07}
This is often traced back to the integrability of these models.~\cite{[{For a discussion of the notion of integrability, see }] caux11}
A number of examples, such as dissipationless energy transport in Heisenberg chains,~\cite{zotos97,heidrichmeisner07}
convincingly demonstrate that integrable systems are ideal candidates in which to search for 
deviations from generic behavior, precisely because of the existence of additional conservation laws.~\cite{zotos97,heidrichmeisner07,sirker11} 
An unambiguous test for ballistic dynamics in condensed matter systems is difficult since one needs to account for, e.g.,
impurities or phonons,~\cite{shimshoni03,louis06,chernychev05} which can in principle break most non-trivial conservation laws.
Nonetheless, there are many intriguing experimental results, in particular for low-dimensional quantum magnets,~\cite{hess07,hlubek10,sologubenko07} that are speculated to be related to the existence of such conservation laws for the underlying spin Hamiltonians.
In contrast, many non-integrable 1D models seem to exhibit dynamics that are compatible with diffusion,~\cite{heidrichmeisner03,hm04,zotos04,jung06,rosch00} although several exceptions have also been identified.~\cite{kirchner99,znidaric13,karrasch12}

In this work we consider the sudden expansion of interacting particles into a homogeneous lattice, sketched in Fig.~\ref{fig1}.
The expansion is induced by suddenly quenching the trapping potential, typically present in all cold atomic gas experiments, to zero.
%Beside a large portion of theoretical studies of  expanding 1D gases,~\cite{ohberg02, rigol04, micheli04, daley05, rigol05, rigol05a, minguzzi05, girardeau06, rodriguez06, hm08, delcampo08, gangardt08, jukic08, jukic09, hm09, gritsev10, lancaster10, kolovsky10, kajala11,karlsson11,langer12, konik12, bolech12, collura13, collura13b, iyer12, iyer13,kessler13}
Our study is motivated by two recent experiments that have realized the sudden expansion in optical lattices.~\cite{schneider12, ronzheimer13}
For fermions in 2D, the experimental data and theoretical modeling suggest that the  dynamics in the high-density regime is diffusive.~\cite{schneider12}
In the case of bosons~\cite{ronzheimer13}, significant differences were observed between strongly interacting particles in 1D versus 2D:
In 1D, the expansion of hard-core bosons is  ballistic (and fast for initial states with one boson per site), whereas in 2D, the atomic cloud barely expands at all,  similar to the diffusive dynamics observed for fermions.
The reason for the ballistic expansion of hard-core bosons in 1D is the fact that this model is indeed integrable and can be mapped to non-interacting fermions.~\cite{cazalilla11} 
Due to an interaction quench performed simultaneously with the removal of the trap, this latter experiment~\cite{ronzheimer13} did not probe the low-energy dynamics at small and intermediate
interaction strengths.
In another recent experiment, the {\it transverse} expansion of bosons in a 3D array of tunnel-coupled 1D tubes was studied.~\cite{reinhard13,li13}
%For theoretical studies of  expanding 1D gases, see, e.g., Refs.~\onlinecite{ohberg02, girardeau06, delcampo08, gangardt08, jukic08, jukic09, gritsev10, kolovsky10, konik12, bolech12, collura13, collura13b, iyer12, iyer13}.

In the sudden expansion, there is always a regime in which one expects ballistic dynamics, namely the limit of long expansion times where the gas becomes dilute and effectively non-interacting.
For this asymptotic regime, one may ask about the properties of the  effective non-interacting Hamiltonian.
For instance, for hard-core bosons on a lattice and the Tonks-Girardeau gas,  the dynamics is controlled by non-interacting fermions.\cite{rigol05, rigol05a, minguzzi05}
This extends to the  Lieb-Liniger model with repulsive interactions,  whereas attractive interactions lead to emergent bosons in the long-time limit.\cite{iyer12,iyer13}
In the cases with repulsive interactions, the existence of exact solutions relies on a mapping from interacting bosons to non-interacting spinless fermions.
In this work we clarify this question for the expansion from ground states of the non-integrable Bose-Hubbard model with repulsive interactions at unit filling, showing that the asymptotic expansion dynamics is also controlled by non-interacting fermions. Moreover, we study the asymptotic behavior of various measures of the expansion velocity and
of the quasi-momentum distribution.
Note that  similar questions have been addressed in the literature, mostly for integrable models.~\cite{ohberg02, girardeau06, delcampo08, gangardt08, jukic08, jukic09, gritsev10, konik12, bolech12, collura13, collura13b, iyer12, iyer13,kolovsky10}

The main goal of our study is to investigate three questions.
First,
we disentangle the effects of the trap opening from the interaction quench by studying the expansion dynamics from the {\it ground state} of the trapped gas.
Surprisingly, we find the same ballistic expansion at sufficiently long times for all initial Mott insulators, different from the
case studied in Ref.~\onlinecite{ronzheimer13}:
Provided the interaction strength exceeds the critical value that separates the bosonic Mott insulators from the superfluid in 1D, the asymptotic expansion velocity becomes independent of the interaction strength, similar to the expansion from a fermionic Mott insulator in 1D.~\cite{langer12}
 It is noteworthy that the expansion velocity approaches its asymptotic value quite fast.
Second,
we argue that in the {\it asymptotic limit} of long expansion times, the density dynamics of all Mott insulators - both bosonic and fermionic -
become identical and give rise to the same density profiles.
The dynamics is then characterized by a flat velocity distribution and therefore a particle-hole symmetric momentum distribution function.
As a result, the asymptotic dynamics is governed by non-interacting fermions.
Third,
we investigate how breaking the integrability of hard-core bosons in 1D affects the ballistic expansion dynamics.
We find that coupling chains to {\it ladders} results in a qualitatively different behavior.

The paper is organized as follows.
In Sec.~\ref{sec:setup} we define the model hamiltonians and observables.
In Sec.~\ref{sec:ballistic} we provide a qualitative discussion of  the differences between ballistic and diffusive expansion dynamics and clarify the notion of the asymptotic regime.
We then focus on the sudden expansion on a single chain in Sec.~\ref{sec:1D}.
We distinguish between
{\it (i)}
hard-core bosons that expand from the product of local Fock states (Sec.~\ref{subsec:HCB});
{\it (ii)}
bosons and fermions at finite interactions that expand from the ground state (Sec.~\ref{subsec:gs});
{\it (iii)}
bosons and fermions at finite interactions that expand from product states, which are not eigenstates (Sec.~\ref{subsec:product}).
In Sec.~\ref{sec:ladder}, we study the sudden expansion of hard-core bosons in the crossover from uncoupled chains to a two-leg ladder, before concluding in Sec.~\ref{sec:conclusion}.
In Appendix~\ref{sec:app1}, we describe details of numerical calculations and the procedure to obtain expansion velocities.
While in the main part of the paper we study the expansion from  initial states with unity filling,
we briefly discuss the sudden expansion from states with lower particle densities in Appendix~\ref{sec:density}.
The results presented in the main article are obtained for the expansion from the box trap.
In Appendix~\ref{sec:harmonic} we show that the main results of our study are also valid in the case of initial harmonic confinement.

%%%%%%%%%%%%%%%%%%%%%%%%%%%%%%%%%%%%%%%%%%%%%%%%%%%%%%%%%%%%%%%

\section{Setup and model} \label{sec:setup}

\subsection{Hamiltonians} \label{sec:hamiltonians}

We investigate the expansion dynamics for both bosons and fermions.
The Bose-Hubbard model (BHM) is defined as
\begin{eqnarray}
H_\sub{\scriptsize BHM} & = & -J\sum_{\langle i,j \rangle_\Vert} (b_i^\dagger b_j + \mbox{h.c.}) -J_{\perp}\sum_{\langle i,j \rangle_\perp} (b_i^\dagger b_j + \mbox{h.c.}) \nonumber \\
 & + & \frac{U}{2}\sum_i n_i (n_i-1) + V_{\mbox{\footnotesize trap}}\theta(-t),\, \label{bh_def}
\end{eqnarray}
where $b_i$ is the boson annihilation operator at site $i$, $n_i=b_i^\dagger b_i$ represents the density, and $U$ is the on-site interaction strength.

This hamiltonian can describe both the single chain studied in the first part of the paper [Fig.~\ref{fig1}(a)], as well as the two-leg ladder studied later [Fig.~\ref{fig1}(b)].
To obtain a chain, we set $J_{\perp}=0$, while $J_{\perp}>0$ corresponds to a two-leg ladder.
The summation index $\langle i,j \rangle_{\Vert(\perp)}$ in Eq.~(\ref{bh_def}) indicates nearest neighbors along (perpendicular to) the chain.
We set the lattice spacing to unity.

We compare our results for the expansion of bosons with the expansion of a two-component Fermi gas.
The Fermi-Hubbard model (FHM) on a chain with length $L$ is defined as
\begin{eqnarray}
H_\sub{\scriptsize FHM} & = & -J\sum_{i=1}^{L-1} \sum_{\sigma\in\{ \uparrow, \downarrow \}} (c_{i,\sigma}^\dagger c_{i+1,\sigma} + \mbox{h.c.}) \nonumber \\
 & + & U\sum_{i=1}^L n_{i,\uparrow} n_{i,\downarrow} + V_{\mbox{\footnotesize trap}}\theta(-t), \label{fh_def}
\end{eqnarray}
where $c_{i,\sigma}$ is the fermion annihilation operator for component $\sigma$ at site $i$, and $n_{i,\sigma} = c_{i,\sigma}^\dagger c_{i,\sigma}$ represents the on-site density of a single fermionic component.

The confining potential $V_{\mbox{\footnotesize trap}}$ at $t<0$ in Eqs.~(\ref{bh_def}) and (\ref{fh_def}) is represented by a box trap.
A box trap for a chain is defined as
\begin{equation} \label{def_boxtrap}
V_{\mbox{\footnotesize trap}}^\sub{(b)} = V_\sub{b} \left( \sum_{i=1}^{i_a} n_i + \sum_{i=i_b}^{L} n_i \right),
\end{equation}
where $V_\sub{b}=10^3J$ and $n_a < n_b$.
In Appendix~\ref{sec:harmonic}, we also present results for an expansion from the harmonic trap.
For bosons and fermions at finite interactions, we calculate the ground state in the trap as well as the time evolution after a sudden removal of the confining trap via the DMRG method~\cite{vidal04,daley04,white04,schollwoeck05,schollwoeck11}
using the time-step $\Delta tJ=1/16$ and a discarded weight $\eta\leq 10^{-4}$.
For non-interacting fermions and hard-core bosons, we use exact diagonalization.~\cite{rigol04a,rigol04}
The time $t$ is measured in units of $1/J$, and we set $\hbar\equiv 1$ hereafter.

Since the main focus of our work is on the expansion from Mott insulators, we fix the initial density to $n=N/L_{\mbox{\footnotesize box}}=1$
(where $N$ represents the total number of particles and the box size, see Eq.~(\ref{def_boxtrap}), is given by $L_{\mbox{\footnotesize box}}=i_b-i_a-1$).
We present results for the expansion velocity for initial densities $n<1$ in Appendix~\ref{sec:density}.

\subsection{Radial, core and average velocity} \label{sec:observables}

We investigate the expansion dynamics in both real and momentum space using different observables.
The time-dependent radius of the density distribution is defined as
\begin{equation}
R^2(t)  = \frac{1}{N} \sum_i \langle n_i(t) \rangle (i-i_0)^2, \label{r_def}
\end{equation}
where $i_0$ represents the center of mass.
For the Fermi-Hubbard model, we define $n_i = \sum_{\sigma} n_{i,\sigma}$.
The corresponding {\it radial} velocity $v_{\mbox{\footnotesize r}}(t)$ is defined through the reduced radius $\tilde R(t)=\sqrt{ R^2(t) - R^2(0)}$ as
\begin{equation}
v_\sub{r}(t) = \frac{\partial \tilde R(t)}{\partial t}.
\end{equation}

%Typically, one associates $\tilde R(t) \propto t$ with ballistic dynamics and $\tilde R(t) \propto \sqrt{t}$ with diffusion.
%In the sudden expansion, since the density is time-dependent, the diffusion equation becomes non-linear, and as a consequence, these criteria are less stringent; see the discussion in Refs.~\onlinecite{schneider12,langer12}.
%For our definition of ballistic dynamics, we require $\tilde R(t) \propto t$ at all times.

In recent experiments on optical lattices,~\cite{schneider12,ronzheimer13} the in-situ density profiles have been measured during the expansion.
In these experiments, the expanding clouds are characterized by the core radius $r_\sub{c}(t)$, which denotes the half-width-at-half-maximum of the density distribution.
In the case of several local density maxima, $r_\sub{c}(t)$ corresponds to half the distance between the outermost points at half the maximal density.
Compared to $R(t)$, the core radius $r_\sub{c}(t)$ has the experimental advantage of being far less sensitive to detection noise.~\cite{schneider12}
Using this core radius, the core velocity $v_\sub{c}$ is defined as
\begin{equation} \label{def_vc}
v_\sub{c}(t) = \frac{\partial r_\sub{c}(t)}{\partial t}.
\end{equation}
We calculate both $v_\sub{r}$ and $v_\sub{c}$ in our study.
In particular, we quantify the expansion in terms of the core velocity $v_\sub{c}$ in the crossover from uncoupled chains to a two-leg ladder and compare to experimental results for the 1D-2D crossover.

Another measure of the expansion velocity is related to the momentum distribution function (MDF).
For the Bose-Hubbard model, the MDF is defined as
\begin{equation}
n_k=\frac{1}{L}\sum_{l,m}e^{-ik(l-m)}\langle b_l^\dagger b_m \rangle,
\end{equation}
while for the Fermi-Hubbard model, it reads
\begin{equation}
n_k=\frac{1}{L}\sum_{l,m,\sigma}e^{-ik(l-m)}\langle c_{l,\sigma}^\dagger c_{m,\sigma} \rangle.
\end{equation}
We define the {\it average} velocity
\begin{equation}
v^2_{\mbox{\footnotesize av}}(t)  = \frac{1}{N} \sum_{k} n_k(t) v_k^2 \label{vav_def}
\end{equation}
as the root mean square velocity, where $v_k = 2J \sin{k}$.
For a non-interacting system, expansion velocities defined through the density distribution and the momentum distribution function are time-independent and identical, therefore,
\begin{equation} \label{vr_vav_noninteracting}
v_\sub{r}(t) = v_\sub{av}(t) = v_\sub{av}(t=0)
\end{equation}
are fully determined from the initial state.

In general, we use the same labels for the expectation values for bosons and fermions.
In the case when we quantitatively compare bosons to fermions, we add an additional index to fermionic quantities (e.g., Sec.~\ref{subsec:HCB} for hard-core bosons).

%%%%%%%%%%%%%%%%%%%
\section{Discussion: Diffusive versus ballistic dynamics and asymptotic dynamics} \label{sec:ballistic}

In classical physics, two prototypical transport mechanisms are ballistic and diffusive transport:
Ballistic transport is characterized by non-decaying currents and the absence of friction.
The prototypical ballistic system consists of non-interacting particles where the individual momenta of all particles, and therefore also all particle currents are conserved due to the absence of collisions. Diffusive systems, on the other hand, are characterized by decaying currents and frequent diffractive collisions, which drive a local thermalization.
Another frequently used scenario of ballistic dynamics relevant in, e.g.,  mesoscopic physics is that where mean-free paths are larger 
than device dimensions. This is not the type of ballistic dynamics that we are interested in (see below). 

While the aforementioned  picture for ballistic dynamics in terms of non-interacting particles carries over to free particles in quantum mechanics, it misses the
possibility of ballistic finite temperature transport in interacting many-body systems, where scattering processes are present but can be  ineffective in causing currents to decay.
This is realized in certain integrable one-dimensional models~\cite{zotos97,heidrichmeisner07,mierzejewski10} and can be traced back to the existence of non-trivial local or quasi-local conservation laws.~\cite{zotos97,prosen11}

A rigorous analysis and definition that encompasses all these cases is usually based on current auto-correlation functions within the framework of linear-response theory.
Ballistic dynamics is realized whenever $C(t)=\langle j(t) j(0)\rangle$ does not decay to zero for $t\to\infty$, see Ref.~\onlinecite{zotos97}.
Expressed in terms of the conductivity,
ballistic transport is defined through a diverging dc-conductivity and is signaled by the presence of a finite Drude weight $D$, defined through the real part of the conductivity
\begin{equation}
\mbox{Re}\,\sigma(\omega) = 2\pi D(T) \delta(\omega) + \sigma_{\rm reg}(\omega),
\end{equation}
where $T$ represents temperature.
We call the dynamics ballistic if $D>0$, irrespective of whether finite-frequency contributions $\sigma_{\rm reg}(\omega)$ exist or not.
Diffusive transport is characterized by a sufficiently fast decay of  current-current correlations, leading to a finite dc-conductivity and a vanishing Drude weight.
In the simplest version of diffusive transport, $\sigma_{\rm reg}(\omega)$  takes the Drude form with a single relaxation time.

Alternative to a steady-state transport experiment in the presence of a gradient in chemical potential or a thermal gradient, we can also probe the qualitative transport behavior by monitoring the expansion of a density perturbation on top of a uniform system.
A ballistic expansion will manifest itself in a linear asymptotic growth of the spatial
variance of the density perturbation, i.e., $R(t)\propto t$, while a diffusive expansion with a fixed diffusion constant $\mathcal{D}$ will lead to the well-known  $R(t)\propto\sqrt{t}$ behavior.
This has been  verified for spin and energy dynamics in the spin-1/2 $XXZ$ chain, both at zero~\cite{langer09,langer11} and finite temperatures.\cite{karrasch-unpub}

In an expansion into the vacuum (i.e., without a constant background), commonly referred to as sudden expansion in the context of quantum gas experiments,
the distinction between diffusive and ballistic dynamics becomes more complicated, since the diffusion constant is typically density-dependent.
Therefore, the diffusion equation becomes nonlinear (see the discussion in Refs.~\onlinecite{schneider12,langer12}) and can also allow for solutions with $R(t)\propto t$.
Another phenomenon that can occur as a consequence of 
interactions and large density gradients is self-trapping.
Recent work suggests that this is not relevant
for the expansion from states with not more than one boson per site.\cite{jreissaty13}

Moreover, as the gas expands and thus becomes very dilute, scattering processes will cease to occur, rendering the gas effectively non-interacting.
One way of visualizing this is to think of the particles becoming velocity-ordered, i.e., the fastest particles will be the ones already furthest outside and no scattering events will occur anymore.
This picture is corroborated by a theoretical analysis of structures in the wave-front of expanding clouds.\cite{eisler13} 
Neglecting the existence of possible bound states (see the discussion in Ref.~\onlinecite{bolech12}), the expansion into the vacuum (in our case, an empty lattice)
can therefore,  in the asymptotic long-time limit, be  described by a non-interacting Hamiltonian:
\begin{equation}
H_{\infty} = \sum_k \epsilon_k n_k^{\infty} \, ,
\end{equation}
where $n_k^{\infty}=n_k(t\to\infty)$
is the asymptotic momentum or quasi-momentum distribution of the particles. In this case, all interaction energy is assumed to be fully converted into
kinetic energy.
It will be one of the main goals of this work to predict both the actual form of $n_k^{\infty}$ and the statistics of the emergent particles for the Bose-Hubbard model.
To give an example, from the fact that hard-core bosons map to free fermions, whose quasi-momentum distribution remains constant, we know  the asymptotic $n_k^{\infty}$
in this case since the physical $n_k$ of the hard-core bosons becomes identical to the one of the underlying non-interacting fermions for $t\to\infty$ through the
so-called dynamical fermionization~\cite{rigol05,minguzzi05} (see Sec.~\ref{subsec:HCB}  for details).
Moreover, it is well-known that 1D bosons with repulsive contact interactions and in the continuum (which is the Lieb-Liniger
model), also map to free fermions\cite{cazalilla11} such that the asymptotic dynamics is also controlled by non-interacting fermions, as was explicitly shown in
Ref.~\onlinecite{iyer12} (see also Refs.~\onlinecite{jukic08,jukic09,gritsev10,gangardt08,delcampo08,girardeau06,ohberg02,bolech12} for further studies addressing the asymptotic form of $n_k$
and other quantities for one-dimensional gases).
Besides the form of $n_k$ at infinite expansion times, it is also interesting
to study the real-space decay of one-body correlations (or other quantities)
in the asymptotic regime. For instance, for hard-core bosons,
one-body correlations still decay with a power law and ground-state
exponents even after infinitely long expansion times.\cite{rigol05}
To summarize, the asymptotic limit of a sudden expansion experiment constitutes in many cases a trivial realization of ballistic dynamics because of infinite
diluteness.
Here, the interesting questions pertain to the form of $n_k^{\infty}$, the statistics of emergent non-interacting particles, and
the question whether the asymptotic properties are controlled by very few generic integrals of motion such as energy per particle, or a larger set of conserved quantities,
as is often the case in the thermalization of integrable one-dimensional systems.\cite{rigol07}

An unambiguous example of ballistic dynamics in the sudden expansion {\it at all times} is the case of hard-core bosons in one dimension. Here the exact mapping to non-interacting
fermions leads to ineffectiveness of interactions in causing diffusion, and the gas expands exactly like a gas of non-interacting fermions as far as density profiles and expansion
velocities are concerned.
This was experimentally observed in Ref.~\onlinecite{ronzheimer13}, constituting a clean realization of ballistic dynamics protected through non-trivial
conservation laws in a strongly interacting integrable model. This model has ballistic transport properties according to the strict linear response definition, i.e.,
\begin{equation} \label{sigma_ballistic}
\mbox{Re}\,\sigma(\omega) = 2\pi D(T) \delta(\omega)\,.
\end{equation}
As a consequence, the diffusion constant related to the dc-conductivity through the Einstein relation $\mathcal{D} =\sigma_{\rm dc}/\chi$ diverges ($\chi$ is the static susceptibility).
In the trivial case of hard-core bosons at density $n=1$,
$\sigma(\omega)$ vanishes entirely, but in the sudden expansion, the density will drop below one, such that the transport coefficients 
for that regime will be probed (i.e., $D=D(n)$ and $\mathcal{D}=\mathcal{D}(n)$ for $n<1$).

The Bose-Hubbard model,  by contrast, is non-integrable for $0<U/J<\infty$ and therefore should have diffusive
transport properties at finite temperatures in the sense of $D=0$ and $\sigma_{\rm dc}<\infty$. The same applies to hard-core
bosons on a ladder, equivalent to an $XX$ spin-1/2 model. In fact, existing studies of spin transport for spin ladders indicate
(i) the absence of a ballistic contribution,\cite{heidrichmeisner03} (ii) diffusive spreading of density perturbations with a finite background density~\cite{langer09,karrasch-unpub}
(for examples of ballistic dynamics on certain spin ladders, see the recent work by \v Znidari\v c in Ref.~\onlinecite{znidaric13}).

An interesting question is therefore whether diffusive dynamics can be seen  in the sudden expansion at intermediate time scales,
before the asymptotic ballistic regime is reached.
This then becomes not only a qualitative, but also a quantitative question, since one expects a diffusive dynamics to become visible only on length scales large compared to the mean-free path between collisions. It is therefore entirely possible that even for a non-integrable model with finite diffusion constants, one may still observe only a ballistic dynamics in the sudden expansion, if the diffusion constants are so large that the mean-free path becomes of the order of the  cloud size.

Our numerical results in Sec.~\ref{subsec:gs} indeed show no deviation from ballistic dynamics for the expansion from the {\it ground state} of bosonic Mott insulators in 1D. 
The picture described in the previous paragraph provides a possible explanation, consistent with the data. 
By contrast, for the expansion from a product state of local Fock states, as realized in Ronzheimer et al., Ref.~\onlinecite{ronzheimer13},
higher-energy scales are probed, where presumably the diffusion constants of the Bose-Hubbard model in the non-integrable regime are much smaller than at low temperatures,
and the observed dynamics at intermediate $U/J$ is different from the $U/J=0$ and $U/J=\infty$ cases.

Our criteria for ballistic dynamics in the sudden expansion are therefore:
(i) $R(t)\propto t$ as a necessary condition (excluding short transient times),
(ii) identical density profiles and expansion velocities when comparing the interacting gas to a non-interacting reference system.
In this work we will compare the density dynamics at finite $U/J$ to the integrable $U/J=\infty$ case.
Alternatively, one can construct fictitious non-interacting reference systems that have the same energy per particle as the interacting gas and ask whether both systems exhibit identical density dynamics.
This is the case for the expansion of {\it fermionic} Mott insulators.~\cite{langer12}
%We will not pursue this strategy in this work but will defer it to a forthcoming publication.~\cite{bolech-unpub}

Indications of the absence of strictly ballistic dynamics at finite expansion times before the gas becomes infinitely dilute
are therefore:
(i) Significant deviations from  $R(t)\propto t$,
(ii) a slower expansion velocity (measured through $v_\sub{r}$ and $v_\sub{c}$),
(iii) the formation of a slowly expanding  high-density core. Point (iii) is motivated by two results:
First, a solution of the non-linear diffusion equation for the sudden expansion of fermions in 2D, which produces a high-density core plus fast ballistic wings, is in agreement with the experimental
results from Ref.~\onlinecite{schneider12}.
Second, analytical and numerical simulations for one-dimensional systems typically find a splitting of the cloud into left- and right-moving portions in the ballistic regime of Mott insulators,~\cite{polini07,langer09,langer11,langer12} but slowly expanding high-density cores in the diffusive case.~\cite{polini07,karrasch-unpub}
All these phenomena are present in the experimental data~\cite{ronzheimer13} for bosons in 1D at intermediate $U/J$ in combination with the interaction quench realized in that experiment,
in agreement with DMRG data for the same initial conditions.~\cite{ronzheimer13}
We shall see here that this behavior also emerges for hard-core bosons on two-leg ladders.

%%%%%%%%%%%%%%%%%%%%%%%%%%%%%%%%%%%%%%%%%%%%%%%%%%%%%%%%%%%%%%%

\section{Expansion on a chain} \label{sec:1D}

\subsection{Expansion of hard-core bosons} \label{subsec:HCB}

\begin{figure}[t]
\includegraphics[width=0.99\columnwidth,clip]{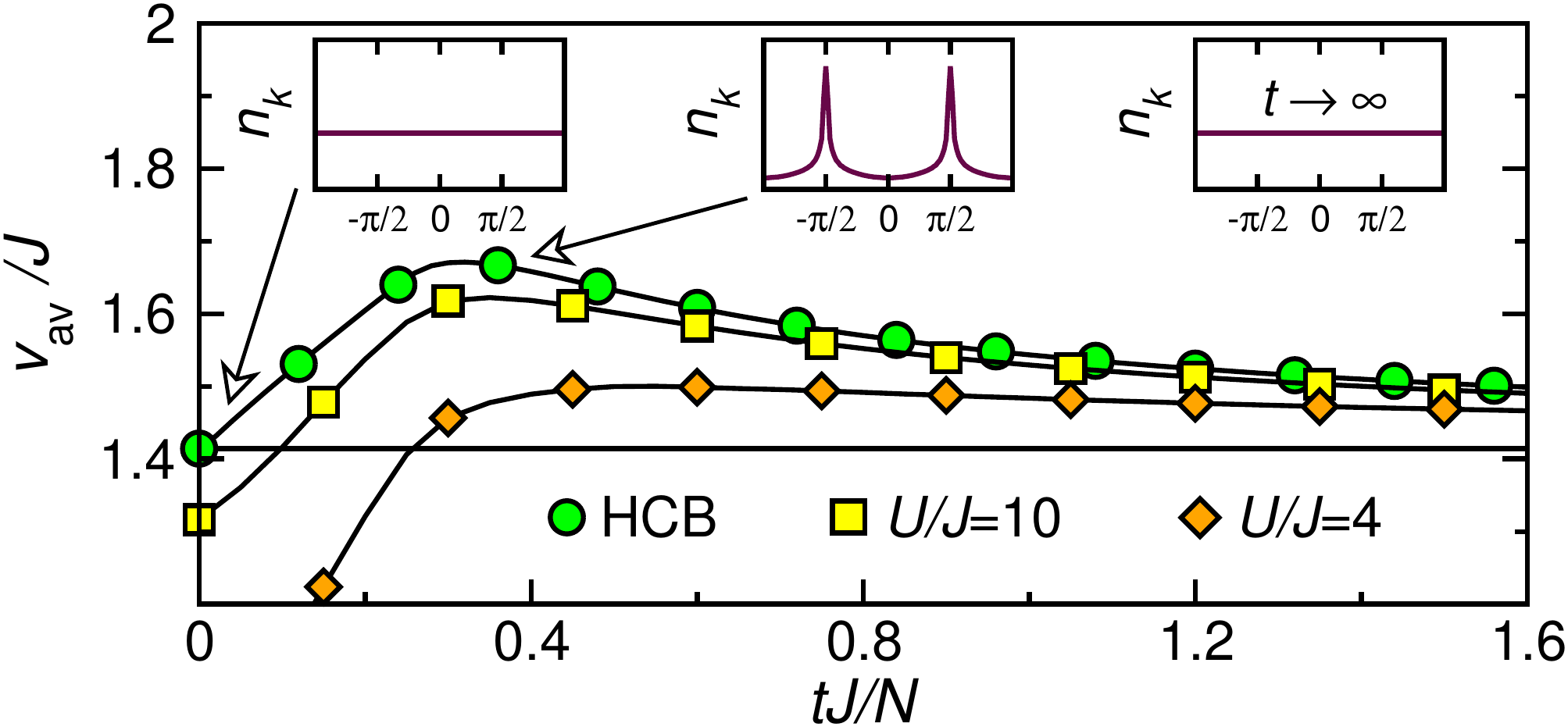}
\caption{(Color online)
{\it
Single chain:
Average velocity of interacting bosons and non-interacting spinless fermions.}
Circles: $v_\sub{av}(t)/J$ for hard-core bosons (HCB) with $N=50$.
Squares and diamonds: $v_\sub{av}(t)/J$ at $U/J=10$ and $4$ using DMRG with $N=10$ particles.
Insets: Corresponding  $n_k(t)$ of hard-core bosons.
Horizontal line: $v_\sub{av}^f(t)/J=\sqrt{2}$ for non-interacting fermions.
}
\label{fig2}
\end{figure}

In order to illustrate the rich phenomenology of this non-equilibrium problem, we first discuss the expansion of hard-core bosons.
They are described by the Bose-Hubbard model with infinitely strong on-site repulsion, i.e., $U/J=\infty$ in Eq.~(\ref{bh_def}).
Contributions from interaction energy are avoided by opposing the condition $(b_i^\dagger)^2=0$, which defines hard-core bosons,  and this  constitutes 
the only integrable point of the 1D Bose-Hubbard model besides $U/J=0$.
At unit filling, the ground state of hard-core bosons in a box is a product of local Fock states
\begin{equation} \label{def_fock}
|\phi_\sub{\tiny Fock}\rangle=\prod_i b_i^\dagger | \emptyset \rangle,
\end{equation}
where $i$ runs over sites within the box.
This state has been realized in a recent experiment,~\cite{ronzheimer13}
which demonstrated that the dynamics of hard-core bosons on a 1D lattice is ballistic.

The origin of ballistic dynamics of 1D hard-core bosons (HCB) stems from the mapping onto non-interacting spinless fermions,~\cite{cazalilla11}
\begin{equation}
H_\sub{HCB}=\sum_k \varepsilon_k n_k^f,
\end{equation}
where $\varepsilon_k = -2J \cos{k}$.
The occupations of fermionic momenta $n_k^f$ are conserved quantities and, as a consequence, the particle current $j = \sum_k v_k n_k^f$ is conserved as well, which indicates ballistic transport.
%In linear response, the conductivity $\mbox{Re}\,\sigma(\omega)$, Eq.~(\ref{sigma_ballistic}), is characterized by a \lev{nonzero} Drude weight $D$ at finite temperatures.
%
%In linear response, the conductivity thus reads $\mbox{Re}\,\sigma(\omega) = 2\pi D \delta(\omega)$ with a nonzero Drude weight $D$ at finite temperatures.
%The diffusion constant is related to the dc-conductivity via the Einstein relation ${\cal D}=\sigma_\sub{\tiny DC}/\chi$ (where $\chi$ represents static susceptibility), and therefore ${\cal D}$ also diverges, as expected for ballistic transport.~\cite{zotos97,heidrichmeisner07}

\begin{figure}[!tb]
\includegraphics[width=0.99\columnwidth,clip]{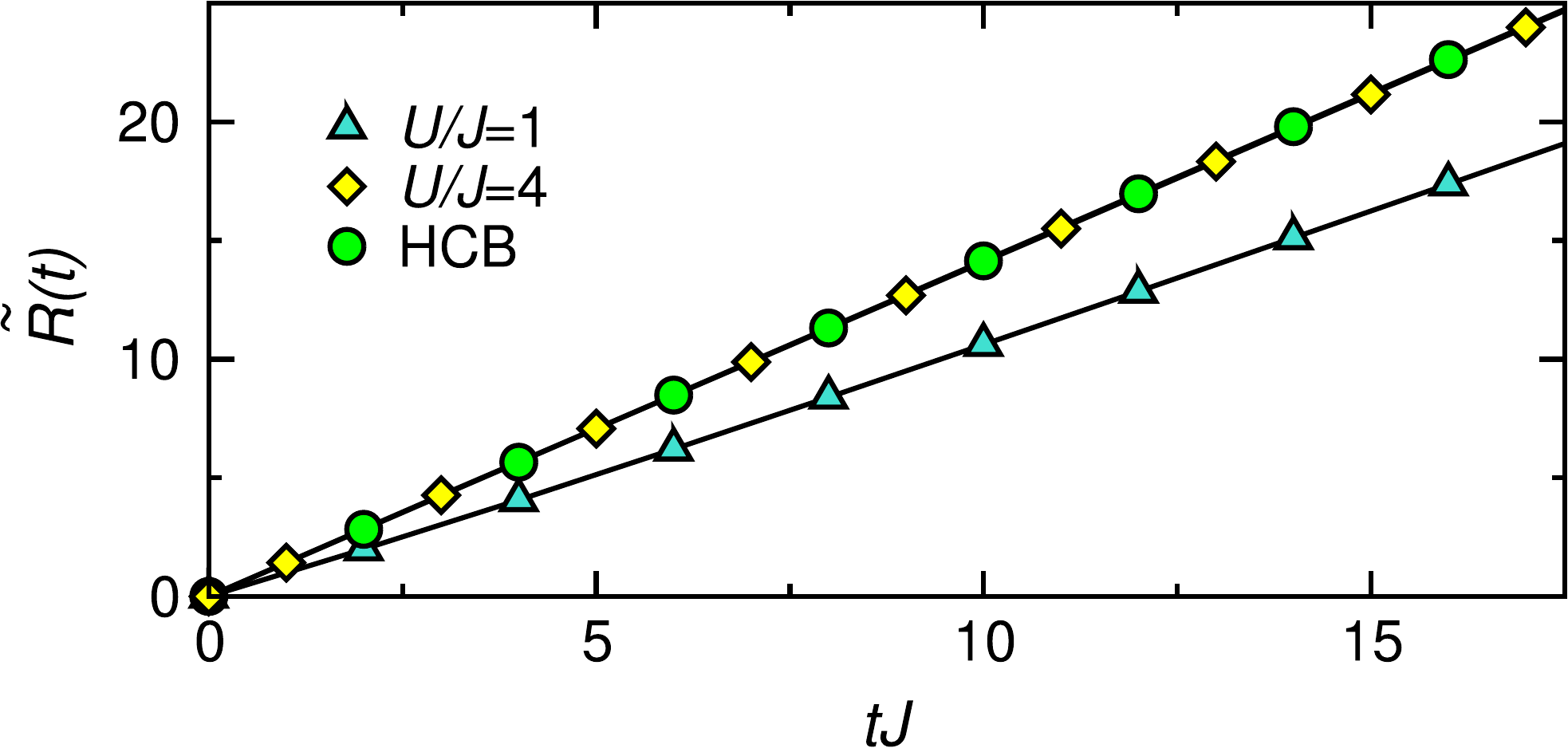}
\caption{(Color online)
{\it
Single chain:
Expansion dynamics of interacting bosons from the ground state.}
Time dependence of the radius $\tilde R(t)$ for $U/J=1$, $4$ and hard-core bosons (HCB).
%, defined in Eq.~(\ref{r_def}).
%We used $(N,N_\sub{max})=(14,5)$ for $U/J=1$ and $(N,N_\sub{max})=(10,4)$ for $U/J=4$.
The expansion of hard-core bosons is described by Eq.~(\ref{rt_hcb}), $\tilde R(t)=\sqrt{2}Jt$.
}\label{figsup1}
\end{figure}

For non-interacting fermions, $\tilde R(t)=v_{\mbox{\footnotesize r}}^f t$,
and expansion velocities defined through the density distribution, $v_{\mbox{\footnotesize r}}^f$, and through the MDF, i.e., $v_{\mbox{\footnotesize av}}^f$, are identical,
 $v_{\mbox{\footnotesize r}}^f=v_{\mbox{\footnotesize av}}^f$.
Both expansion velocities are therefore time-independent quantities fully determined by the initial state.
For the initial density $n=1$, $n_k^f$ is flat and $v_{\mbox{\footnotesize r}}^f=v_\sub{av}^f=\sqrt{2}J$.
This value is plotted as a horizontal line in Fig.~\ref{fig2}.
As a consequence of the mapping to spinless fermions, the density operators for hard-core bosons are identical to the one of fermions,
\begin{equation}
n_i=n_i^f.
\end{equation}
Hard-core bosons hence expand ballistically with
\begin{equation} \label{rt_hcb}
\tilde R(t) = v_{\rm av}^f t,
\end{equation}
as shown in Fig.~\ref{figsup1} (circles).

In contrast to noninteracting spinless fermions, the MDF of hard-core bosons is, however, {\it not} conserved.~\cite{rigol04}
This is illustrated in Fig.~\ref{fig2}, where the non-monotonic behavior of $v_{\mbox{\footnotesize av}}(t)$ [main panel] is a direct consequence of the changes in $n_k(t)$ [insets].
The initial increase of $v_{\mbox{\footnotesize av}}(t)$ reflects the dynamical quasi-condensation at finite momenta $k=\pm \pi/2$, visible in the middle inset.~\cite{rigol04,micheli04,daley05,lancaster10}
The dynamical quasi-condensation can be thought of as a quasi-condensation at $k=0$ in the co-moving frame of the left/right moving hard-core bosons.~\cite{hm08}
At larger times, however, $v_{\mbox{\footnotesize av}}(t)$ decreases again, since in the $t \to \infty$ limit the MDF of hard-core bosons tends to $n^f_k$, the MDF of non-interacting spinless fermions.
This process is  called  dynamical fermionization.~\cite{rigol05, rigol05a, minguzzi05}
The dynamical fermionization results in $v_\sub{av}(t\to\infty)=\sqrt{2}J$ for our initial conditions.
To summarize, the measurement of $v_{\mbox{\footnotesize av}}(t)$ for hard-core bosons is sensitive to the emergence of both dynamical quasi-condensation and fermionization.

%%%%%%%%%%%%%%%%%%%%%%%%%%%%%%%%%%%%%%%%%%%%%%%%%%%%%%%%%%%%%%%

\subsection{Expansion from the ground state} \label{subsec:gs}

Unlike for hard-core bosons, the ground state of interacting bosons and fermions at $U/J<\infty$ is not a simple product state.
The ground state is therefore distinct from the product state, Eq.~(\ref{def_fock}), that was used in recent experiments.
On the other hand, we show in this work that universal features emerge in the expansion when the system in the confining trap is prepared in the ground state.
We therefore organize the discussion along these lines: In this Section, we investigate the expansion from the ground state, while the expansion from product states is investigated in Section~\ref{subsec:product}.

\subsubsection{Bosons} \label{subsubsec:bosons1}

\begin{figure}[!tb]
\includegraphics[width=0.99\columnwidth,clip]{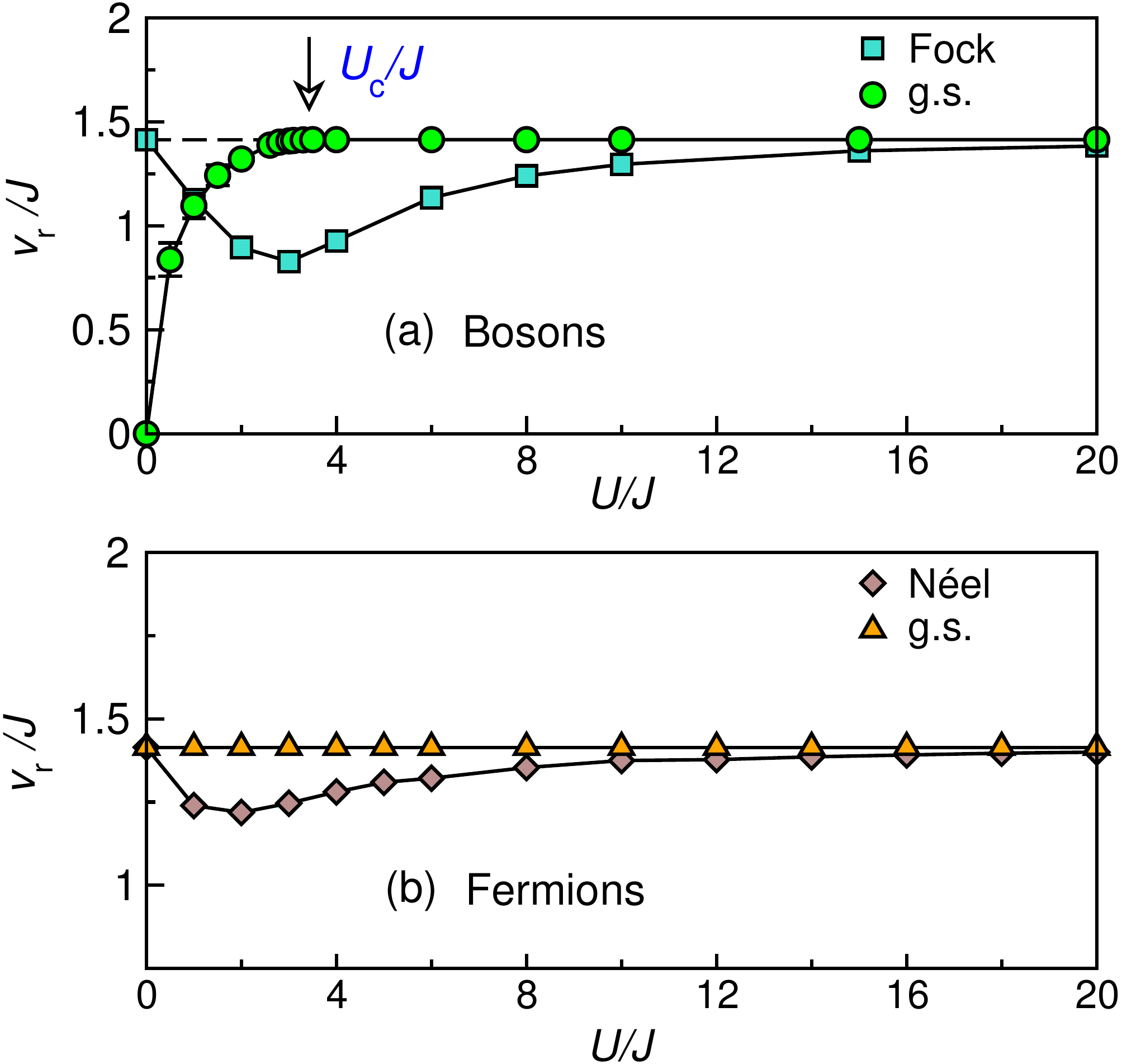}
\caption{(Color online)
{\it
Single chain:
Radial velocity of bosons and fermions.}
(a)
$v_\sub{r}/J$ versus $U/J$ for the Bose-Hubbard model.
Initial states: ground state (g.s.; circles) or $|\phi_\sub{\tiny Fock}\rangle$ (squares, from Ref.~\onlinecite{ronzheimer13}).
(b)
$v_\sub{r}/J$ versus $U/J$ for the Fermi-Hubbard model.
Initial states: ground state (triangles) or  N{\'e}el state $|\phi_\sub{\tiny N}\rangle$ (diamonds).
All data is extrapolated to $N\to\infty$ (see  Appendix~\ref{sec:app1a} for details).
}
\label{fig3}
\end{figure}

\begin{figure*}[!tb]
\includegraphics[width=1.7\columnwidth,clip]{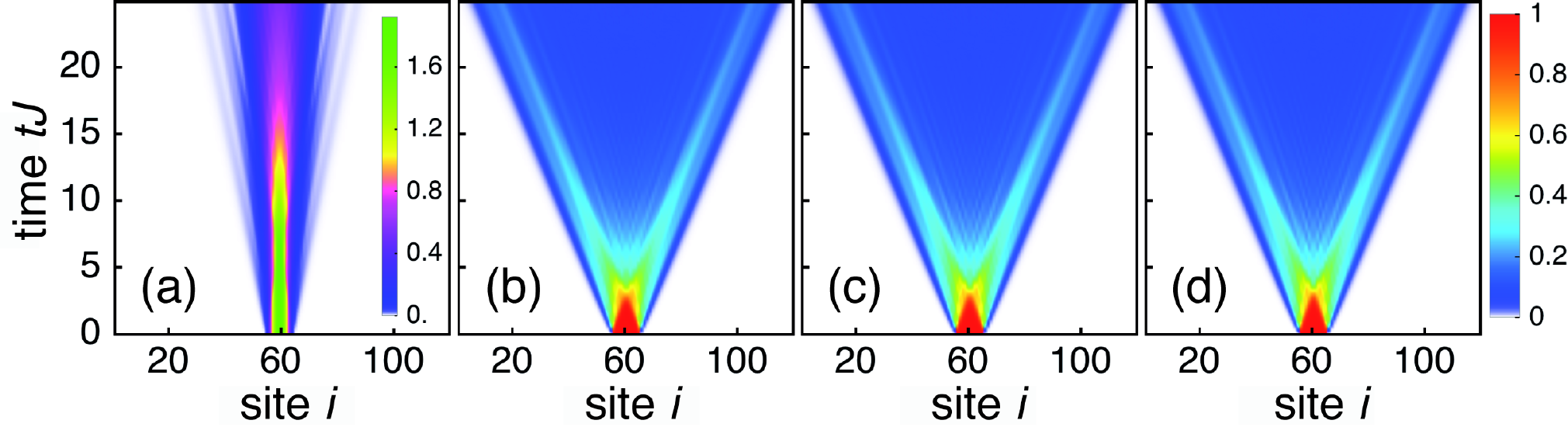}
\caption{(Color online)
{\it
Single chain:
Density profiles of expanding bosons from the ground state.
}
(a)
Expansion from the superfluid phase of the Bose-Hubbard model at $U/J=0$.
(b)-(d)
Expansion from the Mott insulating phase at $U/J=4$, $10$ and hard-core bosons, respectively.
We used $N=10$ particles.
}\label{fig4a}
\end{figure*}

For density $n=1$, there is a quantum phase transition in the 1D Bose-Hubbard model at $U_c/J\sim 3.4$ from the superfluid to the Mott insulator.~\cite{kuhner00}
At $U/J=0$, bosons quasi-condense at $k=0$, leading to $\tilde R(t)=v_rt$ with $v_{\mbox{\footnotesize r}}=v_{\mbox{\footnotesize av}}=0$ for $L_\sub{box}\to \infty$.
Our goal is to investigate the effect of moving away from the integrable points $U/J=0$ and $U/J=\infty$.

The radius $\tilde R(t)$ of the expanding cloud of bosons for different values of $U/J$ is shown in Fig.~\ref{figsup1}.
We see that the radius can be approximated by $\tilde R(t) \propto t$ in a wide time-window, including short times just after the sudden release from the trap.
Such a time dependence of $\tilde R(t)$ is similar to the case of the 1D Fermi-Hubbard model studied in Ref.~\onlinecite{langer12}.
Interestingly, Fig.~\ref{figsup1} reveals that the time-dependence of $\tilde R(t)$ for bosons at $U/J=4$ is virtually identical to the one of hard-core bosons, which expand as $\tilde R(t)=\sqrt{2}J t$.

As a direct consequence of virtually identical radii, we expect that the radial velocity $v_\sub{r}$ should be identical for large enough values of $U/J$.
We estimate the radial velocities $v_\sub{r}$ at $U/J < \infty$ by using the linear fit $\tilde R(t)=v_\sub{r}t$.
The procedure of deducing $v_\sub{r}$ from $\tilde R(t)$ is described in more detail in Appendix~\ref{sec:app1}.
Circles in Fig.~\ref{fig3}(a) show $v_\sub{r}$ as a function of $U/J$ for the expansion from the ground state of the Bose-Hubbard model.
We observe that a specific value of the radial expansion velocity, namely,
\begin{equation}
v_\sub{r}/J=\sqrt{2},
\end{equation}
is characteristic for the entire 1D Mott insulating phase.
This behavior is in clear contrast to that in the superfluid phase, where $v_\sub{r}$ monotonically decreases to zero.
It is remarkable that a quantity measured in a non-equilibrium experiment, namely $v_\sub{r}$, is sensitive to the interaction-driven quantum phase transition in the initial state.
This suggests that because of the ballistic dynamics in the sudden expansion of bosons from the ground state, information about the initial conditions is preserved at asymptotically long times.
It is also interesting to study the dependence of $v_\sub{r}$ on initial densities $n\neq 1$.
Similar to fermions,~\cite{langer12} we find $\tilde R(t) \propto t$ for all $n\leq 1$ and $v_\sub{r}(n)$ depends in a non-monotonic way on the initial density.
We study this dependence in more detail in Appendix~\ref{sec:density}.

Most importantly, it turns out that the identical radii of expanding bosons at moderate and large $U/J$ emerge as a result of a more universal behavior:
After sufficiently long expansion times, the whole {\it density profiles} become indistinguishable from the ones of hard-core bosons.
In Fig.~\ref{fig4a} we plot density profiles of the Bose-Hubbard model at $U/J=0$, $4$, $10$ and hard-core bosons.
For the expansion from the Mott insulating phase, the density profiles look virtually identical at all times, in sharp contrast to the density profile of the superfluid phase.
This indicates that, already for small expansion times, the density dynamics of all expanding Mott insulators is purely ballistic,
since the entire density profiles are identical to the ones of hard-core bosons and therefore to non-interacting fermions.

To quantify this observation, we calculate the deviation of the density profiles at a finite $U/J<\infty$ from those of hard-core bosons (HCB) by defining
\begin{equation}
\xi_{U/J}(t) = \frac{1}{N}  \sum_i \left| \langle n_i(t) \rangle_{U/J} - \langle n_i(t) \rangle_\sub{HCB} \right|. \label{def_xit}
\end{equation}
We find that for Mott insulators, $\xi_{U/J}(t)$ asymptotically decays to zero.
We show $\xi_{U/J}(t)$ in the inset of Fig.~\ref{figsup3}(a) for the Bose-Hubbard model at $U/J=10$.
The small values of $\xi_{U/J}(t)$ are consistent with our observation in Fig.~\ref{fig4a} that the density profiles of bosonic Mott insulators are virtually indistinguishable already at short time.
The absence of strong transient features in the density profiles of expanding Mott insulators is consistent with an extremely weak time-dependence of $v_\sub{r}(t)$, shown in Fig.~\ref{figsup2}(a) in Appendix~\ref{sec:app1}.
$\xi_{U/J}(t)$ reaches its maximum at the similar time when the maximal dynamical quasi-condensation is observed, i.e., $tJ/N \approx 0.3$ (see Fig.~\ref{fig2}).
In the Mott insulating regime of the Bose-Hubbard model, $U>U_c$, the largest deviation of $\xi_{U/J}(t)$ is observed in the vicinity of  $U_c$, but even there, we find typically  $\xi_{U/J}(t) \approx 10^{-2}$ for $U/J=4$ at $tJ=25$.
This result supports our conjecture that the density dynamics of expanding bosonic Mott insulators is governed by the dynamics of non-interacting spinless fermions,
i.e., it cannot be distinguished from that case by measuring only densities or the radius.

Universal features in the expansion can also be detected from observables in momentum space.
Bosons at large but finite $U/J$ share some of the features of hard-core bosons, such as the dynamical quasi-condensation in the transient regime.~\cite{rodriguez06}
An indicator for this behavior is the initial increase  of $v_\sub{av}(t)$, observed in Fig.~\ref{fig2}.
Furthermore, in the asymptotic regime when $v_\sub{av}(t)$ decreases again, our data show that even for $U_c/J<U/J<\infty$,  $v_\sub{av}(t)/J\to \sqrt{2}$ as $t$ increases, very similar to hard-core bosons.
This implies that  the MDF for $t\to \infty$ needs to  be  compatible with this particular value of $v_\sub{av}(t\to\infty)$.
We will further elaborate on this issue in Section~\ref{subsubsec:asymptotic}.

Recent exact results for the Lieb-Liniger model suggest that a dynamical renormalization occurs during the sudden expansion:
The asymptotic dynamics of a repulsively interacting Bose gas is, for any interaction strength, 
governed by the behavior of non-interacting fermions.~\cite{iyer12, iyer13}
Considering that the density decreases as the gas expands, it is conceivable that the dynamics for the expansion from  ground states of the Bose-Hubbard model can be described by the Lieb-Liniger model at long times.
A formal proof of this interpretation is left for future research. 

%%%%%%%%%%%%%%%%%%%%%%%%%%%%%%%%%%%%%%%%%%%%%%%%%%%%%%%%%%%%%%%
\subsubsection{Fermions} \label{subsubsec:fermions1}

\begin{figure}[!t]
\includegraphics[width=0.99\columnwidth,clip]{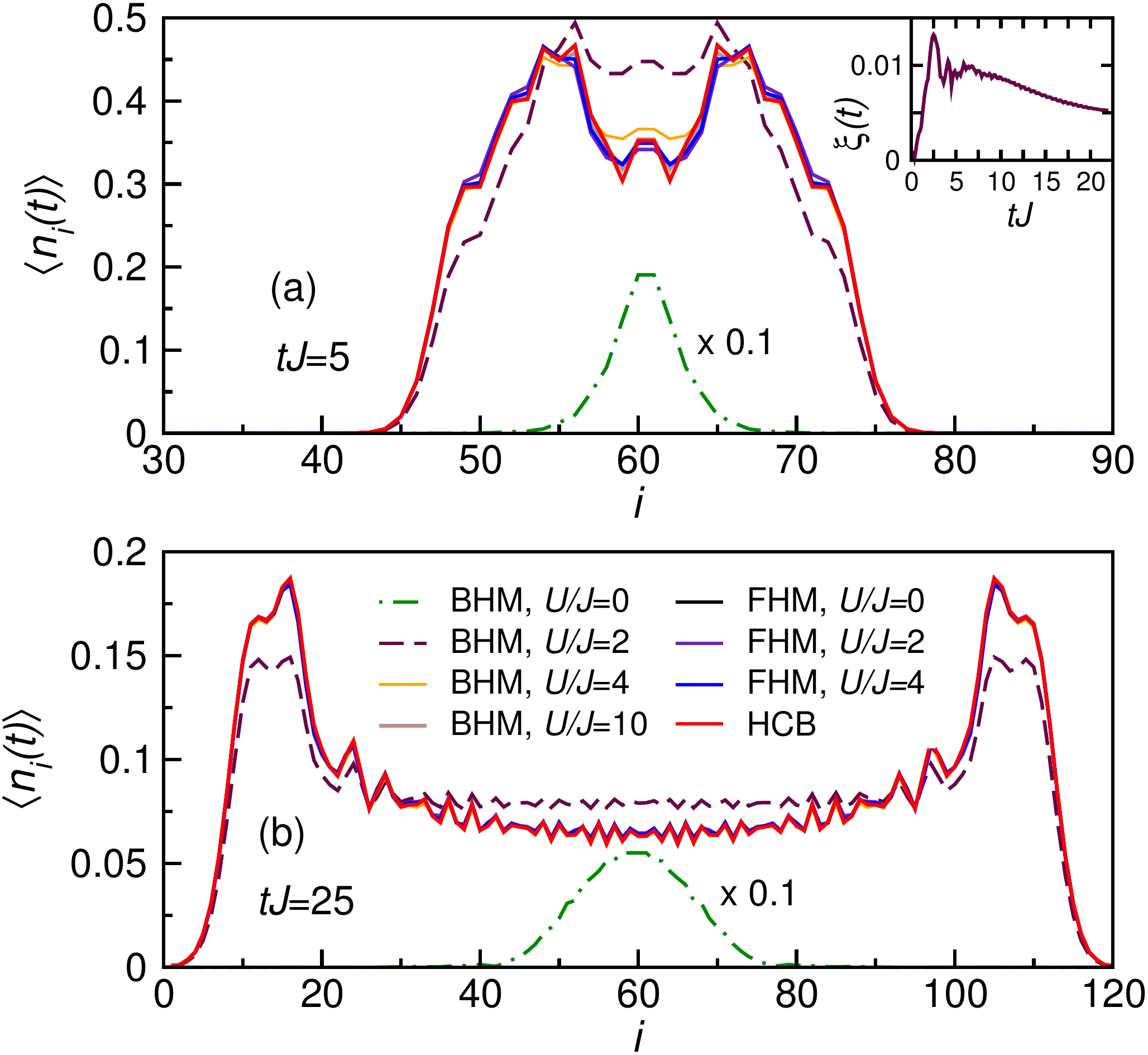}
\caption{(Color online)
{\it
Single chain:
Density profiles for the expansion of bosons and fermions from the ground state.}
Results for $N=10$ particles at different interaction strengths for the Bose-Hubbard model (BHM), Fermi-Hubbard model (FHM) and hard-core bosons (HCB).
Main panel:
Density profiles $\langle n_i \rangle$ at (a) $tJ=5$ and (b) $tJ=25$.
Solid curves in (b) are virtually indistinguishable.
Curves for the BHM at $U/J=0$ are multiplied by a factor of $0.1$
(in the superfluid phase at $U/J=0$, $\langle n_i \rangle$ is not constant but drops to zero at the open boundary, hence $\langle n_i \rangle > 1$ in the center).
The initial trapping potential, Eq.~(\ref{def_boxtrap}), was applied for $i_a=55$ and $i_b=66$.
Inset: $\xi(t)$, defined in Eq.~(\ref{def_xit}), for the Bose-Hubbard model at $U/J=10$.
}\label{figsup3}
\end{figure}

It is very instructive to compare the results for interacting bosons to the Fermi-Hubbard model.
The ground state of the 1D Fermi-Hubbard model is a Mott insulator for any $U/J>0$.
Several aspects of the expansion dynamics of fermions have been studied in Refs.~\onlinecite{hm08,hm09,kajala11,karlsson11,langer12,kessler13}.
Here we focus on density profiles and the comparison to interacting bosons, with a particular interest in the description of asymptotic properties.

We compare the radial velocity $v_\sub{r}$ for the expansion from the ground state of the Fermi-Hubbard model with the results from the Bose-Hubbard model.
It has been shown in Ref.~\onlinecite{langer12} that fermionic Mott insulators expand as $\tilde R(t) = v_\sub{r} t$ with $v_\sub{r}/J=\sqrt{2}$, irrespective of $U/J$.
These results are shown in Fig.~\ref{fig3}(b) (triangles).
Remarkably, both fermionic and bosonic Mott insulators therefore expand with the same expansion velocity.
Similar to the case of bosons, we show in the following that this is a direct consequence of identical density profiles.

In Fig.~\ref{figsup3} we show density profiles $\langle n_i(t) \rangle$ at different interaction strengths for the Bose-Hubbard model, Fermi-Hubbard model and hard-core bosons.
While the density profiles for spinless fermions and hard-core bosons must be identical at all times, our results additionally unveal that they are in fact virtually identical for all fermionic and bosonic Mott insulators, and therefore cannot be distinguished from non-interacting spinless fermions.
Our calculations (not shown) confirm that $\xi_{U/J}(t)$, Eq.~(\ref{def_xit}), asymptotically decays to zero for fermionic  Mott insulators, in analogy with bosonic Mott insulators.
In addition, density profiles of fermionic and bosonic Mott insulators are very similar already at short time, see Fig.~\ref{figsup3}(a) for $tJ=5$.

%%%%%%%%%%%%%%%%%%%%%%%%%%%%%%%%%%%%%%%%%%%%%%%%%%%%%%%%%%%%%%%
\subsubsection{Universal features in the asymptotic regime} \label{subsubsec:asymptotic}

As a main result of the previous investigation, we observed  that the density profiles of bosonic and fermionic Mott insulators become virtually identical during the expansion, hence they become indistinguishable from the one of non-interacting spinless fermions.
Here we provide a general explanation for this behavior in terms of the velocity and momentum distribution functions.
%we conjecture that the asymptotic {\it velocity} distribution $n_{v_k}$ is independent of $v_k$ and identical for all Mott insulators in 1D.

We introduce the velocity distribution function $n_{v_k}$ as a measure of the occupation of states with the group velocity $v_k = 2J \sin{k}$ ($0 \leq |v_k| \leq 2J$), normalized to the total-particle number
\begin{equation}
\sum_k n_k = \sum_{v_k} n_{v_k} = N.
\end{equation}
Since the momentum distribution function of non-interacting fermions does not change during the expansion, their velocity distribution $n_{v_k}$ remains flat for all times for the initial conditions considered here.
For hard-core bosons, dynamical fermionization occurs at asymptotically large times,~\cite{rigol05} hence their velocity distribution $n_{v_k}$ becomes flat when $tJ \to \infty$
(cf. insets in Fig~\ref{fig2}).

For systems with finite $U/J < \infty$, it is not a priori clear how the velocity distribution evolves during the expansion.
Since at long times the system becomes very dilute, the majority of the energy will be converted into kinetic energy and the system becomes effectively non-interacting.
As a consequence, the observation of identical density profiles in this limit requires that in all cases the asymptotic velocity distribution should equal that of non-interacting fermions, which possess a flat $n_{v_k}$.
We therefore conclude that the asymptotic velocity distribution $n_{v_k}$ is independent of $v_k$ and identical for all Mott insulators in 1D,
\begin{equation}
n_{v_k}(t\to\infty) = {\rm const.} \label{nvk_asy}
\end{equation}

\begin{figure}[!t]
\includegraphics[width=0.99\columnwidth,clip]{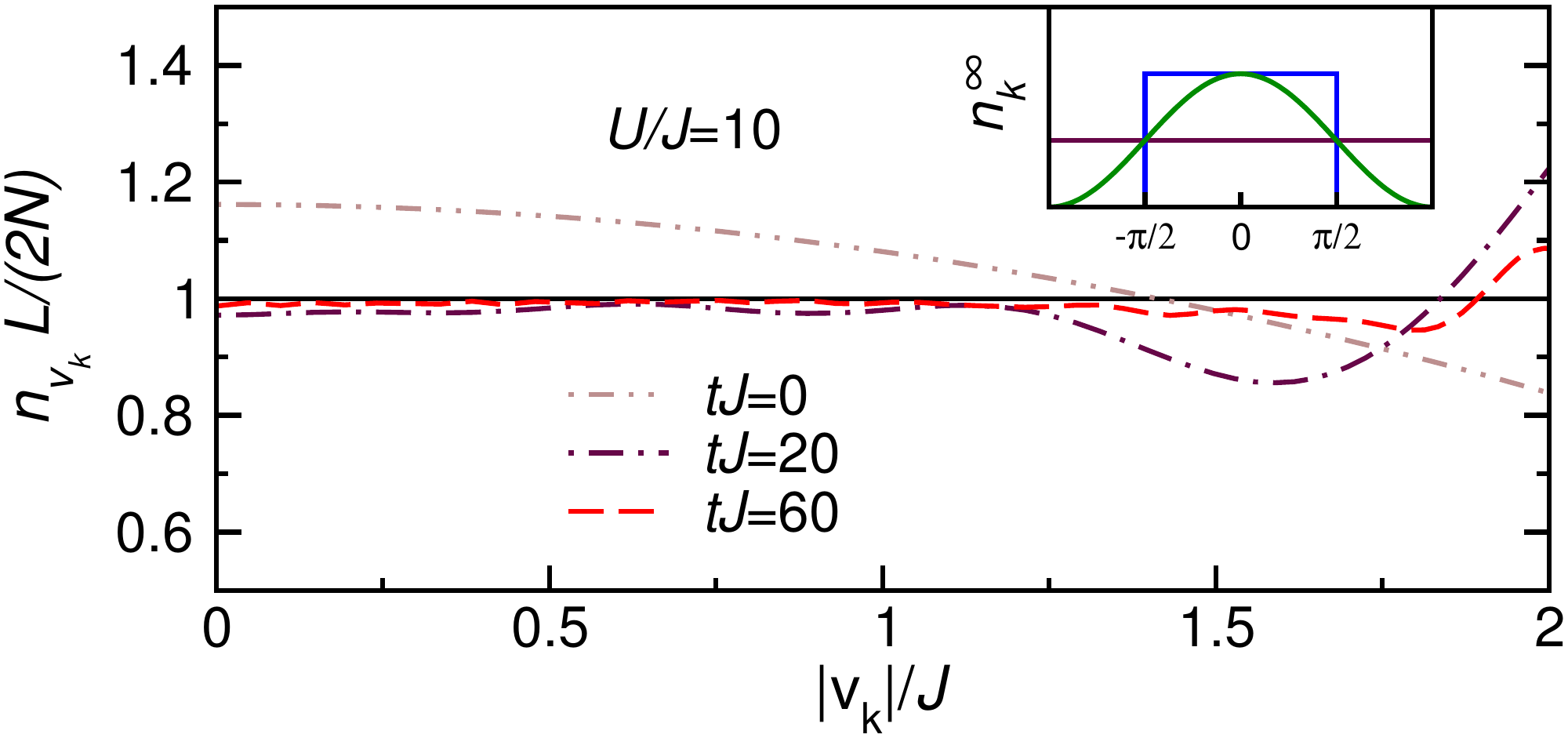}
\caption{(Color online)
{\it
Single chain:
Velocity distribution function $n_{v_k}$ for interacting bosons, expanding from the ground state.}
Main panel: normalized $n_{v_k} L/(2N)$  for $N=4$ and  $U/J=10$ at different times.
Horizontal solid line shows $n_{v_k}$ for non-interacting fermions at any time and hard-core bosons at time $tJ\to \infty$.
The inset depicts possible momentum distribution functions $n_k^{\infty}$  which are particle-hole symmetric and produce a flat $n_{v_k}$.
}\label{figsup3b}
\end{figure}

We check this property with DMRG for a small number of particles and very long times.
Results for $N=4$ and $U/J=10$ are shown in Fig.~\ref{figsup3b} for times up to $tJ=60$.
For increasing expansion times, $n_{v_k}$ indeed converges to a flat function, and thereby provides a numerical confirmation of  Eq.~(\ref{nvk_asy}).

A flat asymptotic velocity distribution $n_{v_k}$ implies that the asymptotic momentum distribution function $n_k^{\infty}$ is particle-hole symmetric.
A particle-hole symmetric MDF is characterized by
\begin{equation}
n_{\pm \frac{\pi}{2}+ \delta k}^{}  + n_{\pm \frac{\pi}{2} - \delta k}^{} = {\rm const.}
\end{equation}
for all $|\delta k| \leq \pi/2$.
In the inset of Fig.~\ref{figsup3b} we show possible $n_k^{\infty}$, all of which are particle-hole symmetric.
In addition, any particle-hole symmetric MDF automatically results in the observed universal asymptotic behavior of $v_{\rm av}(t)$ in Fig.~\ref{fig2}, given by
\begin{equation}
v_\sub{av}(t \to \infty)/J=\sqrt{2}
\end{equation}
for all Mott insulators expanding from the ground state in 1D.

The observation of a particle-hole symmetric asymptotic quasi-momentum distribution function suggests a description
of this regime 
 using a {\it fermionic} non-interacting Hamiltonian. This can be further corroborated by asking whether
the gas at infinite expansion times can be described by a standard equilibrium distribution function. Indeed, it turns
out that this is the case, following the approach of Ref.~\onlinecite{langer12}: By selecting a temperature in order to match the energy per particle and the chemical potential
to account for the fact that the fictitious gas should have originated from the same initial condition of one particle
per site, one realizes that $n_k(t)$,  computed numerically for few particles  
for as long times as possible, can be very well approximated by a Fermi-Dirac distribution (results not shown here).
In order to match the energy per particle of the interacting gas, one needs to choose the fictitious non-interacting system to be a  two-component
Fermi gas.  
The emergence of a typical equilibrium fermionic property in the asymptotic momentum distribution, namely particle-hole symmetry, can be viewed
as a generalization of the dynamical fermionization that was discussed for hard-core bosons and the Tonks-Girardeau gas.\cite{rigol05,minguzzi05}

\subsection{Expansion from product states} \label{subsec:product}

In the previous Section, we investigated the sudden expansion of bosons and fermions from their ground state.
Here we complement these results by investigating the sudden expansion from initial product states of fully localized particles, as studied in a recent experimental and numerical study.~\cite{ronzheimer13}

\subsubsection{Bosons} \label{subsubsec:bosons2}

\begin{figure}[!t]
\includegraphics[width=0.99\columnwidth,clip]{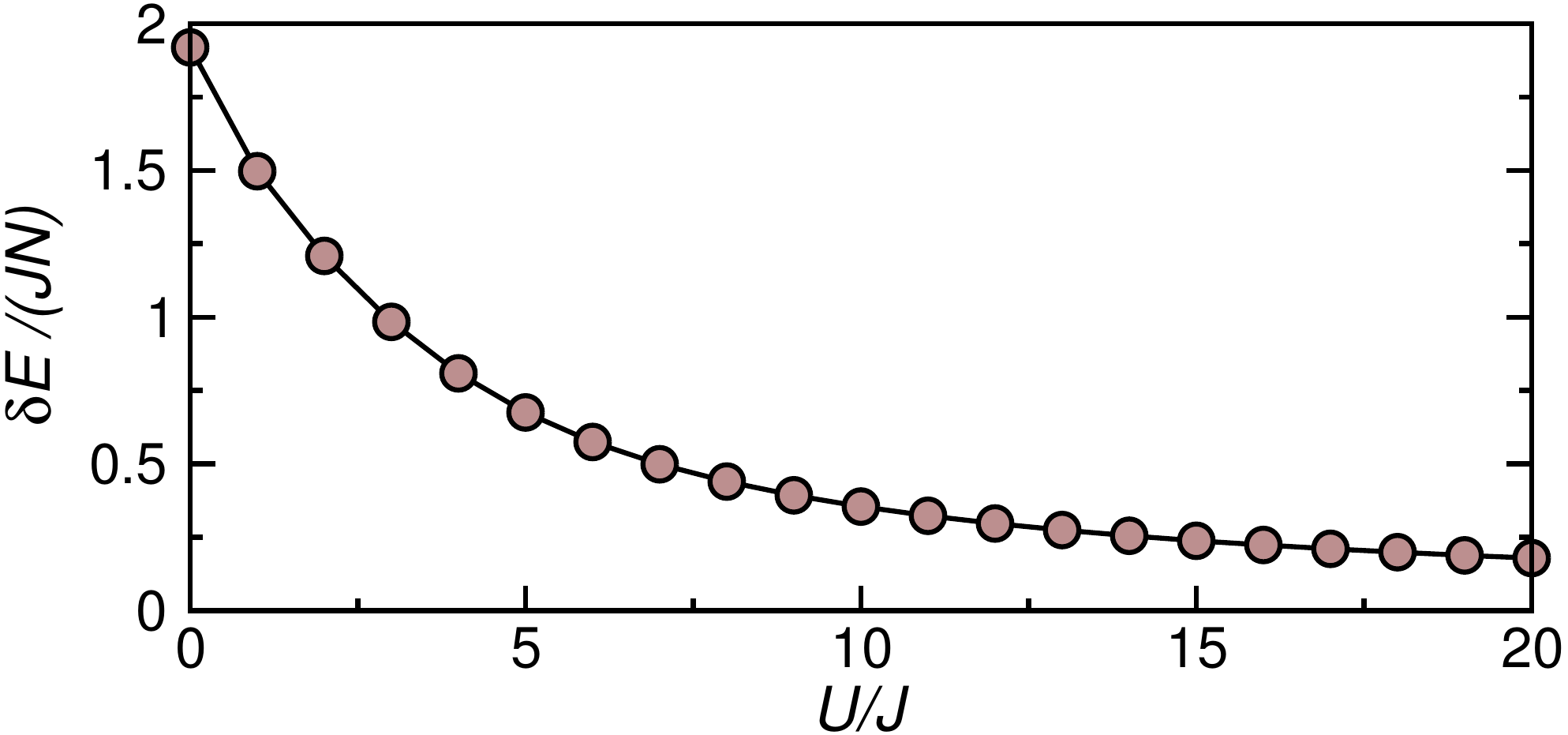}
\caption{(Color online)
{\it
Single chain: Excess energy $\delta E/(JN)$ for  the interaction quench from $U/J=\infty$ to a finite $U/J$ in the Bose-Hubbard model.}
We define $\delta E=E_{\rm exp}-E_0$, where $E_{\mathrm{exp}}$ is computed in the initial state $|\phi_{\rm Fock}\rangle$ and $E_0$ in the ground state.
We used $N=10$ particles.
}\label{figsup4}
\end{figure}

In the aforementioned experimental work, the initial state was $|\phi_\sub{\tiny Fock}\rangle$ defined in Eq.~(\ref{def_fock}), and an additional quench from infinite to finite $U/J$ was performed simultaneously with the removal of the external potential.
Results taken from Ref.~\onlinecite{ronzheimer13} are shown as squares in Fig.~\ref{fig3}(a).
In this case, $v_\sub{r}/J = \sqrt{2}$ at large $U/J$ and at $U/J=0$, with a minimum at $U/J\sim 3\rm{-}4$.
The expansion velocities for the two different  initial states, i.e., the ground state versus $|\phi_\sub{\tiny Fock}\rangle$, exhibit strong differences both in the vicinity of $U\sim U_c$ and at $U/J = 0$.
In the case of $U/J=0$, the difference in expansion velocity is directly related to the different initial $n_k$ through Eqs.~(\ref{vav_def}) and~(\ref{vr_vav_noninteracting}).

One possible measure to quantify the difference between $|\phi_\sub{\tiny Fock}\rangle$  and the ground state of the trapped gas is the excess energy compared to the ground state.
We define the excess energy as $\delta E=E_{\rm exp}-E_0$, where $E_{\mathrm{exp}}$ is the total energy during the expansion and $E_0$ is the ground-state energy for the same $U/J$ and $N$ in the initial trap.
The expansion from a product of local Fock states $|\phi_\sub{\tiny Fock}\rangle$ is characterized by $E_{\mathrm{exp}}=0$, hence $\delta E=|E_0|$.
In Fig.~\ref{figsup4} we plot $\delta E/(JN)$, which is a monotonic function of $U/J$.
In the limit $U/J=0$, $\delta E/N \to 2J$ as $N \to \infty$, while in the opposite limit $U/J \to \infty$, the ground state approaches $|\phi_\sub{\tiny Fock}\rangle$ and so $\delta E \to 0$.
Therefore, in the latter limit the ground state has a large overlap with $|\phi_\sub{\tiny Fock}\rangle$, hence the interaction quench has little effect on the expansion dynamics.

The behavior at $U\sim U_c$ is counter-intuitive: the gas with higher energy per particle, i.e., the expansion from $|\phi_{\rm Fock}\rangle$, expands
{\it slower}.
This behavior can be understood by assuming that in a hotter gas ($\delta E >0$) scattering processes become more efficient in slowing down the expansion. 
This is consistent with the conjecture of Ref.~\onlinecite{ronzheimer13} that for $\delta E > 0$ and $U\sim U_c$, there is diffusion in the Bose-Hubbard model, which, however, still requires more theoretical analysis.

The experimental and numerical results of Ref.~\onlinecite{ronzheimer13} further showed that by preparing the gas at higher energies through the interaction quench, noticeable deviations from ballistic dynamics set in:
$\tilde R(t)\not\propto t$, $v_r <\sqrt{2}J$ (where $v_\sub{r}$ represents radial velocities extracted at long times), and density profiles deviate noticeably from the one of non-interacting particles originating from the same initial state.
We suggest that these differences between the sudden expansion from the ground state versus product states are related to the quantitative values of the diffusion constant, as a function of energy or temperature
(of course, one also has to account for the density dependence).
This is consistent with the qualitative picture discussed in Sec.~\ref{sec:ballistic}.
A quantitative calculation of diffusion constants for the Bose-Hubbard model therefore seems important and is left for future work.

\subsubsection{Fermions} \label{subsubsec:fermions2}

One may conjecture that the dynamics of the Fermi-Hubbard model at intermediate $U/J$ should differ from those of the Bose-Hubbard model for the combination of interaction quench and trap removal with $\delta E>0$, since the former model is integrable and the latter is not.
However, this expectation is not supported by our analysis of expansion velocities when we calculate the expansion dynamics of an initial N\'eel state
\begin{equation}
|\phi_\sub{\tiny N}\rangle = \prod_{i_{\rm odd}} c_{i\uparrow}^\dagger c_{i+1\downarrow}^\dagger | \emptyset \rangle
\end{equation}
in the 1D Fermi-Hubbard model.
This state corresponds to $|\phi_\sub{\tiny Fock} \rangle$ used in the case of the Bose-Hubbard model, and is a ground state only for $U/J =\infty$.
Interestingly, the dependence of $v_\sub{r}/J$ on the interaction strength [diamonds in Fig.~\ref{fig3}(b)] shares striking similarities with that of the 1D Bose-Hubbard model.
Since the 1D Fermi-Hubbard model is integrable for any $U/J$, these results indicate that integrability {\it per se} does not imply a ballistic {\it and} fast expansion.
It is in fact known that not all integrable models have ballistic transport properties at $T>0$ in linear response.~\cite{znidaric11,steinigeweg11,karrasch-unpub}
In addition, the comparison of expansion from ground states versus product states (both for fermions and bosons) demonstrates that the universal asymptotic features that emerge for the expansion from the Mott insulating regime, require $\delta E$ to be close to zero.

The dynamics of fermions is much richer, since the N\'eel state $|\phi_\sub{\tiny N}\rangle$ is only one of many possible degenerate ground states of the Fermi-Hubbard model at $U/J=\infty$.
For the case of two-component fermions and the initial density $n=N/L_\sub{box}=1$, there are $\binom{N}{N_\uparrow}$ possible configurations of the local Fock states with one particle per site, which 
are all degenerate.
We characterize these by the influence of the number $W$ of spin domain walls in the initial states.
An arbitrary product of local Fock states $|\phi_\sub{\tiny W}^{(j)} \rangle$ has $W$ domain walls if
\begin{equation} \label{psiwj_def}
\sum_{i=1}^{L_\sub{\tiny box}-1} \sum_{\sigma\in\{ \uparrow, \downarrow \}} \langle \phi_\sub{\tiny W}^{(j)} | c_{i,-\sigma}^\dagger c_{i,-\sigma} c_{i+1,\sigma}^\dagger c_{i+1,\sigma} | \phi_\sub{\tiny W}^{(j)} \rangle = W,
\end{equation}
where $j$ runs over all states having $W$ domain walls for a fixed density, magnetization and value of $L_\sub{box}$.
To decrease boundary effects we average our results over all states having the same number of domain walls.
%construct the initial states as a superposition of all states having $W$ domain walls
%\begin{equation} \label{psiw_def}
%|\phi_\sub{\tiny W}\rangle = \frac{1}{\sqrt{N_\sub{\tiny W}}} \sum_{j=1}^{N_\sub{\tiny W}} |\phi_\sub{\tiny W}^{(j)} \rangle,
%\end{equation}
%
%where $N_\sub{\tiny W}$ denotes the total number of states having $W$ domain walls for a fixed density and value of $L_\sub{box}$.
For $W_\sub{max}=L_\sub{box}-1$, the initial state is the N\'eel state $|\phi_\sub{\tiny N}\rangle$,
while for $W_\sub{min}=1$, the initial state contains, e.g., a sequence of spin ups followed by a sequence of spin downs.
We investigate systems with a fixed particle number $N=10$ and magnetization zero $(N_{\uparrow}=N_{\downarrow}=5)$.

\begin{figure}[t]
\includegraphics[width=0.99\columnwidth,clip]{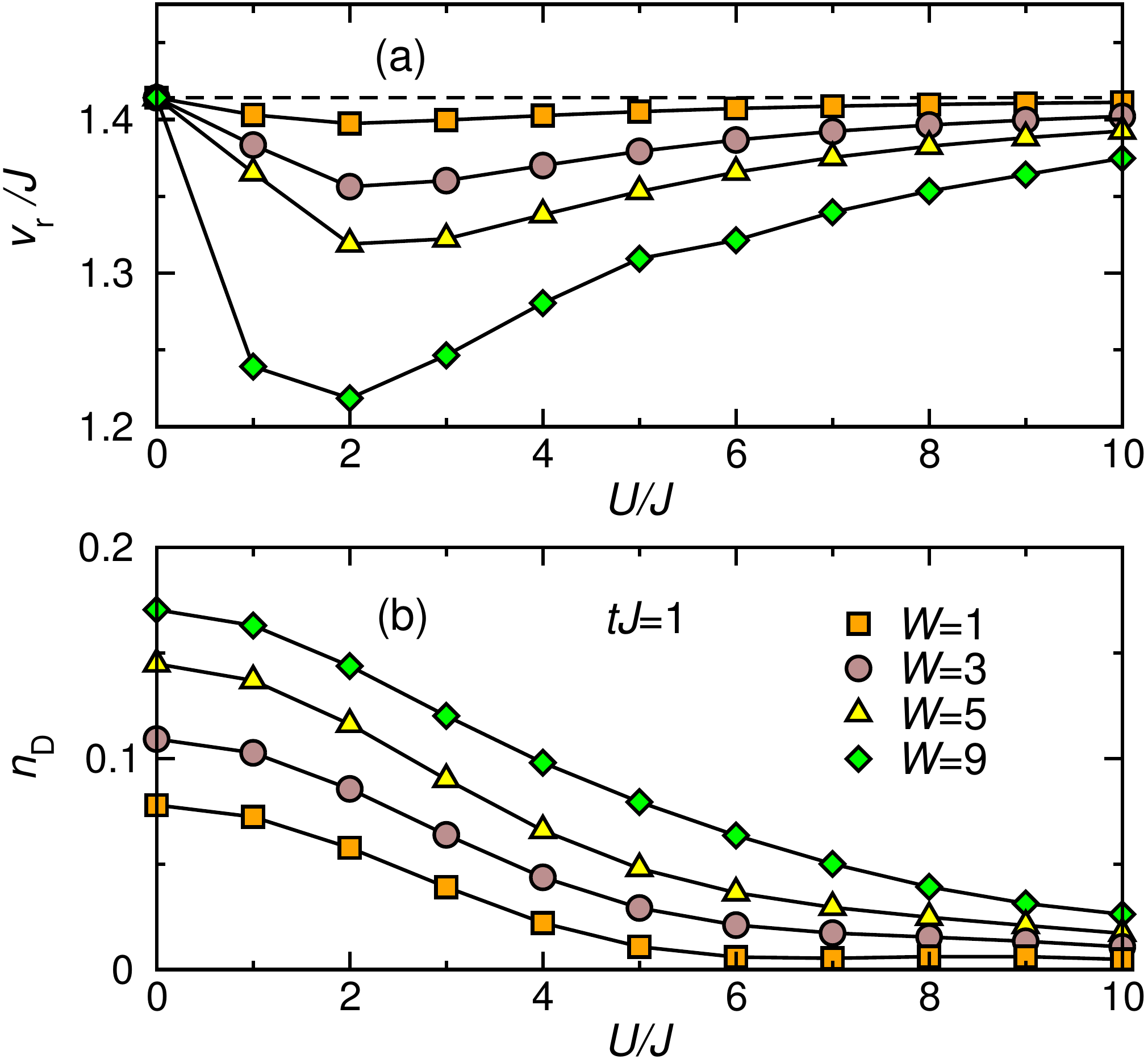}
\caption{(Color online)
{\it
Single chain:
Expansion dynamics of interacting fermions from the states $|\phi_\sub{\tiny W}^{(j)}\rangle$, see Eqs.~(\ref{psiwj_def}).}
The initial state $|\phi_\sub{\tiny W}^{(j)}\rangle$ contains $W$ spin domain walls.
We used $N=10$ particles and zero magnetization $N_{\uparrow}=N_{\downarrow}=5$.
Results are averaged over all states having the same number of domain walls $W$.
(a)
Radial velocities $v_\sub{r}/J$ vs $U/J$;
(b)
Number of double occupancies $n_\sub{{\tiny D}}(t)$ vs $U/J$ at time $tJ=1$, see also Eq.~(\ref{def_double}).
}\label{figsup6}
\end{figure}

Radial velocities $v_\sub{r}$ for the different initial states described above are shown in Fig.~\ref{figsup6}(a).
There are two common limits for all initial states:
{\it (i)}
For non-interacting fermions ($U/J=0$) the radial and average expansion velocities are equal and time-independent, $v_\sub{r}=v_\sub{av}$, see Eq.~(\ref{vr_vav_noninteracting}).
Furthermore, since the initial MDF of all these initial states is flat, it follows that $v_\sub{r}=\sqrt{2}J$;
{\it (ii)}
In the opposite limit $U/J \to \infty$, the density dynamics of the system become again identical for any initial state $|\phi_\sub{\tiny W}^{(j)}\rangle$. 
Moreover, even states with $N_\uparrow \neq N_\downarrow$ yield the same density dynamics including, in particular, the state with maximal magnetization, i.e., single-component (spinless) fermions.
Therefore, the expansion is again ballistic with $v_\sub{r}=v_\sub{av}=\sqrt{2}J$.
These two limits therefore behave analogously to the 1D Bose-Hubbard model as discussed in Sec.~\ref{subsubsec:bosons2} and Ref.~\onlinecite{ronzheimer13}.

The regime of intermediate $U/J$ exhibits the strongest $W$-dependence.
If the initial wavefunction contains states with large ferromagnetic domains, i.e., $W \ll L_\sub{box}$, their local configuration resembles that of spinless fermions (characterized by $v_\sub{r}/J=\sqrt{2}$) for any $U/J$.
Indeed, our results for small $W$ approach this limit.
On the other hand, the minimum of $v_\sub{r}=v_\sub{r}(U/J)$ is the lowest for the initial N\'eel state ($W_\sub{max}=L_\sub{box}-1$).
Note that the energies of all the initial states $|\phi_\sub{\tiny W}^{(j)}\rangle$ are degenerate, $E_\sub{exp}=0$.
For the initial states studied here, the excess energy $\delta E$ monotonically increases with decreasing $U/J$.
In the extreme limits, $\delta E/(JN)=4/\pi$ at $U/J=0$ (as $N \to \infty$), while $\delta E=0$ at $U/J=\infty$.
This effect influences the expansion velocities at large $U/J$, where the deviation of $v_\sub{r}$ from $\sqrt{2}J$ follows the trend of the excess energy given to the system.
In the opposite limit of $U/J\to 0$, the non-interacting point is approached, which again yields $v_\sub{r}=\sqrt{2}J$.
As a consequence, $v_\sub{r}$ has a minimum at intermediate $U/J$ and the dip becomes more pronounced for larger $W$.
Note, though, that the $W$-dependence that is evident in our data shown in Fig.~\ref{figsup6} implies that the dynamics measured through $v_{\rm r}$ does not only
depend on the excess energy $\delta E$ since all the initial states $|\phi_\sub{\tiny W}^{(j)}\rangle$ have the same $\delta E$.

Further insight into the expansion dynamics is provided by the calculation of double occupancies during the expansion,
\begin{equation}
n_\sub{{\tiny D}}(t) = \frac{1}{N} \sum_i \langle c_{i,\uparrow}^\dagger c_{i,\downarrow}^\dagger c_{i,\downarrow} c_{i,\uparrow} \rangle, \label{def_double}
\end{equation}
which obey $n_\sub{{\tiny D}}(t=0)=0$ for an initial state $|\phi_\sub{\tiny W}^{(j)}\rangle$ used in our study.
In Fig.~\ref{figsup6}(b) we plot $n_\sub{{\tiny D}}(t)$ at time $tJ=1$ for different $U/J$.
Since the total energy is conserved during the expansion, the formation of double occupancies is possible only at the expense of reduced kinetic energy.
At a fixed time after opening the trap, $n_\sub{{\tiny D}}$ decreases monotonously as a function of $U/J$.
Furthermore, the formation of double occupancies can, on short time-scales, only occur at sites with antiparallel neighboring spins, which in turn decreases the expansion velocity at finite $U/J$. 
Hence, $n_\sub{{\tiny D}}(t)$ increases as a function of $W,$ as observed in Fig.~\ref{figsup6}(b).
Our results shown in Fig.~\ref{figsup6}(a) suggest that $v_\sub{r}$ can be used as a probe of the quality of state preparation in experiments.

%%%%%%%%%%%%%%%%%%%%%%%%%%%%%%%%%%%%%%%%%%%%%%%%%%%%%%%%%%%%%%%
\section{Expansion of hard-core bosons on a two-leg ladder} \label{sec:ladder}

The main result of the previous Section was the observation of asymptotic universality for bosons and fermions expanding from their respective Mott insulating ground state on a single chain.
We showed that the density profiles of all Mott insulators become virtually indistinguishable from hard-core bosons (or equivalently non-interacting fermions) in the asymptotic limit.
Moreover, the breaking of integrability by going from 1D hard-core bosons to 1D bosons with finite interactions $U/J < \infty$ does not influence the density dynamics for the expansion from the ground state.
%(while it is for product states, Ref.~\onlinecite{ronzheimer13}).
In this Section, we focus on breaking the integrability of hard-core bosons in a different way:
While they are integrable on 1D chains, this is no longer true for any higher-dimensional or coupled system.
We investigate the expansion on a two-leg ladder, sketched in Fig.~\ref{fig1}(b), as a function of the perpendicular hopping parameter $J_\perp$.
Properties of interacting bosons in the  crossover from 1D to higher-dimensional lattice~\cite{pollet13} as well as the sudden expansion in 2D and 3D systems~\cite{hen10,jreissaty11,jreissaty13} represent a very timely topic. 

\subsection{Mapping to interacting spinless fermions} \label{subsec:mapping}

\begin{figure}[!tb]
\includegraphics[width=0.99\columnwidth,clip]{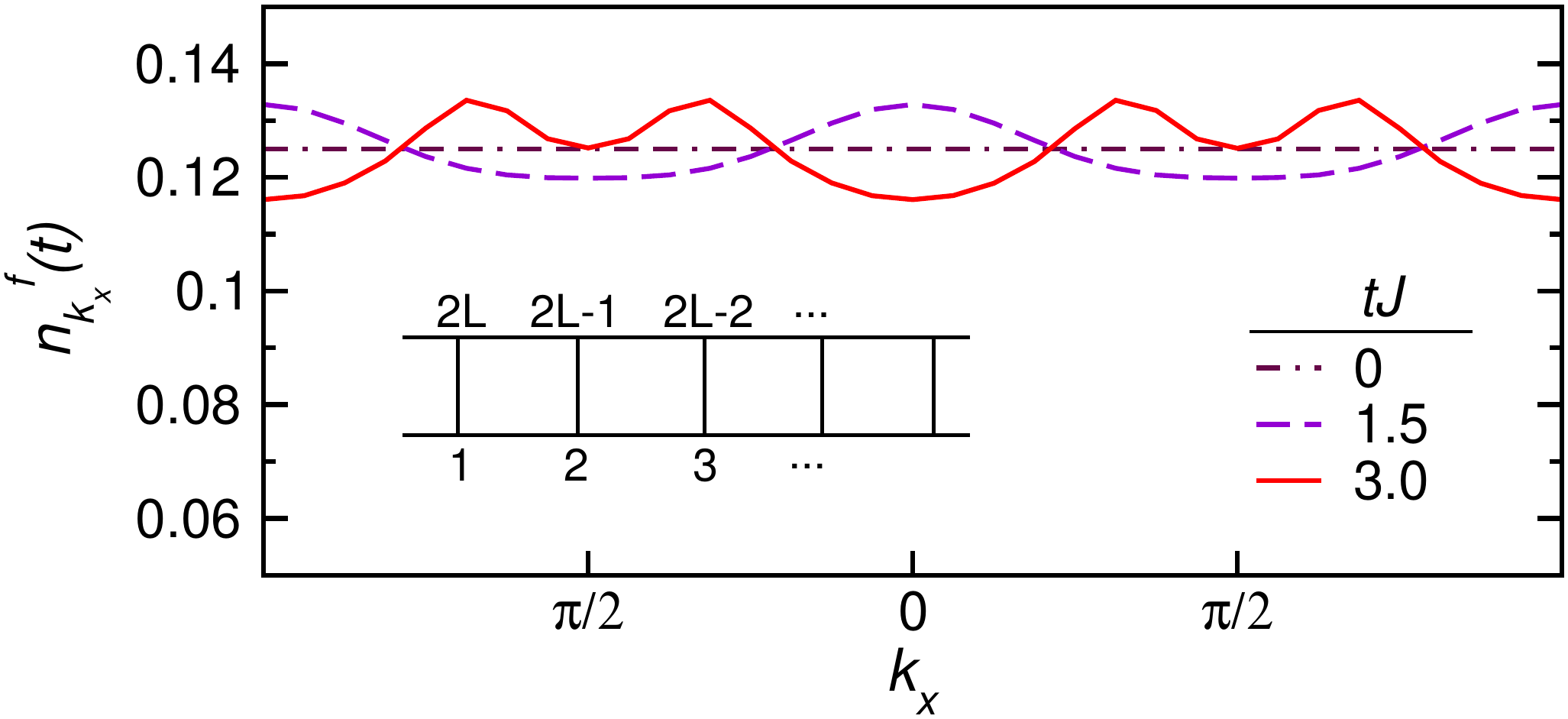}
\caption{(Color online)
{\it
Two-leg ladder:}
Momentum distribution function $n_{k_x}^f(t)$, Eq.~(\ref{def_nkladd}), for interacting spinless fermions (which can be mapped to hard-core bosons), Eq.~(\ref{def_hamisf}), at different times.
The inset shows the fermionic counting applied in calculations on a two-leg ladder with $L$ rungs and $J_\perp = J$.
We used $N=4$ and $L=16$.
}\label{figsup9}
\end{figure}

\begin{figure*}[!tb]
\includegraphics[width=1.7\columnwidth,clip]{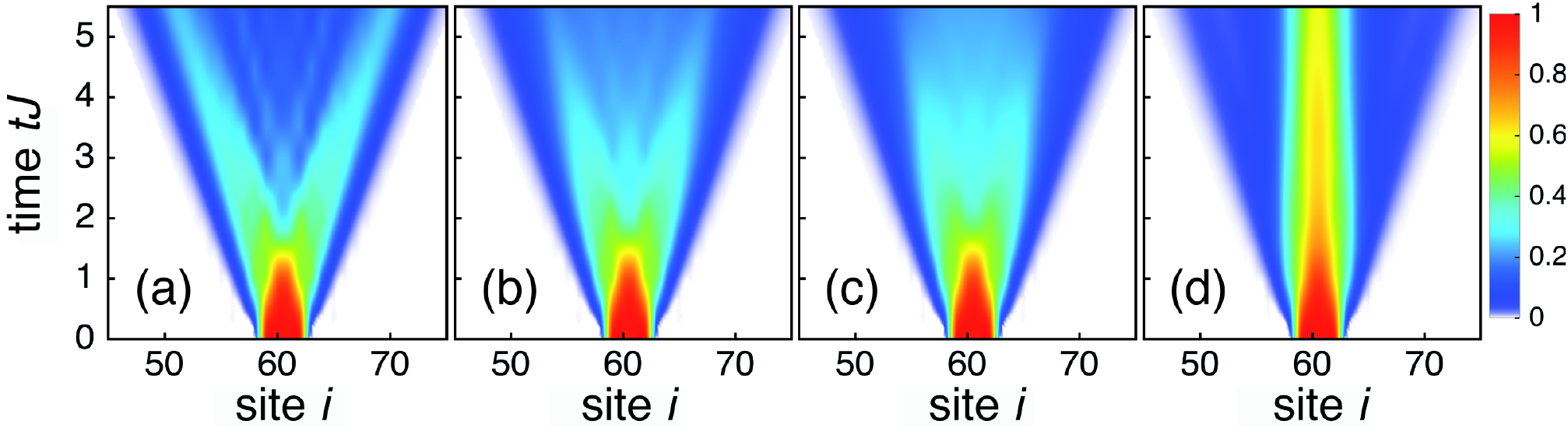}
\caption{(Color online)
{\it
Two-leg ladder:
Density profiles of hard-core bosons.
}
(a)-(d): Expansion from $|\phi_\sub{\tiny Fock}\rangle$ for $J_{\perp}/J=0.1$,~$0.4$,~$0.5$,~$1$, respectively.
$\langle n_i(t)\rangle$ is measured along one of the legs. We used $N=8$ particles.
}\label{fig4b}
\end{figure*}

For 1D systems of hard-core bosons the conservation of fermionic $n_k^f$ is the core reason for the fast ballistic dynamics of strongly interacting particles.
In contrast to 1D systems, hard-core bosons on a ladder can only be mapped to {\it interacting} spinless fermions, since the sign of the hopping matrix element depends on the occupation number of other sites.
In our calculation, we follow the numbering indicated in Fig.~\ref{figsup9}.
The corresponding Hamiltonian reads
\begin{eqnarray}
H & = & -J \sum_{i=1}^{2L-1} (c_i^\dagger c_{i+1} + \mbox{h.c.}) \nonumber \\
 & - & J_\perp \sum_{i=1}^{L} (c_i^\dagger \left[ \prod_{j=i+1}^{2L-i} (1-2n_j) \right] c_{2L+1-i} + \mbox{h.c.}). \label{def_hamisf}
\end{eqnarray}
We define the momentum distribution function $n_k^f$ on a two-leg ladder with $L$ rungs as
\begin{eqnarray} \label{def_nkladd}
n_{k_x}^f & \equiv & \frac{1}{2} \left( n^f_{[k_x,k_y=0]} + n^f_{[k_x,k_y=\pi]}\right), \\
n_{\bf k}^f & = & \frac{1}{2L} \sum_{{\bf r'}, {\bf r}} e^{i {\bf (r - r')\cdot k}} \langle c_{\bf r'}^\dagger c_{\bf r} \rangle,
\end{eqnarray}
where ${\bf r}$ and ${\bf r'}$ represent vectors associated to sites on the ladder.
As the main difference in relation to the 1D system, $n_k^f$ are not conserved during the expansion, and the model is non-integrable as soon as $J_{\perp}/J>0$.
As an example, we show $n_{k_x}^f$ for $J_{\perp}=J$ at different times in Fig.~\ref{figsup9}.

%%%%%%%%%%%%%%%%%%%%%%%%%%
\subsection{Crossover: Coupling chains to a two-leg ladder} \label{subsec:crossover}

%In this work we study the expansion in the crossover from two uncoupled chains to a two-leg ladder, which is still quasi-1D, but numerically tractable.
We now show that coupling chains to a two-leg ladder has a dramatic effect for hard-core bosons expanding from $|\phi_\sub{\tiny Fock}\rangle$, Eq.~(\ref{def_fock}).
In Figs.~\ref{fig4b}(a)-(d) we present density profiles $\langle n_i(t) \rangle$ for different values of $J_\perp/J$.
At $J_{\perp}/J\ll 1$, the expanding cloud develops two well-defined wings, as expected for ballistic dynamics in 1D.~\cite{polini07,langer09,langer11}
At $J_{\perp}/J=1$, on the other hand, there remains a stable core of particles in the center of the lattice which barely delocalizes.
This suggests that the expansion is qualitatively different from the one observed for strongly interacting particles in 1D, indicative of diffusion.~\cite{polini07,schneider12}

\begin{figure}[!tb]
\includegraphics[width=0.99\columnwidth,clip]{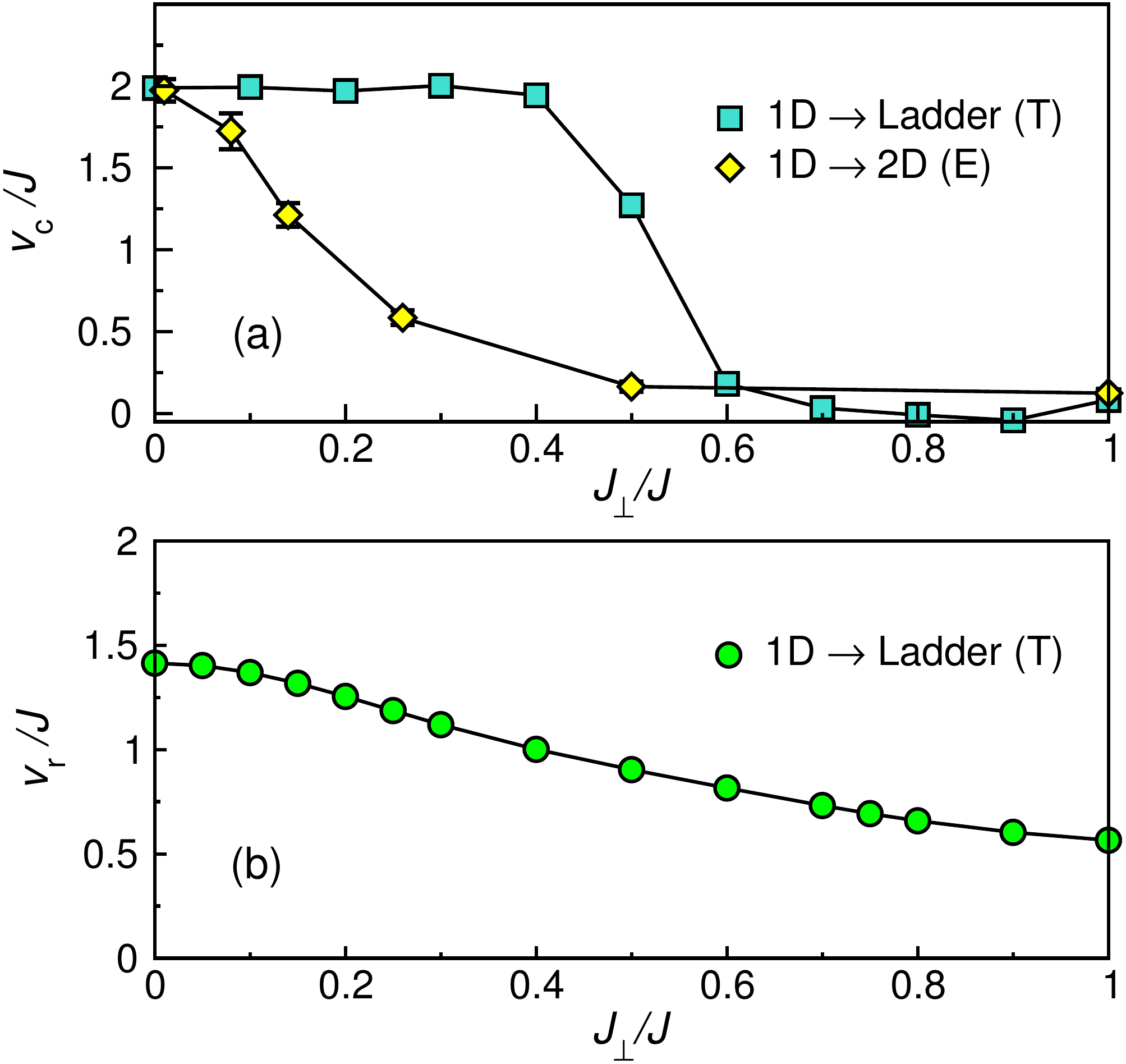}
\caption{(Color online)
{\it
Two-leg ladder: Expansion velocities as a function of $J_{\perp}/J$.
}
(a)
Core velocitiy $v_{\mbox{c}}/J$.
Squares: Theoretical (T) results for hard-core bosons on a two-leg ladder ($N=12$).
Diamonds: Experimental (E) results~\cite{ronzheimer13} for the 1D-2D crossover at $U/J=20$.
(b)
Radial velocitiy $v_\sub{r}/J$.
Circles: Theoretical (T) results for hard-core bosons on a two-leg ladder ($N=12$).
}\label{fig5}
\end{figure}

%Next we study how the expansion velocity changes as $J_\perp/J$ is varied continuously from $0$ to $1$.
In Fig.~\ref{fig5}(a) (squares) we show the {\it core} expansion velocity $v_\sub{c}$, Eq.~(\ref{def_vc}).
It is derived from the half-width-at-half-maximum of the density distribution, $r_\sub{c}(t)$, which is shown in Fig.~\ref{figsup7}(a) of Appendix~\ref{sec:app1b}.
Remarkably, $v_\sub{c}$ exhibits a sharp drop at intermediate $J_\perp/J\approx 0.5 $ from $v_\sub{c}/J\approx 2$ to a vanishing $v_\sub{c}/J\approx 0$ at large $J_\perp/J$.
For sufficiently small perpendicular hopping, $J_\perp/J \lesssim 0.4$, $v_\sub{c}$ detects the fast wings observed in Fig.~\ref{fig4b}(a), which expand with $v_k/J\approx 2$.
For larger values of the perpendicular hopping, $J_\perp/J \gtrsim 0.6$, however, the formation of a stable core in the density distribution dominates $v_\sub{c}$ and renders it small.
We compare $v_\sub{c}$ for different particle numbers $N$ in Fig.~\ref{figsup8} of Appendix~\ref{sec:app1b}.
The result suggests that a sharp drop persists around $J_\perp/J \approx 0.5$ as $N$ increases.
It indicates that at this particular value of $J_\perp/J$ the stable central core becomes higher than the maxima in the wings.

We also calculated the radial velocity $v_\sub{r}$.
Results in Fig.~\ref{fig5}(b) reveal that $v_\sub{r}$ exhibits a smooth dependence on $J_\perp/J$.
Since $\tilde R(t)$ is a sum over the whole density profile, $v_\sub{r}$ measures how the relative density of the fast ballistic wings decreases when $J_\perp/J$ increases.
Due to the formation of the high-density core for $J_{\perp}\gtrsim J/2$, as shown in Fig.~\ref{fig4b}(d), $v_\sub{r}$ decreases by a factor of $\sim 2.5$ with respect to the expansion on uncoupled chains at $J_{\perp}=0$.
In notable contrast to 1D hard-core bosons, on a ladder, $\tilde R(t)$ is {\it not} linear in time but undergoes transient dynamics, see Fig.~\ref{figsup7}(b).
We measure $v_\sub{r}$ as a linear fit to $\tilde R(t)$ at the longest times, see Appendix~\ref{sec:app1b}.
The comparison between $v_\sub{c}$ and $v_\sub{r}$ makes transparent that the sharp drop of $v_\sub{c}$ at $J_\perp/J\approx 0.5$ is specific to $v_\sub{c}$, but
not indicative of a qualitative change in the dynamics  (e.g. from ballistic to diffusive) occurring at this particular value of $J_\perp/J$.

The recent experiment Ref.~\onlinecite{ronzheimer13} studied the sudden expansion in the crossover from an array of uncoupled chains to a square lattice.
The experimental results for the core velocity $v_\sub{c}$ are included in Fig.~\ref{fig5}(a) (diamonds) for comparison.
Since a two-leg ladder represents a quasi-1D system while a square lattice is two-dimensional, there is no reason that the two curves in Fig.~\ref{fig5}(a) should be quantitatively similar.
It is nevertheless very intriguing that in both cases, a two-leg ladder and a square lattice, a sufficiently large $J_\perp/J$ results in a very slow expansion and a stable high-density core.
Despite the quantitative differences concerning the $v_\sub{c} = v_\sub{c}(J_\perp/J)$ curves, the conservation of the fermionic MDF $n_k^f$ of strictly 1D hard-core bosons is violated as soon as $J_\perp/J>0$ in both cases.
We argue that this gives rise to the changes of the expansion dynamics compared to 1D hard-core bosons.

Our results identify hard-core bosons on a ladder as an ideal testbed to study the effects of integrability breaking in experiments.
The required homogeneous ladder potentials can be readily realized in the experiment by combining the superlattice technique of Ref.~\onlinecite{chen11} with the control of the external confinement demonstrated in Refs.~\onlinecite{schneider12,ronzheimer13}, provided that the transverse potential created by the superlattice is overall anticonfining.
This can easily be fulfilled by using a blue-detuned short period lattice and suitable beam waists.

\section{Conclusions} \label{sec:conclusion}

We studied the expansion dynamics of bosons and fermions on a chain and of hard-core bosons on a two-leg ladder, focusing on the initial density $n=1$.
Remarkably, we observe that on a chain - starting from the correlated ground state - both bosonic and fermionic Mott insulators expand with the same fast expansion velocity and show virtually identical density profiles, independent of the interaction strength $U/J>U_c/J$.
As a consequence, both systems share the same flat velocity distribution in the asymptotic regime.
This requires a particle-hole symmetric MDF and implies that the asymptotic dynamics is controlled by non-interacting fermions.
In that sense, the non-integrable Bose-Hubbard model with repulsive interactions therefore exhibits an asymptotic dynamics with emergent free fermions.
This is similar to integrable bosonic models (hard-core bosons on a lattice, the Tonks-Girardeau gas and the Lieb-Liniger model with repulsive interactions), for which an exact mapping to free fermions exists, and our results may be viewed as a generalization of the dynamical fermionization of 
the bosonic quasi-momentum distribution function of hard-core bosons and the Tonks-Girardeau gas\cite{rigol05,minguzzi05}
to the Bose-Hubbard model with repulsive interactions.

Our results show that the measurement of the radial expansion velocity $v_\sub{r}$ is sensitive to the interaction-driven quantum phase transition in the initially confined system.
A comparison of the expansion from the box trap with the expansion from the harmonic trap suggests that the main observations are robust against the choice of the initial confining potential.

We further compare the expansion from ground states to the expansion from a product of local Fock states with a single boson per site, as realized in Ref.~\onlinecite{ronzheimer13}:
If the system on a chain is prepared in the product state with an energy above the ground-state energy (i.e., at intermediate $U/J$), the universal expansion dynamics reported above is lost.
While for the expansion from ground states we observe ballistic dynamics, the excess energy results in a slower and presumably diffusive dynamics.~\cite{ronzheimer13}
In other words, the system with higher energy per particle expands slower.
Our work calls for future studies of diffusion and transport coefficients of the Bose-Hubbard model at finite temperatures.
We carried out the analogous calculation for fermions and arrived at similar conclusions:
By preparing the system in a non-eigenstate with a an excess energy, e.g., in the product of local Fock states, the expansion is slowed down for $0<U/J<\infty$.

Another notable result emerges in the crossover from uncoupled chains to a two-leg ladder.
While the fast and ballistic dynamics of integrable 1D hard-core bosons persist to the expansion from the ground state at finite values of $U/J<\infty$, coupling chains to a ladder changes the behavior qualitatively.
The core reason for this behavior is the breaking of the conservation of the fermionic MDF, which is conserved only for hard-core bosons on a chain.
Since a two-leg ladder potential can be realized experimentally in optical lattices, it would be interesting to observe this phenomenon with ultra-cold atoms.
Moreover, hard-core bosons on a ladder are equivalent to $XX$ spin-1/2 models, connecting our work to studies of spin transport in 1D quantum magnets.~\cite{heidrichmeisner03, heidrichmeisner07,karrasch12,znidaric13}

\section*{Acknowledgement} \label{sec:ack}

We thank N. Andrei, C. Bolech, and M. Rigol for fruitful discussions,
and we thank S. Thwaite, M. Rigol,  and J. P. Ronzheimer for a critical reading of the manuscript.
L.V. is supported by the Alexander von Humboldt Foundation.
S.L. was financially supported by AFOSR grant FA9550-12-1-0057.
I.P.M. acknowledges support from the Australian Research Council Centre of Excellence for Engineered Quantum Systems and the Discovery
Projects funding scheme (Project No.~DP1092513).
F.H.-M., U.S., and U.S. acknowledge support from the Deutsche Forschungsgemeinschaft (DFG) through FOR 801.

%%%%%%%%%%%%%%%%%%%%%%%%%%%%%%%%%%%%%%%%%%%%%%%%%%%%%%%%%%%%%%%
\appendix

%%%%%%%%%%%%%%%%%%%%%%%%%%%%%%%%%
\section{Extraction of radial velocities} \label{sec:app1}

%%%%%%%%%%%%%%%%%%%%%%%%%%%%%%%%%
\subsection{Interacting bosons on a single chain} \label{sec:app1a}

In this Appendix we clarify how the radial velocities $v_\sub{r}$ of the 1D Bose-Hubbard model, presented in Fig.~\ref{fig3}, are extracted from real-time calculations using the DMRG method with a discarded weight $\eta\leq 10^{-4}$.
In all our data, we used a time step $\Delta tJ=1/16$ and different values of $N_\sub{max}$ to verify that the results are independent of $N_\sub{max}$
($N_\sub{max}$ denotes the maximal number of bosons per site used in DMRG calculations).
We limit the analysis to the expansion from the ground state, while the analysis of the expansion from a product of local Fock states was performed in Ref.~\onlinecite{ronzheimer13}.

\begin{figure}[!t]
\includegraphics[width=0.99\columnwidth,clip]{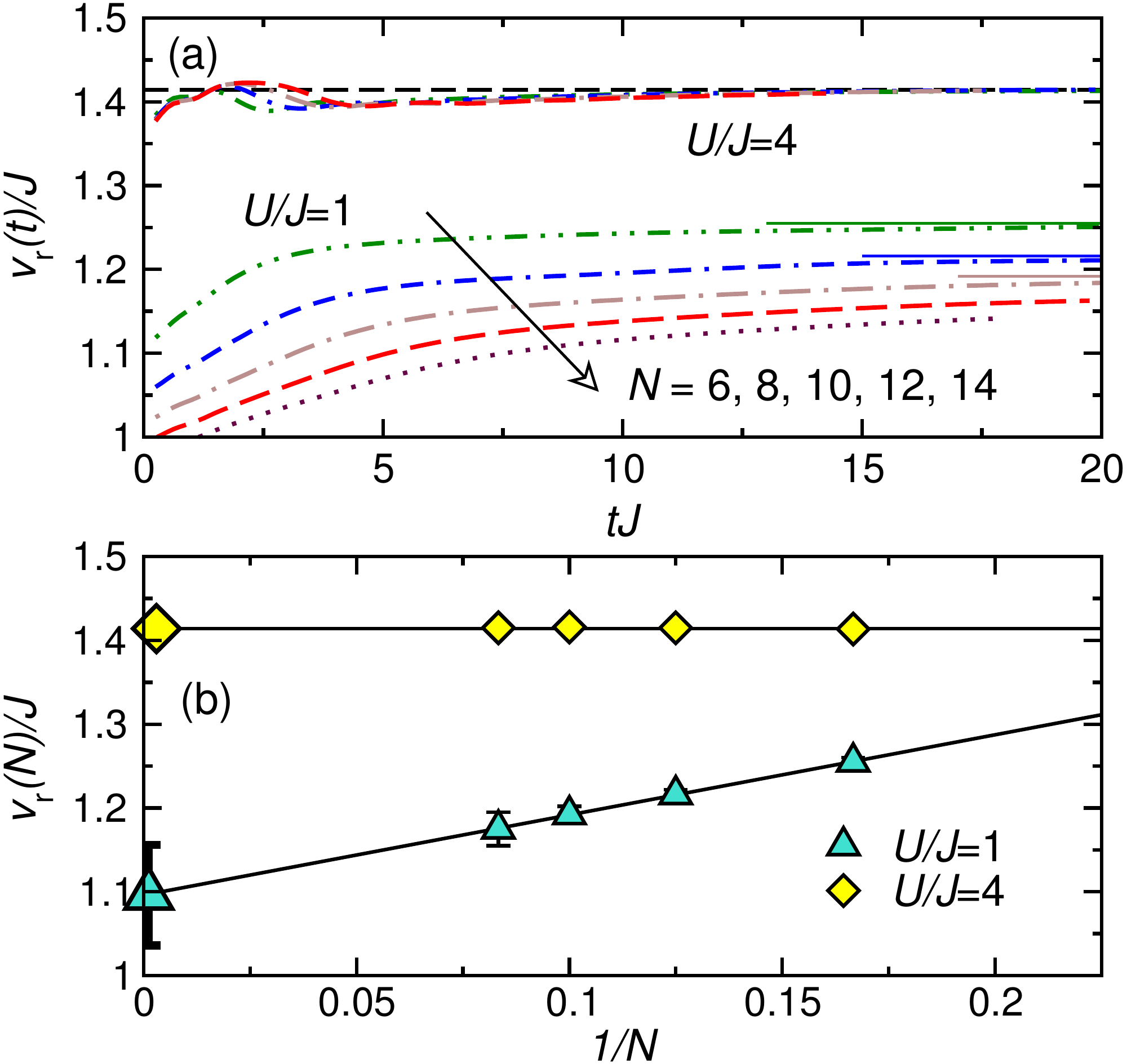}
\caption{(Color online)
{\it
Single chain:
Expansion dynamics of interacting bosons from the ground state.}
(a)
Time-dependence of the radial velocity $v_\sub{r}(t)=\partial \tilde R(t)/\partial t$.
At $U/J=1$, we show results for $N=6,8,10,12,14$ particles with $N_\sub{max}=5$.
At $U/J=4$, the four nearly overlapping curves show results for $N=6,8,10,12$ particles with $N_\sub{max}=5,5,4,3$, respectively.
For the data shown in this figure, a discarded weight $\eta=10^{-5}$ was used. 
Horizontal solid lines on the right-hand-side of the figure indicate the extrapolated values of $v_\sub{r}(t)$ at $tJ\gg 1$.
These values (with the corresponding error bars) are used in (b) where the finite-size scaling is performed.
(b)
$v_\sub{r}(N)$ for different number of particles $N$.
Large symbols at $1/N = 0$ denote the extrapolated values when $N\to \infty$.
These values (with the corresponding error bars) are used in Fig.~\ref{fig3}(a) (circles) of the main text where $v_\sub{r}$ vs $U/J$ is plotted.
}\label{figsup2}
\end{figure}

We estimate $v_\sub{r}$ by using the linear fit $\tilde R(t)=v_\sub{r}t$.
Nevertheless, to further increase the accuracy of our results by clarifying the dependence of the expansion dynamics on the initial number of particles $N$,
we study the time-dependence of the radial velocity, $v_\sub{r}(t)=\partial \tilde R(t)/\partial t$.
Results are shown in Fig.~\ref{figsup2} for the same values of $U/J$ as in Fig.~\ref{figsup1}.
At large $U/J$, $v_\sub{r}(t)$ is almost time-independent, while for smaller $U/J$ a transient time dependence becomes more pronounced.
In particular, the time before $v_\sub{r}(t)$ becomes stationary increases for smaller $U/J$.
We define the radial velocity, presented in Fig.~\ref{fig3}(a) as the asymptotic value $v_\sub{r} = v_\sub{r}(t\to \infty)$ when $N\to \infty$.

We pursue the following two-step process to calculate $v_\sub{r}$:
{\it (i)}
For a fixed number of particles $N$, we obtain $v_\sub{r}(N) = v_\sub{r}(N;t\to \infty)$ by taking the value at the largest time available from our simulations, provided that the change of $v_\sub{r}(t)$ in the last few time units (typically 5-10 time units) is below $1\%$.
This is shown in Fig.~\ref{figsup2}(a).
In addition, Fig.~\ref{figsup2}(a) reveals that $v_\sub{r}(N)$ is essentially $N$-independent for $U/J=4$, while this is no longer the case for $U/J=1$;
{\it (ii)}
We perform a fit $v_\sub{r}(N) = \alpha \frac{1}{N} + \beta$, yielding the desired value $v_\sub{r} = \beta$.
The fits are shown in Fig.~\ref{figsup2}(b).
Both steps produce an uncertainty which is then assigned to $v_\sub{r}$.
In case no error bar is indicated in the figures, it implies that it is smaller than the size of the symbol.

%%%%%%%%%%%%%%%%%%%%%%%%%%%%%%%%%
\subsection{Hard-core bosons on a two-leg ladder} \label{sec:app1b}

%%%
\begin{figure}[!tb]
\includegraphics[width=0.99\columnwidth,clip]{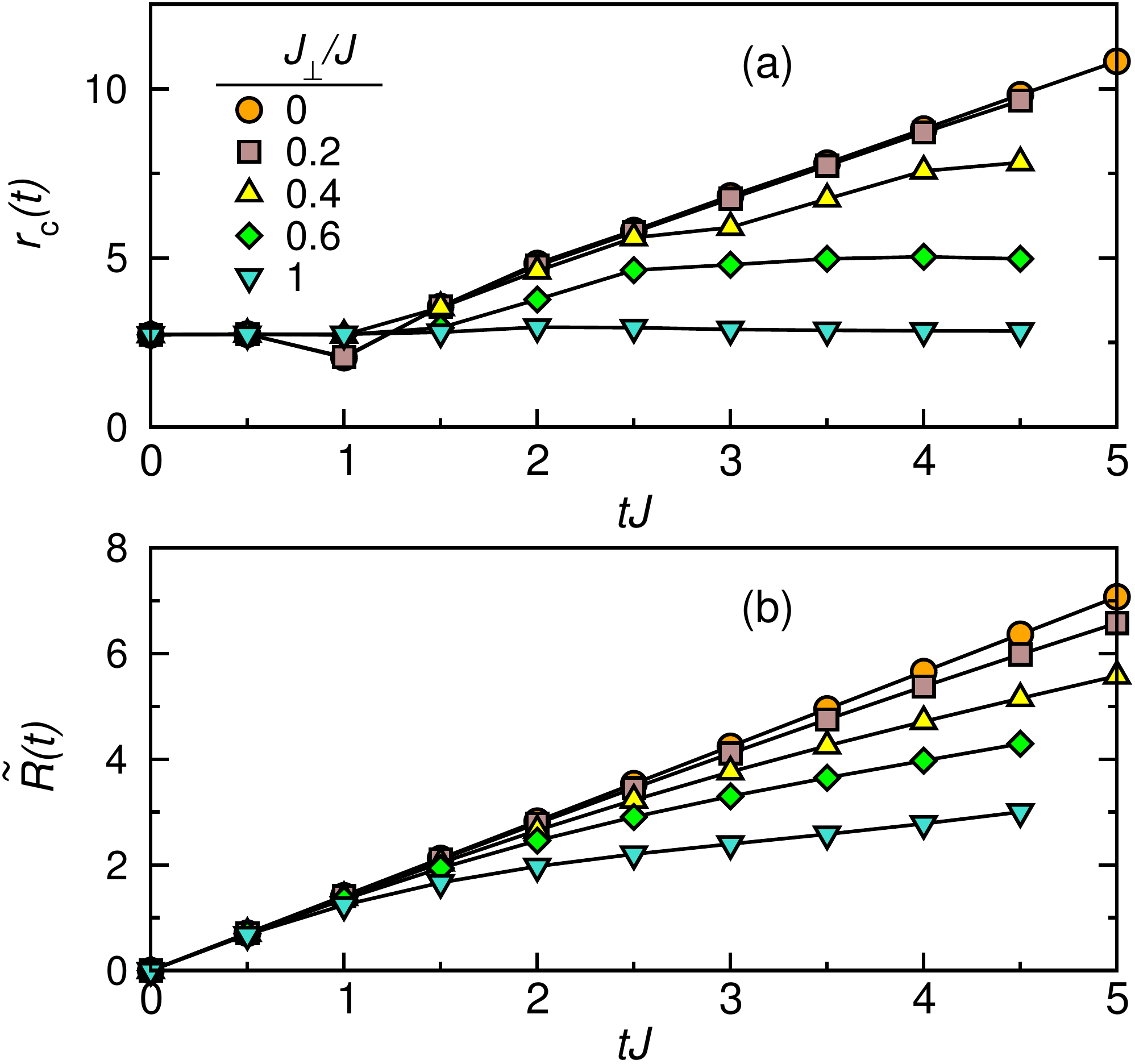}
\caption{(Color online)
{\it
Two-leg ladder:
Expansion dynamics of hard-core bosons from a local product of Fock states, $|\phi_\sub{\tiny Fock}\rangle$.}
(a)
Time dependence of the half-width-at-half-maximum $r_\sub{c}(t)$ for different $J_\perp/J$.
The corresponding velocity $v_\sub{c}$ is shown in Fig.~\ref{fig5}(a).
(b)
Time dependence of the radius $\tilde R(t)$ for different $J_\perp/J$.
The corresponding velocity $v_\sub{r}$ vs $J_\perp/J$ is shown in Fig.~\ref{fig5}(b).
Both velocities are extracted from a linear fit of $r_\sub{c}(t)$ and $\tilde R(t)$ in the time interval $2\leq tJ< 5$.
We used $N=12$ particles.
We show $v_\sub{c}$ vs $J_\perp/J$ for different particle numbers $N$ in Fig.~\ref{figsup8}.
}\label{figsup7}
\end{figure}
%%%

For a two-leg ladder, we study two different expansion velocities originating from complementary measures of the expanding cloud of particles.
Besides the  radius $\tilde R(t)$, we also calculate $r_\sub{c}(t)$, which is defined as the half-width-at-half-maximum of the density distribution $\langle n_i(t) \rangle$.
In the case when the half of the maximal local density is measured at more than two sites, $r_\sub{c}$ corresponds to the outermost site. 
This definition follows Ref.~\onlinecite{ronzheimer13}.

\begin{figure}[!tb]
\includegraphics[width=0.99\columnwidth,clip]{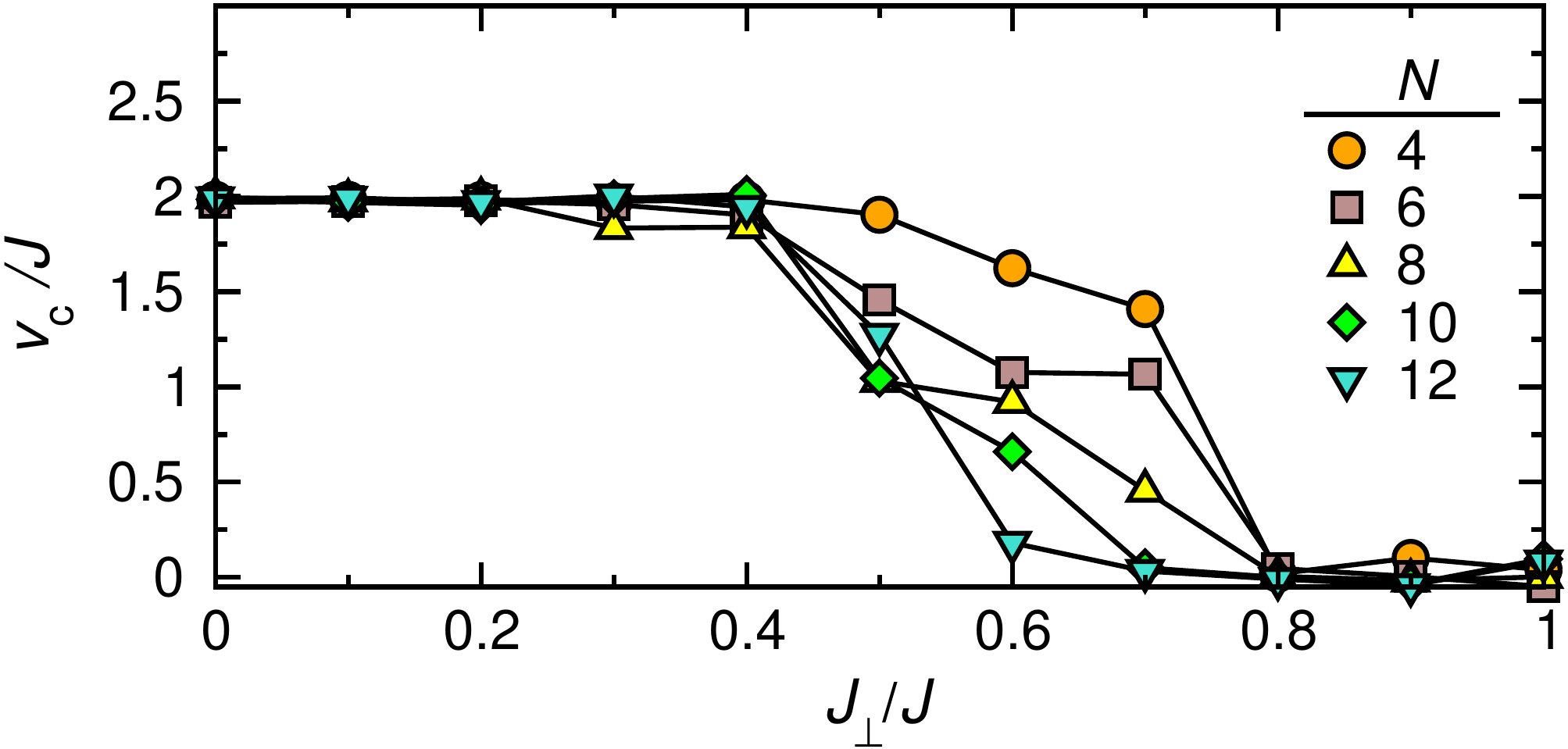}
\caption{(Color online)
{\it
Two-leg ladder:
Expansion dynamics of hard-core bosons from a local product of Fock states, $|\phi_\sub{\tiny Fock}\rangle$.}
Core expansion velocity $v_c$ as a function of $J_{\perp}/J$ for different particle numbers $N$.
}\label{figsup8}
\end{figure}

Figure~\ref{figsup7}(a) shows $r_\sub{c}(t)$ for different values of $J_\perp/J$.
After some short transient dynamics, considerable differences occur in the time dependence of $r_\sub{c}$.
This can be understood from the structure of the density profiles $\langle n_i(t) \rangle$ shown in Figs.~\ref{fig4b}(a)-(d).
We use the fitting function $r_\sub{c}(t) = v_\sub{c}t + \gamma$ in the time interval $2\leq tJ< 5$ to avoid transient dynamics and the kinks in $r_\sub{c}(t)$.
As a result, the core velocity $v_\sub{c}$, plotted in Fig.~\ref{fig5}(a), exhibits a strong dependence on $J_\perp/J$, ranging from $v_\sub{c}/J\approx 2$ at $J_\perp/J=0$, to $v_\sub{c}/J\approx 0$ at $J_\perp/J=1$.

For comparison we show $\tilde R(t)$ in Fig.~\ref{figsup7}(b), which essentially exhibits the same properties, however, with a less dramatic drop of the expansion velocity as a function of $J_\perp/J$.
In contrast to 1D hard-core bosons, $\tilde R(t)$ is not linear in time but undergoes transient dynamics.
We obtain the radial velocity $v_\sub{r}$ from fitting $\tilde R(t) = v_\sub{r} t + \delta$ to the numerical data in the time interval $2\leq tJ< 5$.
The radial velocity is shown in Fig.~\ref{fig5}(b).

In Fig.~\ref{figsup8}, we show the dependence of the core velocity $v_\sub{c}$ on the particle number $N$.
Finite-size effects disappear as we approach the integrable limit $J_{\perp}=0$ and become very small for $J_{\perp}\to J$.
For the available values of $N$, the drop of $v_c$ from $2J$ to zero occurs at $J_\perp/J\approx 0.5$.

%%%%%%%%%%%%%%%%%%%%%%%%%%%%%%%%%
\section{Expansion for densities $n<1$} \label{sec:density}

We now discuss the influence of the initial density $n$ on the expansion dynamics from the ground state of the Bose-Hubbard model on a chain for different interaction strengths $U/J$.
In the main part of the paper, we focused on $n=1$, while here we discuss in more detail results for $n<1$.

\begin{figure}[t]
\includegraphics[width=0.99\columnwidth,clip]{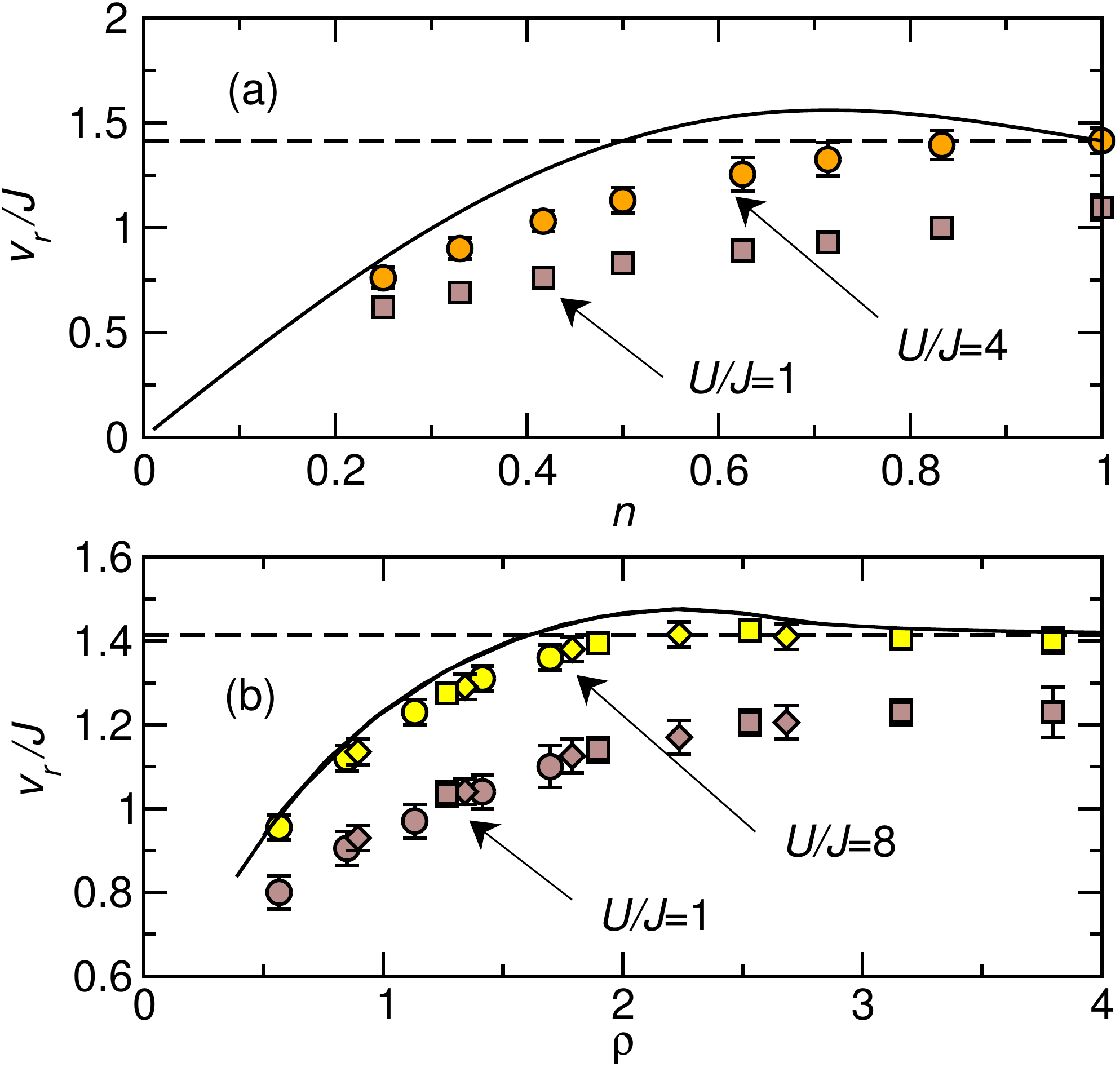}
\caption{(Color online)
{\it
Single chain:
Expansion dynamics of interacting bosons from the initial ground state.}
(a) Box trap:
Radial expansion velocity $v_\sub{r}/J$ vs the initial boson density $n=N/L_\sub{box}$, for two different interaction strengths $U/J=1$ (squares) and $U/J=4$ (circles).
Solid line denotes results for hard-core bosons, Eq.~(\ref{vr_n_hcb}).
(b) Harmonic trap:
$v_\sub{r}/J$ vs $\rho$ for two different interaction strengths $U/J=1$ and $U/J=8$.
The initial boson density is defined as $\rho=N\sqrt{V_\sub{h}}$.
Circles, diamonds and squares represent results for $V_\sub{h}=0.02J$, $V_\sub{h}=0.05J$ and $V_\sub{h}=0.10J$, respectively.
Solid line denotes results for hard-core bosons calculated by exact diagonalization.~\cite{rigol04a,rigol04}
}\label{figsup40}
\end{figure}

Hard-core bosons can be mapped to noninteracting spinless fermions.
As a consequence, we argued in Sec.~\ref{subsec:HCB} that the bosonic radial velocity is equal to the fermionic one,
\begin{equation}
v_{\mbox{\footnotesize r}}=v_{\mbox{\footnotesize r}}^f=v_\sub{av}^f.
\end{equation}
This implies that $v_\sub{r}/J$ is fully determined by the corresponding initial Fermi momentum $k_\sub{F}$, which is in turn related to the initial density $n$.
The time evolution of noninteracting fermions can be calculated analytically and yields~\cite{langer12}
\begin{equation} \label{vr_n_hcb}
v_\sub{r}/J = \sqrt{2 \left[ 1 - \frac{\sin{(n\pi)}\cos{(n\pi)}} {n\pi} \right]}.
\end{equation}
The result consistently describes the intuitive limits $v_\sub{r} \xrightarrow{n\to 0} 0$ and $v_\sub{r} \xrightarrow{n\to 1} \sqrt{2}J$.
%$n \xrightarrow[under]{over} v$
However, $v_\sub{r}(n)$ is not monotonic for hard-core bosons since it has a maximum at an incommensurate initial density $n$. 
The solid line in Fig.~\ref{figsup40}(a) presents $v_\sub{r}$ versus $n$ as given by Eq.~(\ref{vr_n_hcb}).

When the interaction strength $U/J$ is decreased to finite values, an overall decrease of $v_\sub{r}$ is observed, see Fig.~\ref{figsup40}(a).
Moreover, $v_\sub{r}$ becomes monotonic as a function of $n$.

%%%%%%%%%%%%%%%%%%%%%%%%%%%%%%%%%
\section{Expansion from a harmonic trap} \label{sec:harmonic}

We briefly discuss the influence of the choice of the trapping potential for $t<0$.
In the main part of the paper, we considered the expansion from a box trap only.
Here we focus on a harmonic trap, since it models the situation commonly realized in experiments on optical lattices.~\cite{schneider12,ronzheimer13}
A harmonic trap for a chain is defined as
\begin{equation} \label{def_harmonictrap}
V_{\mbox{\footnotesize trap}}^\sub{(h)} = V_\sub{h} \sum_{i=1}^{L} n_i \left( i - \frac{L+1}{2} \right)^2.
\end{equation}
The effective density of particles is defined as $\rho = N\sqrt{V_\sub{h}}$.
In analogy to the studies for the box trap, we are interested in values of $\rho$ such that the initial ground state is either a superfluid state with $\langle n_i \rangle<1$ for any site $i$, or a Mott insulating state with $\langle n_i \rangle = 1$ in the center of the trap and $\langle n_i \rangle<1$ on the edges.

\begin{figure}[!t]
\includegraphics[width=0.99\columnwidth,clip]{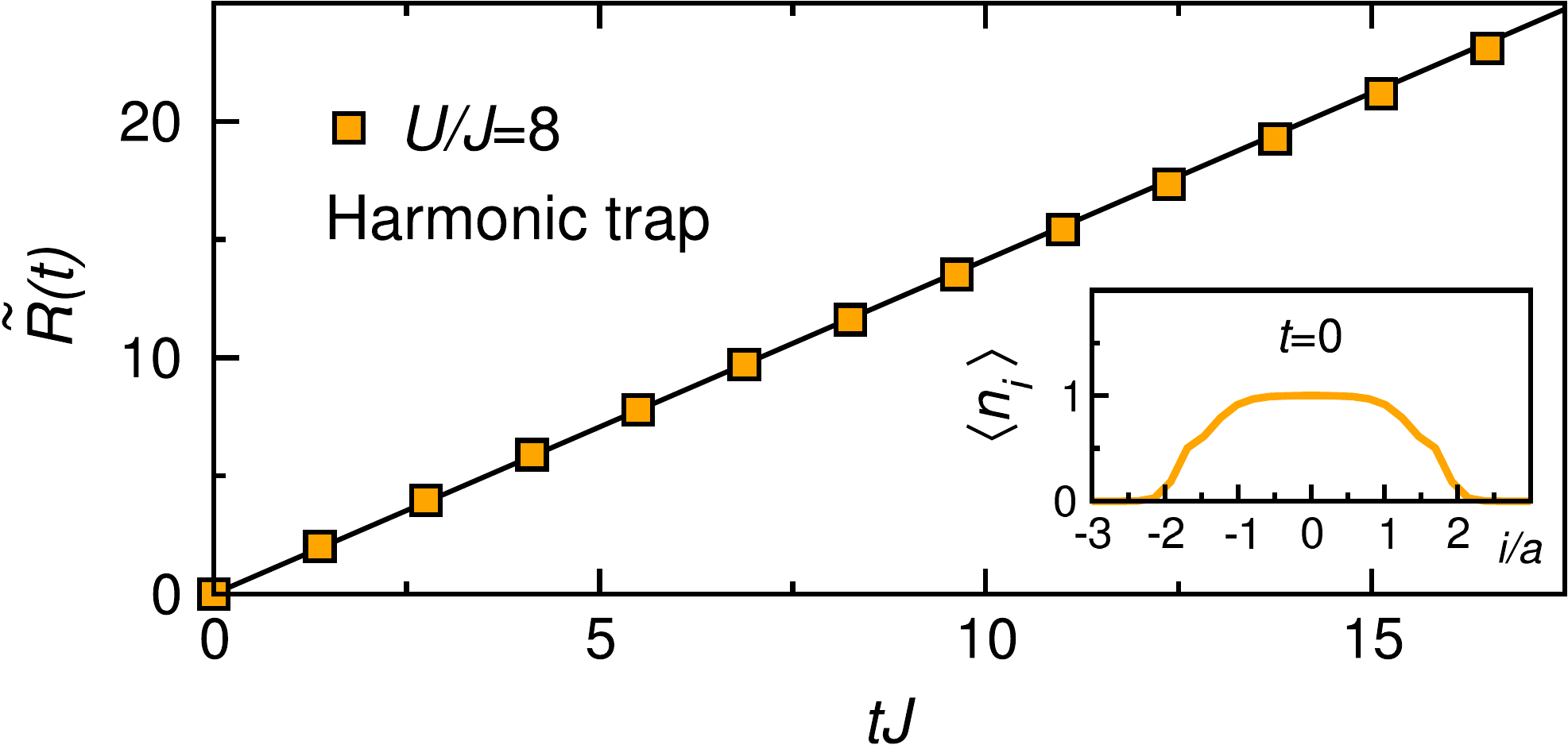}
\caption{(Color online)
{\it
Single chain:
Expansion dynamics of interacting bosons from the ground state in the harmonic trap.}
We used $U/J=8$, $\rho=3.16$, $V_\sub{h}/J=0.05102$ and $(N,N_\sub{max})=(14,4)$.
Main panel:
Time dependence of the radius $\tilde R(t)$ (squares).
Solid line represents a fit $\tilde R(t)=\sqrt{2}J t$.
Inset:
Density profile $\langle n_i \rangle$ at time $t=0$.
We rescale site units to $i/a$, where $a=1/\sqrt{V_\sub{h}}$.
}\label{figsup5}
\end{figure}

In Fig.~\ref{figsup5} we present the time-dependence of the radius $\tilde R(t)$ for $U/J=8$ and $\rho = 3.16$.
For these parameter values, the ground state is a Mott insulator with $\langle n_i \rangle=1$ in the center of the trap (the initial density profile is shown in the inset of Fig.~\ref{figsup5}).
The radius increases linearly in time, $\tilde R(t) \propto t$, in quantitative agreement with expansion from the box trap in Fig.~\ref{figsup1}.
As a main result, we show that the characteristic expansion velocity $v_\sub{r}/J=\sqrt{2}$ of the Mott insulator emerges both in the box trap as well as in the harmonic trap.
In the main panel of Fig.~\ref{figsup5}, squares represent the numerical data while the solid line represents $\tilde R(t)=\sqrt{2}J \,t$.

Our study reveals that the main observations, which we made for the box trap, carry over to the expansion from the harmonic trap. 
In particular, the Mott insulating phase is characterized by $v_\sub{r}/J=\sqrt{2}$.
In fact, the Mott plateau with $v_\sub{r}/J=\sqrt{2}$ can be observed in the case when $\rho$ is fixed and $U/J$ varied, as well as in the opposite case when $U/J$ is fixed and $\rho$ varied.
The latter case is presented in Fig.~\ref{figsup40}(b).
Even though any Mott insulating state within the harmonic trap contains some superfluid fraction at the edge, this does not break the universality in expansion dynamics observed in terms of expansion velocity (see also Ref.~\onlinecite{langer12}).

Results for hard-core bosons [solid line in Fig.~\ref{figsup40}(b)], show that the plateau at $v_\sub{r}/J=\sqrt{2}$, characterizing the expansion from a Mott insulator, emerges roughly at initial density $\rho>\rho^* \sim 3$.
In general, the lower bound for the density $\rho^*$ to observe the Mott insulating plateau $\langle n_i \rangle = 1$ in the center of the trap is a function of interaction strength $U/J$.
However, at large $U/J$, $\rho^*$ is almost independent of $U/J$ and roughly given by $\rho^* \sim 2.7$, see Refs.~\onlinecite{kollath04,rigol09}.
Our results show that an estimate for $\rho^*$ at fixed $U/J$ can also be obtained by measuring the deviation of $v_\sub{r}(\rho)/J$ from $\sqrt{2}$.
For hard-core bosons, $v_\sub{r}$ is a non-monotonic function of $\rho$.
This result is consistent with the case of the box trap shown in Fig.~\ref{figsup40}(a) and Eq.~(\ref{vr_n_hcb}).
At finite $U/J\lesssim 8$, see Fig.~\ref{figsup40}(b), $v_\sub{r}$ monotonically decreases with decreasing $\rho$.
In both cases, $v_\sub{r}(\rho)/J = \sqrt{2}$ is a rough measure of the presence of the Mott insulator in the initial state.

The Mott-insulator-to-superfluid transition also leaves its fingerprints on the expansion dynamics if $U/J$ is varied at a fixed $\rho > \rho^*$.
In the Mott-insulating phase when $U/J=8$, Fig.~\ref{figsup40}(b) shows that $v_\sub{r}(\rho)/J = \sqrt{2}$, while in the superfluid phase when $U/J=1$, we get $v_\sub{r}(\rho)/J < \sqrt{2}$.
Even though the differences in expansion velocities are not dramatic, our results indicate that $v_\sub{r}$ could be used as a measure of the presence of Mott state within the harmonic trap, and complements calculations of other quantities which also detect the Mott state such as local density fluctuations and compressibility.~\cite{batrouni02,wessel04,rigol09}
Moreover, the density dependence of $v_\sub{r}$ is similar to a two-component Fermi gas.~\cite{langer12}

%%%%%%%%%%%%%%%%%%%%%%%%%%%%%%%%%%%%%%%%  Bibliography

\bibliography{references}

\begin{thebibliography}{92}%
\makeatletter
\providecommand \@ifxundefined [1]{%
 \@ifx{#1\undefined}
}%
\providecommand \@ifnum [1]{%
 \ifnum #1\expandafter \@firstoftwo
 \else \expandafter \@secondoftwo
 \fi
}%
\providecommand \@ifx [1]{%
 \ifx #1\expandafter \@firstoftwo
 \else \expandafter \@secondoftwo
 \fi
}%
\providecommand \natexlab [1]{#1}%
\providecommand \enquote  [1]{``#1''}%
\providecommand \bibnamefont  [1]{#1}%
\providecommand \bibfnamefont [1]{#1}%
\providecommand \citenamefont [1]{#1}%
\providecommand \href@noop [0]{\@secondoftwo}%
\providecommand \href [0]{\begingroup \@sanitize@url \@href}%
\providecommand \@href[1]{\@@startlink{#1}\@@href}%
\providecommand \@@href[1]{\endgroup#1\@@endlink}%
\providecommand \@sanitize@url [0]{\catcode `\\12\catcode `\$12\catcode
  `\&12\catcode `\#12\catcode `\^12\catcode `\_12\catcode `\%12\relax}%
\providecommand \@@startlink[1]{}%
\providecommand \@@endlink[0]{}%
\providecommand \url  [0]{\begingroup\@sanitize@url \@url }%
\providecommand \@url [1]{\endgroup\@href {#1}{\urlprefix }}%
\providecommand \urlprefix  [0]{URL }%
\providecommand \Eprint [0]{\href }%
\providecommand \doibase [0]{http://dx.doi.org/}%
\providecommand \selectlanguage [0]{\@gobble}%
\providecommand \bibinfo  [0]{\@secondoftwo}%
\providecommand \bibfield  [0]{\@secondoftwo}%
\providecommand \translation [1]{[#1]}%
\providecommand \BibitemOpen [0]{}%
\providecommand \bibitemStop [0]{}%
\providecommand \bibitemNoStop [0]{.\EOS\space}%
\providecommand \EOS [0]{\spacefactor3000\relax}%
\providecommand \BibitemShut  [1]{\csname bibitem#1\endcsname}%
\let\auto@bib@innerbib\@empty
%</preamble>
\bibitem [{\citenamefont {Bloch}\ \emph {et~al.}(2008)\citenamefont {Bloch},
  \citenamefont {Dalibard},\ and\ \citenamefont {Zwerger}}]{bloch08}%
  \BibitemOpen
  \bibfield  {author} {\bibinfo {author} {\bibfnamefont {I.}~\bibnamefont
  {Bloch}}, \bibinfo {author} {\bibfnamefont {J.}~\bibnamefont {Dalibard}}, \
  and\ \bibinfo {author} {\bibfnamefont {W.}~\bibnamefont {Zwerger}},\
  }\href@noop {} {\bibfield  {journal} {\bibinfo  {journal} {Rev. Mod. Phys.}\
  }\textbf {\bibinfo {volume} {80}},\ \bibinfo {pages} {885} (\bibinfo {year}
  {2008})}\BibitemShut {NoStop}%
\bibitem [{\citenamefont {Greiner}\ \emph {et~al.}(2002)\citenamefont
  {Greiner}, \citenamefont {Mandel}, \citenamefont {H\"ansch},\ and\
  \citenamefont {Bloch}}]{greiner02}%
  \BibitemOpen
  \bibfield  {author} {\bibinfo {author} {\bibfnamefont {M.}~\bibnamefont
  {Greiner}}, \bibinfo {author} {\bibfnamefont {O.}~\bibnamefont {Mandel}},
  \bibinfo {author} {\bibfnamefont {T.}~\bibnamefont {H\"ansch}}, \ and\
  \bibinfo {author} {\bibfnamefont {I.}~\bibnamefont {Bloch}},\ }\href@noop {}
  {\bibfield  {journal} {\bibinfo  {journal} {Nature (London)}\ }\textbf
  {\bibinfo {volume} {419}},\ \bibinfo {pages} {51} (\bibinfo {year}
  {2002})}\BibitemShut {NoStop}%
\bibitem [{\citenamefont {Kinoshita}\ \emph {et~al.}(2006)\citenamefont
  {Kinoshita}, \citenamefont {Wenger},\ and\ \citenamefont
  {Weiss}}]{kinoshita06}%
  \BibitemOpen
  \bibfield  {author} {\bibinfo {author} {\bibfnamefont {T.}~\bibnamefont
  {Kinoshita}}, \bibinfo {author} {\bibfnamefont {T.}~\bibnamefont {Wenger}}, \
  and\ \bibinfo {author} {\bibfnamefont {D.~S.}\ \bibnamefont {Weiss}},\
  }\href@noop {} {\bibfield  {journal} {\bibinfo  {journal} {Nature}\ }\textbf
  {\bibinfo {volume} {440}},\ \bibinfo {pages} {900} (\bibinfo {year}
  {2006})}\BibitemShut {NoStop}%
\bibitem [{\citenamefont {Hofferberth}\ \emph {et~al.}(2007)\citenamefont
  {Hofferberth}, \citenamefont {Lesanovsky}, \citenamefont {Fisher},
  \citenamefont {Schumm},\ and\ \citenamefont {Schmiedmayer}}]{hofferberth07}%
  \BibitemOpen
  \bibfield  {author} {\bibinfo {author} {\bibfnamefont {S.}~\bibnamefont
  {Hofferberth}}, \bibinfo {author} {\bibfnamefont {I.}~\bibnamefont
  {Lesanovsky}}, \bibinfo {author} {\bibfnamefont {B.}~\bibnamefont {Fisher}},
  \bibinfo {author} {\bibfnamefont {T.}~\bibnamefont {Schumm}}, \ and\ \bibinfo
  {author} {\bibfnamefont {J.}~\bibnamefont {Schmiedmayer}},\ }\href@noop {}
  {\bibfield  {journal} {\bibinfo  {journal} {Nature}\ }\textbf {\bibinfo
  {volume} {449}},\ \bibinfo {pages} {324} (\bibinfo {year}
  {2007})}\BibitemShut {NoStop}%
\bibitem [{\citenamefont {Trotzky}\ \emph {et~al.}(2012)\citenamefont
  {Trotzky}, \citenamefont {Chen}, \citenamefont {Flesch}, \citenamefont
  {McCulloch}, \citenamefont {Schollw\"ock}, \citenamefont {Eisert},\ and\
  \citenamefont {Bloch}}]{trotzky12}%
  \BibitemOpen
  \bibfield  {author} {\bibinfo {author} {\bibfnamefont {S.}~\bibnamefont
  {Trotzky}}, \bibinfo {author} {\bibfnamefont {Y.-A.}\ \bibnamefont {Chen}},
  \bibinfo {author} {\bibfnamefont {A.}~\bibnamefont {Flesch}}, \bibinfo
  {author} {\bibfnamefont {I.~P.}\ \bibnamefont {McCulloch}}, \bibinfo {author}
  {\bibfnamefont {U.}~\bibnamefont {Schollw\"ock}}, \bibinfo {author}
  {\bibfnamefont {J.}~\bibnamefont {Eisert}}, \ and\ \bibinfo {author}
  {\bibfnamefont {I.}~\bibnamefont {Bloch}},\ }\href@noop {} {\bibfield
  {journal} {\bibinfo  {journal} {Nature Phys.}\ }\textbf {\bibinfo {volume}
  {8}},\ \bibinfo {pages} {325} (\bibinfo {year} {2012})}\BibitemShut {NoStop}%
\bibitem [{\citenamefont {Meinert}\ \emph {et~al.}(2013)\citenamefont
  {Meinert}, \citenamefont {Mark}, \citenamefont {Kirilov}, \citenamefont
  {Lauber}, \citenamefont {Weinmann}, \citenamefont {Daley},\ and\
  \citenamefont {N\"agerl}}]{meinert13}%
  \BibitemOpen
  \bibfield  {author} {\bibinfo {author} {\bibfnamefont {F.}~\bibnamefont
  {Meinert}}, \bibinfo {author} {\bibfnamefont {M.~J.}\ \bibnamefont {Mark}},
  \bibinfo {author} {\bibfnamefont {E.}~\bibnamefont {Kirilov}}, \bibinfo
  {author} {\bibfnamefont {K.}~\bibnamefont {Lauber}}, \bibinfo {author}
  {\bibfnamefont {P.}~\bibnamefont {Weinmann}}, \bibinfo {author}
  {\bibfnamefont {A.~J.}\ \bibnamefont {Daley}}, \ and\ \bibinfo {author}
  {\bibfnamefont {H.-C.}\ \bibnamefont {N\"agerl}},\ }\href@noop {} {\bibfield
  {journal} {\bibinfo  {journal} {Phys. Rev. Lett.}\ }\textbf {\bibinfo
  {volume} {111}},\ \bibinfo {pages} {053003} (\bibinfo {year}
  {2013})}\BibitemShut {NoStop}%
\bibitem [{\citenamefont {Fertig}\ \emph {et~al.}(2005)\citenamefont {Fertig},
  \citenamefont {O'Hara}, \citenamefont {Huckans}, \citenamefont {Rolston},
  \citenamefont {Phillips},\ and\ \citenamefont {Porto}}]{fertig05}%
  \BibitemOpen
  \bibfield  {author} {\bibinfo {author} {\bibfnamefont {C.~D.}\ \bibnamefont
  {Fertig}}, \bibinfo {author} {\bibfnamefont {K.~M.}\ \bibnamefont {O'Hara}},
  \bibinfo {author} {\bibfnamefont {J.~H.}\ \bibnamefont {Huckans}}, \bibinfo
  {author} {\bibfnamefont {S.~L.}\ \bibnamefont {Rolston}}, \bibinfo {author}
  {\bibfnamefont {W.~D.}\ \bibnamefont {Phillips}}, \ and\ \bibinfo {author}
  {\bibfnamefont {J.~V.}\ \bibnamefont {Porto}},\ }\href@noop {} {\bibfield
  {journal} {\bibinfo  {journal} {Phys. Rev. Lett.}\ }\textbf {\bibinfo
  {volume} {94}},\ \bibinfo {pages} {120403} (\bibinfo {year}
  {2005})}\BibitemShut {NoStop}%
\bibitem [{\citenamefont {Brantut}\ \emph {et~al.}(2012)\citenamefont
  {Brantut}, \citenamefont {Meineke}, \citenamefont {Stadler}, \citenamefont
  {Krinner},\ and\ \citenamefont {Esslinger}}]{brantut12}%
  \BibitemOpen
  \bibfield  {author} {\bibinfo {author} {\bibfnamefont {J.-P.}\ \bibnamefont
  {Brantut}}, \bibinfo {author} {\bibfnamefont {J.}~\bibnamefont {Meineke}},
  \bibinfo {author} {\bibfnamefont {D.}~\bibnamefont {Stadler}}, \bibinfo
  {author} {\bibfnamefont {S.}~\bibnamefont {Krinner}}, \ and\ \bibinfo
  {author} {\bibfnamefont {T.}~\bibnamefont {Esslinger}},\ }\href@noop {}
  {\bibfield  {journal} {\bibinfo  {journal} {Science}\ }\textbf {\bibinfo
  {volume} {337}},\ \bibinfo {pages} {1069} (\bibinfo {year}
  {2012})}\BibitemShut {NoStop}%
\bibitem [{\citenamefont {Stadler}\ \emph {et~al.}(2012)\citenamefont
  {Stadler}, \citenamefont {Krinner}, \citenamefont {Meineke}, \citenamefont
  {Brantut},\ and\ \citenamefont {Esslinger}}]{stadler12}%
  \BibitemOpen
  \bibfield  {author} {\bibinfo {author} {\bibfnamefont {D.}~\bibnamefont
  {Stadler}}, \bibinfo {author} {\bibfnamefont {S.}~\bibnamefont {Krinner}},
  \bibinfo {author} {\bibfnamefont {J.}~\bibnamefont {Meineke}}, \bibinfo
  {author} {\bibfnamefont {J.-P.}\ \bibnamefont {Brantut}}, \ and\ \bibinfo
  {author} {\bibfnamefont {T.}~\bibnamefont {Esslinger}},\ }\href@noop {}
  {\bibfield  {journal} {\bibinfo  {journal} {Nature}\ }\textbf {\bibinfo
  {volume} {491}},\ \bibinfo {pages} {736} (\bibinfo {year}
  {2012})}\BibitemShut {NoStop}%
\bibitem [{\citenamefont {Schneider}\ \emph {et~al.}(2012)\citenamefont
  {Schneider}, \citenamefont {Hackerm\"uller}, \citenamefont {Ronzheimer},
  \citenamefont {Will}, \citenamefont {Braun}, \citenamefont {Best},
  \citenamefont {Bloch}, \citenamefont {Demler}, \citenamefont {Mandt},
  \citenamefont {Rasch},\ and\ \citenamefont {Rosch}}]{schneider12}%
  \BibitemOpen
  \bibfield  {author} {\bibinfo {author} {\bibfnamefont {U.}~\bibnamefont
  {Schneider}}, \bibinfo {author} {\bibfnamefont {L.}~\bibnamefont
  {Hackerm\"uller}}, \bibinfo {author} {\bibfnamefont {J.~P.}\ \bibnamefont
  {Ronzheimer}}, \bibinfo {author} {\bibfnamefont {S.}~\bibnamefont {Will}},
  \bibinfo {author} {\bibfnamefont {S.}~\bibnamefont {Braun}}, \bibinfo
  {author} {\bibfnamefont {T.}~\bibnamefont {Best}}, \bibinfo {author}
  {\bibfnamefont {I.}~\bibnamefont {Bloch}}, \bibinfo {author} {\bibfnamefont
  {E.}~\bibnamefont {Demler}}, \bibinfo {author} {\bibfnamefont
  {S.}~\bibnamefont {Mandt}}, \bibinfo {author} {\bibfnamefont
  {D.}~\bibnamefont {Rasch}}, \ and\ \bibinfo {author} {\bibfnamefont
  {A.}~\bibnamefont {Rosch}},\ }\href@noop {} {\bibfield  {journal} {\bibinfo
  {journal} {Nature Phys.}\ }\textbf {\bibinfo {volume} {8}},\ \bibinfo {pages}
  {213} (\bibinfo {year} {2012})}\BibitemShut {NoStop}%
\bibitem [{\citenamefont {Ronzheimer}\ \emph {et~al.}(2013)\citenamefont
  {Ronzheimer}, \citenamefont {Schreiber}, \citenamefont {Braun}, \citenamefont
  {Hodgman}, \citenamefont {Langer}, \citenamefont {McCulloch}, \citenamefont
  {Heidrich-Meisner}, \citenamefont {Bloch},\ and\ \citenamefont
  {Schneider}}]{ronzheimer13}%
  \BibitemOpen
  \bibfield  {author} {\bibinfo {author} {\bibfnamefont {J.}~\bibnamefont
  {Ronzheimer}}, \bibinfo {author} {\bibfnamefont {M.}~\bibnamefont
  {Schreiber}}, \bibinfo {author} {\bibfnamefont {S.}~\bibnamefont {Braun}},
  \bibinfo {author} {\bibfnamefont {S.}~\bibnamefont {Hodgman}}, \bibinfo
  {author} {\bibfnamefont {S.}~\bibnamefont {Langer}}, \bibinfo {author}
  {\bibfnamefont {I.}~\bibnamefont {McCulloch}}, \bibinfo {author}
  {\bibfnamefont {F.}~\bibnamefont {Heidrich-Meisner}}, \bibinfo {author}
  {\bibfnamefont {I.}~\bibnamefont {Bloch}}, \ and\ \bibinfo {author}
  {\bibfnamefont {U.}~\bibnamefont {Schneider}},\ }\href@noop {} {\bibfield
  {journal} {\bibinfo  {journal} {Phys. Rev. Lett.}\ }\textbf {\bibinfo
  {volume} {110}},\ \bibinfo {pages} {205301} (\bibinfo {year}
  {2013})}\BibitemShut {NoStop}%
\bibitem [{\citenamefont {Reinhard}\ \emph {et~al.}(2013)\citenamefont
  {Reinhard}, \citenamefont {Riou}, \citenamefont {Zundel}, \citenamefont
  {Weiss}, \citenamefont {Li}, \citenamefont {Rey},\ and\ \citenamefont
  {Hipolito}}]{reinhard13}%
  \BibitemOpen
  \bibfield  {author} {\bibinfo {author} {\bibfnamefont {A.}~\bibnamefont
  {Reinhard}}, \bibinfo {author} {\bibfnamefont {J.-F.}\ \bibnamefont {Riou}},
  \bibinfo {author} {\bibfnamefont {L.~A.}\ \bibnamefont {Zundel}}, \bibinfo
  {author} {\bibfnamefont {D.~S.}\ \bibnamefont {Weiss}}, \bibinfo {author}
  {\bibfnamefont {S.}~\bibnamefont {Li}}, \bibinfo {author} {\bibfnamefont
  {A.~M.}\ \bibnamefont {Rey}}, \ and\ \bibinfo {author} {\bibfnamefont
  {R.}~\bibnamefont {Hipolito}},\ }\href@noop {} {\bibfield  {journal}
  {\bibinfo  {journal} {Phys. Rev. Lett.}\ }\textbf {\bibinfo {volume} {110}},\
  \bibinfo {pages} {033001} (\bibinfo {year} {2013})}\BibitemShut {NoStop}%
\bibitem [{\citenamefont {Li}\ \emph {et~al.}(2013)\citenamefont {Li},
  \citenamefont {Manmana}, \citenamefont {Rey}, \citenamefont {Hipolito},
  \citenamefont {Reinhard}, \citenamefont {Riou}, \citenamefont {Zundel},\ and\
  \citenamefont {Weiss}}]{li13}%
  \BibitemOpen
  \bibfield  {author} {\bibinfo {author} {\bibfnamefont {S.}~\bibnamefont
  {Li}}, \bibinfo {author} {\bibfnamefont {S.}~\bibnamefont {Manmana}},
  \bibinfo {author} {\bibfnamefont {A.~M.}\ \bibnamefont {Rey}}, \bibinfo
  {author} {\bibfnamefont {R.}~\bibnamefont {Hipolito}}, \bibinfo {author}
  {\bibfnamefont {A.}~\bibnamefont {Reinhard}}, \bibinfo {author}
  {\bibfnamefont {J.-F.}\ \bibnamefont {Riou}}, \bibinfo {author}
  {\bibfnamefont {L.~A.}\ \bibnamefont {Zundel}}, \ and\ \bibinfo {author}
  {\bibfnamefont {D.~S.}\ \bibnamefont {Weiss}},\ }\href@noop {} {\bibfield
  {journal} {\bibinfo  {journal} {Phys. Rev. A}\ }\textbf {\bibinfo {volume}
  {88}},\ \bibinfo {pages} {023419} (\bibinfo {year} {2013})}\BibitemShut
  {NoStop}%
\bibitem [{\citenamefont {Polkovnikov}\ \emph {et~al.}(2011)\citenamefont
  {Polkovnikov}, \citenamefont {Sengupta}, \citenamefont {Silva},\ and\
  \citenamefont {Vengalattore}}]{polkovnikov11}%
  \BibitemOpen
  \bibfield  {author} {\bibinfo {author} {\bibfnamefont {A.}~\bibnamefont
  {Polkovnikov}}, \bibinfo {author} {\bibfnamefont {K.}~\bibnamefont
  {Sengupta}}, \bibinfo {author} {\bibfnamefont {A.}~\bibnamefont {Silva}}, \
  and\ \bibinfo {author} {\bibfnamefont {M.}~\bibnamefont {Vengalattore}},\
  }\href@noop {} {\bibfield  {journal} {\bibinfo  {journal} {Rev. Mod. Phys}\
  }\textbf {\bibinfo {volume} {83}},\ \bibinfo {pages} {863} (\bibinfo {year}
  {2011})}\BibitemShut {NoStop}%
\bibitem [{\citenamefont {Rigol}\ \emph {et~al.}(2008)\citenamefont {Rigol},
  \citenamefont {Dunjko},\ and\ \citenamefont {Olshanii}}]{rigol08}%
  \BibitemOpen
  \bibfield  {author} {\bibinfo {author} {\bibfnamefont {M.}~\bibnamefont
  {Rigol}}, \bibinfo {author} {\bibfnamefont {V.}~\bibnamefont {Dunjko}}, \
  and\ \bibinfo {author} {\bibfnamefont {M.}~\bibnamefont {Olshanii}},\
  }\href@noop {} {\bibfield  {journal} {\bibinfo  {journal} {Nature}\ }\textbf
  {\bibinfo {volume} {452}},\ \bibinfo {pages} {854} (\bibinfo {year}
  {2008})}\BibitemShut {NoStop}%
\bibitem [{\citenamefont {Zotos}\ \emph {et~al.}(1997)\citenamefont {Zotos},
  \citenamefont {Naef},\ and\ \citenamefont {Prelov{\v{s}}ek}}]{zotos97}%
  \BibitemOpen
  \bibfield  {author} {\bibinfo {author} {\bibfnamefont {X.}~\bibnamefont
  {Zotos}}, \bibinfo {author} {\bibfnamefont {F.}~\bibnamefont {Naef}}, \ and\
  \bibinfo {author} {\bibfnamefont {P.}~\bibnamefont {Prelov{\v{s}}ek}},\
  }\href@noop {} {\bibfield  {journal} {\bibinfo  {journal} {Phys. Rev. B}\
  }\textbf {\bibinfo {volume} {55}},\ \bibinfo {pages} {11029} (\bibinfo {year}
  {1997})}\BibitemShut {NoStop}%
\bibitem [{\citenamefont {Prosen}(2011)}]{prosen11}%
  \BibitemOpen
  \bibfield  {author} {\bibinfo {author} {\bibfnamefont {T.}~\bibnamefont
  {Prosen}},\ }\href@noop {} {\bibfield  {journal} {\bibinfo  {journal} {Phys.
  Rev. Lett.}\ }\textbf {\bibinfo {volume} {106}},\ \bibinfo {pages} {217206}
  (\bibinfo {year} {2011})}\BibitemShut {NoStop}%
\bibitem [{\citenamefont {Sirker}\ \emph {et~al.}(2011)\citenamefont {Sirker},
  \citenamefont {Pereira},\ and\ \citenamefont {Affleck}}]{sirker11}%
  \BibitemOpen
  \bibfield  {author} {\bibinfo {author} {\bibfnamefont {J.}~\bibnamefont
  {Sirker}}, \bibinfo {author} {\bibfnamefont {R.~G.}\ \bibnamefont {Pereira}},
  \ and\ \bibinfo {author} {\bibfnamefont {I.}~\bibnamefont {Affleck}},\
  }\href@noop {} {\bibfield  {journal} {\bibinfo  {journal} {Phys. Rev. B}\
  }\textbf {\bibinfo {volume} {83}},\ \bibinfo {pages} {035115} (\bibinfo
  {year} {2011})}\BibitemShut {NoStop}%
\bibitem [{\citenamefont {Heidrich-Meisner}\ \emph {et~al.}(2007)\citenamefont
  {Heidrich-Meisner}, \citenamefont {Honecker},\ and\ \citenamefont
  {Brenig}}]{heidrichmeisner07}%
  \BibitemOpen
  \bibfield  {author} {\bibinfo {author} {\bibfnamefont {F.}~\bibnamefont
  {Heidrich-Meisner}}, \bibinfo {author} {\bibfnamefont {A.}~\bibnamefont
  {Honecker}}, \ and\ \bibinfo {author} {\bibfnamefont {W.}~\bibnamefont
  {Brenig}},\ }\href@noop {} {\bibfield  {journal} {\bibinfo  {journal} {Eur.
  Phys. J. Special Topics}\ }\textbf {\bibinfo {volume} {151}},\ \bibinfo
  {pages} {135} (\bibinfo {year} {2007})}\BibitemShut {NoStop}%
\bibitem [{\citenamefont {Peredes}\ \emph {et~al.}(2004)\citenamefont
  {Peredes}, \citenamefont {Widera}, \citenamefont {Murg}, \citenamefont
  {Mandel}, \citenamefont {F\"olling}, \citenamefont {Cirac}, \citenamefont
  {Shlyapnikov}, \citenamefont {H\"ansch},\ and\ \citenamefont
  {Bloch}}]{peredes04}%
  \BibitemOpen
  \bibfield  {author} {\bibinfo {author} {\bibfnamefont {B.}~\bibnamefont
  {Peredes}}, \bibinfo {author} {\bibfnamefont {A.}~\bibnamefont {Widera}},
  \bibinfo {author} {\bibfnamefont {V.}~\bibnamefont {Murg}}, \bibinfo {author}
  {\bibfnamefont {O.}~\bibnamefont {Mandel}}, \bibinfo {author} {\bibfnamefont
  {S.}~\bibnamefont {F\"olling}}, \bibinfo {author} {\bibfnamefont
  {I.}~\bibnamefont {Cirac}}, \bibinfo {author} {\bibfnamefont {G.~V.}\
  \bibnamefont {Shlyapnikov}}, \bibinfo {author} {\bibfnamefont {T.~W.}\
  \bibnamefont {H\"ansch}}, \ and\ \bibinfo {author} {\bibfnamefont
  {I.}~\bibnamefont {Bloch}},\ }\href@noop {} {\bibfield  {journal} {\bibinfo
  {journal} {Nature}\ }\textbf {\bibinfo {volume} {429}},\ \bibinfo {pages}
  {277} (\bibinfo {year} {2004})}\BibitemShut {NoStop}%
\bibitem [{\citenamefont {Kinoshita}\ \emph {et~al.}(2004)\citenamefont
  {Kinoshita}, \citenamefont {Wenger},\ and\ \citenamefont
  {Weiss}}]{kinoshita04}%
  \BibitemOpen
  \bibfield  {author} {\bibinfo {author} {\bibfnamefont {T.}~\bibnamefont
  {Kinoshita}}, \bibinfo {author} {\bibfnamefont {T.}~\bibnamefont {Wenger}}, \
  and\ \bibinfo {author} {\bibfnamefont {D.~S.}\ \bibnamefont {Weiss}},\
  }\href@noop {} {\bibfield  {journal} {\bibinfo  {journal} {Science}\ }\textbf
  {\bibinfo {volume} {305}},\ \bibinfo {pages} {1125} (\bibinfo {year}
  {2004})}\BibitemShut {NoStop}%
\bibitem [{\citenamefont {Caux}\ and\ \citenamefont {Mossel}()}]{caux11}%
  \BibitemOpen
  \bibfield  {author} {\bibinfo {author} {\bibfnamefont {J.~S.}\ \bibnamefont
  {Caux}}\ and\ \bibinfo {author} {\bibfnamefont {J.}~\bibnamefont {Mossel}},\
  }\href@noop {} {\bibfield  {journal} {\bibinfo  {journal} {J. Stat. Mech.}\
  }\textbf {\bibinfo {volume} {{\rm\bf(2011)}}},\ \bibinfo {pages}
  {P02023}}\BibitemShut {NoStop}%
\bibitem [{\citenamefont {Rigol}\ \emph {et~al.}(2007)\citenamefont {Rigol},
  \citenamefont {Dunjko}, \citenamefont {Yurovsky},\ and\ \citenamefont
  {Olshanii}}]{rigol07}%
  \BibitemOpen
  \bibfield  {author} {\bibinfo {author} {\bibfnamefont {M.}~\bibnamefont
  {Rigol}}, \bibinfo {author} {\bibfnamefont {V.}~\bibnamefont {Dunjko}},
  \bibinfo {author} {\bibfnamefont {V.}~\bibnamefont {Yurovsky}}, \ and\
  \bibinfo {author} {\bibfnamefont {M.}~\bibnamefont {Olshanii}},\ }\href@noop
  {} {\bibfield  {journal} {\bibinfo  {journal} {Phys. Rev. Lett.}\ }\textbf
  {\bibinfo {volume} {98}},\ \bibinfo {pages} {050405} (\bibinfo {year}
  {2007})}\BibitemShut {NoStop}%
\bibitem [{\citenamefont {Manmana}\ \emph {et~al.}(2007)\citenamefont
  {Manmana}, \citenamefont {Wessel}, \citenamefont {Noack},\ and\ \citenamefont
  {Muramatsu}}]{manmana07}%
  \BibitemOpen
  \bibfield  {author} {\bibinfo {author} {\bibfnamefont {S.}~\bibnamefont
  {Manmana}}, \bibinfo {author} {\bibfnamefont {S.}~\bibnamefont {Wessel}},
  \bibinfo {author} {\bibfnamefont {R.}~\bibnamefont {Noack}}, \ and\ \bibinfo
  {author} {\bibfnamefont {A.}~\bibnamefont {Muramatsu}},\ }\href@noop {}
  {\bibfield  {journal} {\bibinfo  {journal} {Phys. Rev. Lett.}\ }\textbf
  {\bibinfo {volume} {98}},\ \bibinfo {pages} {210405} (\bibinfo {year}
  {2007})}\BibitemShut {NoStop}%
\bibitem [{\citenamefont {Barthel}\ and\ \citenamefont
  {Schollw\"{o}ck}(2008)}]{barthel08}%
  \BibitemOpen
  \bibfield  {author} {\bibinfo {author} {\bibfnamefont {T.}~\bibnamefont
  {Barthel}}\ and\ \bibinfo {author} {\bibfnamefont {U.}~\bibnamefont
  {Schollw\"{o}ck}},\ }\href@noop {} {\bibfield  {journal} {\bibinfo  {journal}
  {Phys. Rev. Lett.}\ }\textbf {\bibinfo {volume} {100}},\ \bibinfo {eid}
  {100601} (\bibinfo {year} {2008})}\BibitemShut {NoStop}%
\bibitem [{\citenamefont {Shimshoni}\ \emph {et~al.}(2003)\citenamefont
  {Shimshoni}, \citenamefont {Andrei},\ and\ \citenamefont
  {Rosch}}]{shimshoni03}%
  \BibitemOpen
  \bibfield  {author} {\bibinfo {author} {\bibfnamefont {E.}~\bibnamefont
  {Shimshoni}}, \bibinfo {author} {\bibfnamefont {N.}~\bibnamefont {Andrei}}, \
  and\ \bibinfo {author} {\bibfnamefont {A.}~\bibnamefont {Rosch}},\
  }\href@noop {} {\bibfield  {journal} {\bibinfo  {journal} {Phys. Rev. B}\
  }\textbf {\bibinfo {volume} {68}},\ \bibinfo {pages} {104401} (\bibinfo
  {year} {2003})}\BibitemShut {NoStop}%
\bibitem [{\citenamefont {Louis}\ \emph {et~al.}(2006)\citenamefont {Louis},
  \citenamefont {Prelov\ifmmode~\check{s}\else \v{s}\fi{}ek},\ and\
  \citenamefont {Zotos}}]{louis06}%
  \BibitemOpen
  \bibfield  {author} {\bibinfo {author} {\bibfnamefont {K.}~\bibnamefont
  {Louis}}, \bibinfo {author} {\bibfnamefont {P.}~\bibnamefont
  {Prelov\ifmmode~\check{s}\else \v{s}\fi{}ek}}, \ and\ \bibinfo {author}
  {\bibfnamefont {X.}~\bibnamefont {Zotos}},\ }\href@noop {} {\bibfield
  {journal} {\bibinfo  {journal} {Phys. Rev. B}\ }\textbf {\bibinfo {volume}
  {74}},\ \bibinfo {pages} {235118} (\bibinfo {year} {2006})}\BibitemShut
  {NoStop}%
\bibitem [{\citenamefont {Chernyshev}\ and\ \citenamefont
  {Rozhkov}(2005)}]{chernychev05}%
  \BibitemOpen
  \bibfield  {author} {\bibinfo {author} {\bibfnamefont {A.~L.}\ \bibnamefont
  {Chernyshev}}\ and\ \bibinfo {author} {\bibfnamefont {A.~V.}\ \bibnamefont
  {Rozhkov}},\ }\href@noop {} {\bibfield  {journal} {\bibinfo  {journal} {Phys.
  Rev. B}\ }\textbf {\bibinfo {volume} {72}},\ \bibinfo {pages} {104423}
  (\bibinfo {year} {2005})}\BibitemShut {NoStop}%
\bibitem [{\citenamefont {Hess}(2007)}]{hess07}%
  \BibitemOpen
  \bibfield  {author} {\bibinfo {author} {\bibfnamefont {C.}~\bibnamefont
  {Hess}},\ }\href@noop {} {\bibfield  {journal} {\bibinfo  {journal} {Eur.
  Phys. J. Spec. Topics}\ }\textbf {\bibinfo {volume} {151}},\ \bibinfo {pages}
  {73} (\bibinfo {year} {2007})}\BibitemShut {NoStop}%
\bibitem [{\citenamefont {Hlubek}\ \emph {et~al.}(2010)\citenamefont {Hlubek},
  \citenamefont {Ribeiro}, \citenamefont {Saint-Martin}, \citenamefont
  {Revcolevschi}, \citenamefont {Roth}, \citenamefont {Behr}, \citenamefont
  {B\"uchner},\ and\ \citenamefont {Hess}}]{hlubek10}%
  \BibitemOpen
  \bibfield  {author} {\bibinfo {author} {\bibfnamefont {N.}~\bibnamefont
  {Hlubek}}, \bibinfo {author} {\bibfnamefont {P.}~\bibnamefont {Ribeiro}},
  \bibinfo {author} {\bibfnamefont {R.}~\bibnamefont {Saint-Martin}}, \bibinfo
  {author} {\bibfnamefont {A.}~\bibnamefont {Revcolevschi}}, \bibinfo {author}
  {\bibfnamefont {G.}~\bibnamefont {Roth}}, \bibinfo {author} {\bibfnamefont
  {G.}~\bibnamefont {Behr}}, \bibinfo {author} {\bibfnamefont {B.}~\bibnamefont
  {B\"uchner}}, \ and\ \bibinfo {author} {\bibfnamefont {C.}~\bibnamefont
  {Hess}},\ }\href@noop {} {\bibfield  {journal} {\bibinfo  {journal} {Phys.
  Rev. B}\ }\textbf {\bibinfo {volume} {81}},\ \bibinfo {pages} {020405}
  (\bibinfo {year} {2010})}\BibitemShut {NoStop}%
\bibitem [{\citenamefont {Sologubenko}\ \emph {et~al.}(2007)\citenamefont
  {Sologubenko}, \citenamefont {Lorenz}, \citenamefont {Ott},\ and\
  \citenamefont {Freimuth}}]{sologubenko07}%
  \BibitemOpen
  \bibfield  {author} {\bibinfo {author} {\bibfnamefont {A.~V.}\ \bibnamefont
  {Sologubenko}}, \bibinfo {author} {\bibfnamefont {T.}~\bibnamefont {Lorenz}},
  \bibinfo {author} {\bibfnamefont {H.~R.}\ \bibnamefont {Ott}}, \ and\
  \bibinfo {author} {\bibfnamefont {A.}~\bibnamefont {Freimuth}},\ }\href@noop
  {} {\bibfield  {journal} {\bibinfo  {journal} {J. Low Temp. Phys.}\ }\textbf
  {\bibinfo {volume} {147}},\ \bibinfo {pages} {387} (\bibinfo {year}
  {2007})}\BibitemShut {NoStop}%
\bibitem [{\citenamefont {Heidrich-Meisner}\ \emph {et~al.}(2003)\citenamefont
  {Heidrich-Meisner}, \citenamefont {Honecker}, \citenamefont {Cabra},\ and\
  \citenamefont {Brenig}}]{heidrichmeisner03}%
  \BibitemOpen
  \bibfield  {author} {\bibinfo {author} {\bibfnamefont {F.}~\bibnamefont
  {Heidrich-Meisner}}, \bibinfo {author} {\bibfnamefont {A.}~\bibnamefont
  {Honecker}}, \bibinfo {author} {\bibfnamefont {D.~C.}\ \bibnamefont {Cabra}},
  \ and\ \bibinfo {author} {\bibfnamefont {W.}~\bibnamefont {Brenig}},\
  }\href@noop {} {\bibfield  {journal} {\bibinfo  {journal} {Phys. Rev. B}\
  }\textbf {\bibinfo {volume} {68}},\ \bibinfo {pages} {134436} (\bibinfo
  {year} {2003})}\BibitemShut {NoStop}%
\bibitem [{\citenamefont {Heidrich-Meisner}\ \emph {et~al.}(2004)\citenamefont
  {Heidrich-Meisner}, \citenamefont {Honecker}, \citenamefont {Cabra},\ and\
  \citenamefont {Brenig}}]{hm04}%
  \BibitemOpen
  \bibfield  {author} {\bibinfo {author} {\bibfnamefont {F.}~\bibnamefont
  {Heidrich-Meisner}}, \bibinfo {author} {\bibfnamefont {A.}~\bibnamefont
  {Honecker}}, \bibinfo {author} {\bibfnamefont {D.~C.}\ \bibnamefont {Cabra}},
  \ and\ \bibinfo {author} {\bibfnamefont {W.}~\bibnamefont {Brenig}},\
  }\href@noop {} {\bibfield  {journal} {\bibinfo  {journal} {Phys. Rev. Lett.}\
  }\textbf {\bibinfo {volume} {92}},\ \bibinfo {pages} {069703} (\bibinfo
  {year} {2004})}\BibitemShut {NoStop}%
\bibitem [{\citenamefont {Zotos}(2004)}]{zotos04}%
  \BibitemOpen
  \bibfield  {author} {\bibinfo {author} {\bibfnamefont {X.}~\bibnamefont
  {Zotos}},\ }\href@noop {} {\bibfield  {journal} {\bibinfo  {journal} {Phys.
  Rev. Lett.}\ }\textbf {\bibinfo {volume} {92}},\ \bibinfo {pages} {067202}
  (\bibinfo {year} {2004})}\BibitemShut {NoStop}%
\bibitem [{\citenamefont {Jung}\ \emph {et~al.}(2006)\citenamefont {Jung},
  \citenamefont {Helmes},\ and\ \citenamefont {Rosch}}]{jung06}%
  \BibitemOpen
  \bibfield  {author} {\bibinfo {author} {\bibfnamefont {P.}~\bibnamefont
  {Jung}}, \bibinfo {author} {\bibfnamefont {R.~W.}\ \bibnamefont {Helmes}}, \
  and\ \bibinfo {author} {\bibfnamefont {A.}~\bibnamefont {Rosch}},\
  }\href@noop {} {\bibfield  {journal} {\bibinfo  {journal} {Phys. Rev. Lett.}\
  }\textbf {\bibinfo {volume} {96}},\ \bibinfo {pages} {067202} (\bibinfo
  {year} {2006})}\BibitemShut {NoStop}%
\bibitem [{\citenamefont {Rosch}\ and\ \citenamefont {Andrei}(2000)}]{rosch00}%
  \BibitemOpen
  \bibfield  {author} {\bibinfo {author} {\bibfnamefont {A.}~\bibnamefont
  {Rosch}}\ and\ \bibinfo {author} {\bibfnamefont {N.}~\bibnamefont {Andrei}},\
  }\href@noop {} {\bibfield  {journal} {\bibinfo  {journal} {Phys. Rev. Lett.}\
  }\textbf {\bibinfo {volume} {85}},\ \bibinfo {pages} {1092} (\bibinfo {year}
  {2000})}\BibitemShut {NoStop}%
\bibitem [{\citenamefont {Kirchner}\ \emph {et~al.}(1999)\citenamefont
  {Kirchner}, \citenamefont {Evertz},\ and\ \citenamefont
  {Hanke}}]{kirchner99}%
  \BibitemOpen
  \bibfield  {author} {\bibinfo {author} {\bibfnamefont {S.}~\bibnamefont
  {Kirchner}}, \bibinfo {author} {\bibfnamefont {H.~G.}\ \bibnamefont
  {Evertz}}, \ and\ \bibinfo {author} {\bibfnamefont {W.}~\bibnamefont
  {Hanke}},\ }\href@noop {} {\bibfield  {journal} {\bibinfo  {journal} {Phys.
  Rev. B}\ }\textbf {\bibinfo {volume} {59}},\ \bibinfo {pages} {1825}
  (\bibinfo {year} {1999})}\BibitemShut {NoStop}%
\bibitem [{\citenamefont {\v{Z}nidari\v{c}}(2013)}]{znidaric13}%
  \BibitemOpen
  \bibfield  {author} {\bibinfo {author} {\bibfnamefont {M.}~\bibnamefont
  {\v{Z}nidari\v{c}}},\ }\href@noop {} {\bibfield  {journal} {\bibinfo
  {journal} {Phys. Rev. Lett.}\ }\textbf {\bibinfo {volume} {110}},\ \bibinfo
  {pages} {070602} (\bibinfo {year} {2013})}\BibitemShut {NoStop}%
\bibitem [{\citenamefont {Karrasch}\ \emph {et~al.}(2013)\citenamefont
  {Karrasch}, \citenamefont {Ilan},\ and\ \citenamefont {Moore}}]{karrasch12}%
  \BibitemOpen
  \bibfield  {author} {\bibinfo {author} {\bibfnamefont {C.}~\bibnamefont
  {Karrasch}}, \bibinfo {author} {\bibfnamefont {R.}~\bibnamefont {Ilan}}, \
  and\ \bibinfo {author} {\bibfnamefont {J.~E.}\ \bibnamefont {Moore}},\
  }\href@noop {} {\bibfield  {journal} {\bibinfo  {journal} {Phys. Rev. B}\
  }\textbf {\bibinfo {volume} {88}},\ \bibinfo {pages} {195129} (\bibinfo
  {year} {2013})}\BibitemShut {NoStop}%
\bibitem [{\citenamefont {Cazalilla}\ \emph {et~al.}(2011)\citenamefont
  {Cazalilla}, \citenamefont {Citro}, \citenamefont {Giamarchi}, \citenamefont
  {Orignac},\ and\ \citenamefont {Rigol}}]{cazalilla11}%
  \BibitemOpen
  \bibfield  {author} {\bibinfo {author} {\bibfnamefont {M.~A.}\ \bibnamefont
  {Cazalilla}}, \bibinfo {author} {\bibfnamefont {R.}~\bibnamefont {Citro}},
  \bibinfo {author} {\bibfnamefont {T.}~\bibnamefont {Giamarchi}}, \bibinfo
  {author} {\bibfnamefont {E.}~\bibnamefont {Orignac}}, \ and\ \bibinfo
  {author} {\bibfnamefont {M.}~\bibnamefont {Rigol}},\ }\href@noop {}
  {\bibfield  {journal} {\bibinfo  {journal} {Rev. Mod. Phys.}\ }\textbf
  {\bibinfo {volume} {83}},\ \bibinfo {pages} {1405} (\bibinfo {year}
  {2011})}\BibitemShut {NoStop}%
\bibitem [{\citenamefont {Rigol}\ and\ \citenamefont
  {Muramatsu}(2005{\natexlab{a}})}]{rigol05}%
  \BibitemOpen
  \bibfield  {author} {\bibinfo {author} {\bibfnamefont {M.}~\bibnamefont
  {Rigol}}\ and\ \bibinfo {author} {\bibfnamefont {A.}~\bibnamefont
  {Muramatsu}},\ }\href@noop {} {\bibfield  {journal} {\bibinfo  {journal}
  {Phys. Rev. Lett.}\ }\textbf {\bibinfo {volume} {94}},\ \bibinfo {pages}
  {240403} (\bibinfo {year} {2005}{\natexlab{a}})}\BibitemShut {NoStop}%
\bibitem [{\citenamefont {Rigol}\ and\ \citenamefont
  {Muramatsu}(2005{\natexlab{b}})}]{rigol05a}%
  \BibitemOpen
  \bibfield  {author} {\bibinfo {author} {\bibfnamefont {M.}~\bibnamefont
  {Rigol}}\ and\ \bibinfo {author} {\bibfnamefont {A.}~\bibnamefont
  {Muramatsu}},\ }\href@noop {} {\bibfield  {journal} {\bibinfo  {journal}
  {Mod. Phys. Lett. B}\ }\textbf {\bibinfo {volume} {19}},\ \bibinfo {pages}
  {861} (\bibinfo {year} {2005}{\natexlab{b}})}\BibitemShut {NoStop}%
\bibitem [{\citenamefont {Minguzzi}\ and\ \citenamefont
  {Gangardt}(2005)}]{minguzzi05}%
  \BibitemOpen
  \bibfield  {author} {\bibinfo {author} {\bibfnamefont {A.}~\bibnamefont
  {Minguzzi}}\ and\ \bibinfo {author} {\bibfnamefont {D.~M.}\ \bibnamefont
  {Gangardt}},\ }\href@noop {} {\bibfield  {journal} {\bibinfo  {journal}
  {Phys. Rev. Lett.}\ }\textbf {\bibinfo {volume} {94}},\ \bibinfo {pages}
  {240404} (\bibinfo {year} {2005})}\BibitemShut {NoStop}%
\bibitem [{\citenamefont {Iyer}\ and\ \citenamefont {Andrei}(2012)}]{iyer12}%
  \BibitemOpen
  \bibfield  {author} {\bibinfo {author} {\bibfnamefont {D.}~\bibnamefont
  {Iyer}}\ and\ \bibinfo {author} {\bibfnamefont {N.}~\bibnamefont {Andrei}},\
  }\href {\doibase 10.1103/PhysRevLett.109.115304} {\bibfield  {journal}
  {\bibinfo  {journal} {Phys. Rev. Lett.}\ }\textbf {\bibinfo {volume} {109}},\
  \bibinfo {pages} {115304} (\bibinfo {year} {2012})}\BibitemShut {NoStop}%
\bibitem [{\citenamefont {Iyer}\ \emph {et~al.}(2013)\citenamefont {Iyer},
  \citenamefont {Guan},\ and\ \citenamefont {Andrei}}]{iyer13}%
  \BibitemOpen
  \bibfield  {author} {\bibinfo {author} {\bibfnamefont {D.}~\bibnamefont
  {Iyer}}, \bibinfo {author} {\bibfnamefont {H.}~\bibnamefont {Guan}}, \ and\
  \bibinfo {author} {\bibfnamefont {N.}~\bibnamefont {Andrei}},\ }\href@noop {}
  {\bibfield  {journal} {\bibinfo  {journal} {Phys. Rev. A}\ }\textbf {\bibinfo
  {volume} {87}},\ \bibinfo {pages} {053628} (\bibinfo {year}
  {2013})}\BibitemShut {NoStop}%
\bibitem [{\citenamefont {\"Ohberg}\ and\ \citenamefont
  {Santos}(2002)}]{ohberg02}%
  \BibitemOpen
  \bibfield  {author} {\bibinfo {author} {\bibfnamefont {P.}~\bibnamefont
  {\"Ohberg}}\ and\ \bibinfo {author} {\bibfnamefont {L.}~\bibnamefont
  {Santos}},\ }\href {\doibase 10.1103/PhysRevLett.89.240402} {\bibfield
  {journal} {\bibinfo  {journal} {Phys. Rev. Lett.}\ }\textbf {\bibinfo
  {volume} {89}},\ \bibinfo {pages} {240402} (\bibinfo {year}
  {2002})}\BibitemShut {NoStop}%
\bibitem [{\citenamefont {Girardeau}\ and\ \citenamefont
  {Minguzzi}(2006)}]{girardeau06}%
  \BibitemOpen
  \bibfield  {author} {\bibinfo {author} {\bibfnamefont {M.~D.}\ \bibnamefont
  {Girardeau}}\ and\ \bibinfo {author} {\bibfnamefont {A.}~\bibnamefont
  {Minguzzi}},\ }\href@noop {} {\bibfield  {journal} {\bibinfo  {journal}
  {Phys. Rev. Lett.}\ }\textbf {\bibinfo {volume} {96}},\ \bibinfo {pages}
  {080404} (\bibinfo {year} {2006})}\BibitemShut {NoStop}%
\bibitem [{\citenamefont {del Campo}(2008)}]{delcampo08}%
  \BibitemOpen
  \bibfield  {author} {\bibinfo {author} {\bibfnamefont {A.}~\bibnamefont {del
  Campo}},\ }\href@noop {} {\bibfield  {journal} {\bibinfo  {journal} {Phys.
  Rev. A}\ }\textbf {\bibinfo {volume} {78}},\ \bibinfo {pages} {045602}
  (\bibinfo {year} {2008})}\BibitemShut {NoStop}%
\bibitem [{\citenamefont {Gangardt}\ and\ \citenamefont
  {Pustilnik}(2008)}]{gangardt08}%
  \BibitemOpen
  \bibfield  {author} {\bibinfo {author} {\bibfnamefont {D.~M.}\ \bibnamefont
  {Gangardt}}\ and\ \bibinfo {author} {\bibfnamefont {M.}~\bibnamefont
  {Pustilnik}},\ }\href@noop {} {\bibfield  {journal} {\bibinfo  {journal}
  {Phys. Rev. A}\ }\textbf {\bibinfo {volume} {77}},\ \bibinfo {pages}
  {041604(R)} (\bibinfo {year} {2008})}\BibitemShut {NoStop}%
\bibitem [{\citenamefont {Juki\ifmmode~\acute{c}\else \'{c}\fi{}}\ \emph
  {et~al.}(2008)\citenamefont {Juki\ifmmode~\acute{c}\else \'{c}\fi{}},
  \citenamefont {Pezer}, \citenamefont {Gasenzer},\ and\ \citenamefont
  {Buljan}}]{jukic08}%
  \BibitemOpen
  \bibfield  {author} {\bibinfo {author} {\bibfnamefont {D.}~\bibnamefont
  {Juki\ifmmode~\acute{c}\else \'{c}\fi{}}}, \bibinfo {author} {\bibfnamefont
  {R.}~\bibnamefont {Pezer}}, \bibinfo {author} {\bibfnamefont
  {T.}~\bibnamefont {Gasenzer}}, \ and\ \bibinfo {author} {\bibfnamefont
  {H.}~\bibnamefont {Buljan}},\ }\href {\doibase 10.1103/PhysRevA.78.053602}
  {\bibfield  {journal} {\bibinfo  {journal} {Phys. Rev. A}\ }\textbf {\bibinfo
  {volume} {78}},\ \bibinfo {pages} {053602} (\bibinfo {year}
  {2008})}\BibitemShut {NoStop}%
\bibitem [{\citenamefont {Juki\ifmmode~\acute{c}\else \'{c}\fi{}}\ \emph
  {et~al.}(2009)\citenamefont {Juki\ifmmode~\acute{c}\else \'{c}\fi{}},
  \citenamefont {Klajn},\ and\ \citenamefont {Buljan}}]{jukic09}%
  \BibitemOpen
  \bibfield  {author} {\bibinfo {author} {\bibfnamefont {D.}~\bibnamefont
  {Juki\ifmmode~\acute{c}\else \'{c}\fi{}}}, \bibinfo {author} {\bibfnamefont
  {B.}~\bibnamefont {Klajn}}, \ and\ \bibinfo {author} {\bibfnamefont
  {H.}~\bibnamefont {Buljan}},\ }\href@noop {} {\bibfield  {journal} {\bibinfo
  {journal} {Phys. Rev. A}\ }\textbf {\bibinfo {volume} {79}},\ \bibinfo
  {pages} {033612} (\bibinfo {year} {2009})}\BibitemShut {NoStop}%
\bibitem [{\citenamefont {Gritsev}\ \emph {et~al.}(2010)\citenamefont
  {Gritsev}, \citenamefont {Barmettler},\ and\ \citenamefont
  {Demler}}]{gritsev10}%
  \BibitemOpen
  \bibfield  {author} {\bibinfo {author} {\bibfnamefont {V.}~\bibnamefont
  {Gritsev}}, \bibinfo {author} {\bibfnamefont {P.}~\bibnamefont {Barmettler}},
  \ and\ \bibinfo {author} {\bibfnamefont {E.}~\bibnamefont {Demler}},\
  }\href@noop {} {\bibfield  {journal} {\bibinfo  {journal} {New. J. Phys.}\
  }\textbf {\bibinfo {volume} {12}},\ \bibinfo {pages} {113005} (\bibinfo
  {year} {2010})}\BibitemShut {NoStop}%
\bibitem [{\citenamefont {Caux}\ and\ \citenamefont {Konik}(2012)}]{konik12}%
  \BibitemOpen
  \bibfield  {author} {\bibinfo {author} {\bibfnamefont {J.-S.}\ \bibnamefont
  {Caux}}\ and\ \bibinfo {author} {\bibfnamefont {R.~M.}\ \bibnamefont
  {Konik}},\ }\href {\doibase 10.1103/PhysRevLett.109.175301} {\bibfield
  {journal} {\bibinfo  {journal} {Phys. Rev. Lett.}\ }\textbf {\bibinfo
  {volume} {109}},\ \bibinfo {pages} {175301} (\bibinfo {year}
  {2012})}\BibitemShut {NoStop}%
\bibitem [{\citenamefont {Bolech}\ \emph {et~al.}(2012)\citenamefont {Bolech},
  \citenamefont {Heidrich-Meisner}, \citenamefont {Langer}, \citenamefont
  {McCulloch}, \citenamefont {Orso},\ and\ \citenamefont {Rigol}}]{bolech12}%
  \BibitemOpen
  \bibfield  {author} {\bibinfo {author} {\bibfnamefont {C.~J.}\ \bibnamefont
  {Bolech}}, \bibinfo {author} {\bibfnamefont {F.}~\bibnamefont
  {Heidrich-Meisner}}, \bibinfo {author} {\bibfnamefont {S.}~\bibnamefont
  {Langer}}, \bibinfo {author} {\bibfnamefont {I.~P.}\ \bibnamefont
  {McCulloch}}, \bibinfo {author} {\bibfnamefont {G.}~\bibnamefont {Orso}}, \
  and\ \bibinfo {author} {\bibfnamefont {M.}~\bibnamefont {Rigol}},\
  }\href@noop {} {\bibfield  {journal} {\bibinfo  {journal} {Phys. Rev. Lett.}\
  }\textbf {\bibinfo {volume} {109}},\ \bibinfo {pages} {110602} (\bibinfo
  {year} {2012})}\BibitemShut {NoStop}%
\bibitem [{\citenamefont {Collura}\ \emph {et~al.}(2013)\citenamefont
  {Collura}, \citenamefont {Sotiriadis},\ and\ \citenamefont
  {Calabrese}}]{collura13}%
  \BibitemOpen
  \bibfield  {author} {\bibinfo {author} {\bibfnamefont {M.}~\bibnamefont
  {Collura}}, \bibinfo {author} {\bibfnamefont {S.}~\bibnamefont {Sotiriadis}},
  \ and\ \bibinfo {author} {\bibfnamefont {P.}~\bibnamefont {Calabrese}},\
  }\href@noop {} {\bibfield  {journal} {\bibinfo  {journal} {Phys. Rev. Lett.}\
  }\textbf {\bibinfo {volume} {110}},\ \bibinfo {pages} {245301} (\bibinfo
  {year} {2013})}\BibitemShut {NoStop}%
\bibitem [{\citenamefont {Collura}\ \emph {et~al.}()\citenamefont {Collura},
  \citenamefont {Sotiriadis},\ and\ \citenamefont {Calabrese}}]{collura13b}%
  \BibitemOpen
  \bibfield  {author} {\bibinfo {author} {\bibfnamefont {M.}~\bibnamefont
  {Collura}}, \bibinfo {author} {\bibfnamefont {S.}~\bibnamefont {Sotiriadis}},
  \ and\ \bibinfo {author} {\bibfnamefont {P.}~\bibnamefont {Calabrese}},\
  }\href@noop {} {\bibfield  {journal} {\bibinfo  {journal} {J. Stat. Mech.}\
  }\textbf {\bibinfo {volume} {{\rm\bf(2013)}}},\ \bibinfo {pages}
  {P09025}}\BibitemShut {NoStop}%
\bibitem [{\citenamefont {Kolovsky}\ and\ \citenamefont
  {Korsch}(2010)}]{kolovsky10}%
  \BibitemOpen
  \bibfield  {author} {\bibinfo {author} {\bibfnamefont {A.~R.}\ \bibnamefont
  {Kolovsky}}\ and\ \bibinfo {author} {\bibfnamefont {H.~J.}\ \bibnamefont
  {Korsch}},\ }\href@noop {} {\bibfield  {journal} {\bibinfo  {journal} {J. of
  Siberian Federal University: Mathematics \& Physics}\ }\textbf {\bibinfo
  {volume} {3}},\ \bibinfo {pages} {311} (\bibinfo {year} {2010})}\BibitemShut
  {NoStop}%
\bibitem [{\citenamefont {Langer}\ \emph {et~al.}(2012)\citenamefont {Langer},
  \citenamefont {Schuetz}, \citenamefont {McCulloch}, \citenamefont
  {Schollw\"ock},\ and\ \citenamefont {Heidrich-Meisner}}]{langer12}%
  \BibitemOpen
  \bibfield  {author} {\bibinfo {author} {\bibfnamefont {S.}~\bibnamefont
  {Langer}}, \bibinfo {author} {\bibfnamefont {M.~J.~A.}\ \bibnamefont
  {Schuetz}}, \bibinfo {author} {\bibfnamefont {I.~P.}\ \bibnamefont
  {McCulloch}}, \bibinfo {author} {\bibfnamefont {U.}~\bibnamefont
  {Schollw\"ock}}, \ and\ \bibinfo {author} {\bibfnamefont {F.}~\bibnamefont
  {Heidrich-Meisner}},\ }\href@noop {} {\bibfield  {journal} {\bibinfo
  {journal} {Phys. Rev. A}\ }\textbf {\bibinfo {volume} {85}},\ \bibinfo
  {pages} {043618} (\bibinfo {year} {2012})}\BibitemShut {NoStop}%
\bibitem [{\citenamefont {Vidal}(2004)}]{vidal04}%
  \BibitemOpen
  \bibfield  {author} {\bibinfo {author} {\bibfnamefont {G.}~\bibnamefont
  {Vidal}},\ }\href@noop {} {\bibfield  {journal} {\bibinfo  {journal} {Phys.
  Rev. Lett.}\ }\textbf {\bibinfo {volume} {93}},\ \bibinfo {pages} {040502}
  (\bibinfo {year} {2004})}\BibitemShut {NoStop}%
\bibitem [{\citenamefont {Daley}\ \emph {et~al.}()\citenamefont {Daley},
  \citenamefont {Kollath}, \citenamefont {Schollw\"ock},\ and\ \citenamefont
  {Vidal}}]{daley04}%
  \BibitemOpen
  \bibfield  {author} {\bibinfo {author} {\bibfnamefont {A.}~\bibnamefont
  {Daley}}, \bibinfo {author} {\bibfnamefont {C.}~\bibnamefont {Kollath}},
  \bibinfo {author} {\bibfnamefont {U.}~\bibnamefont {Schollw\"ock}}, \ and\
  \bibinfo {author} {\bibfnamefont {G.}~\bibnamefont {Vidal}},\ }\href@noop {}
  {\bibfield  {journal} {\bibinfo  {journal} {J. Stat. Mech.: Theory Exp.}\
  }\textbf {\bibinfo {volume} {{\rm\bf(2004)}}},\ \bibinfo {pages}
  {P04005}}\BibitemShut {NoStop}%
\bibitem [{\citenamefont {White}\ and\ \citenamefont
  {Feiguin}(2004)}]{white04}%
  \BibitemOpen
  \bibfield  {author} {\bibinfo {author} {\bibfnamefont {S.~R.}\ \bibnamefont
  {White}}\ and\ \bibinfo {author} {\bibfnamefont {A.~E.}\ \bibnamefont
  {Feiguin}},\ }\href@noop {} {\bibfield  {journal} {\bibinfo  {journal} {Phys.
  Rev. Lett.}\ }\textbf {\bibinfo {volume} {93}},\ \bibinfo {pages} {076401}
  (\bibinfo {year} {2004})}\BibitemShut {NoStop}%
\bibitem [{\citenamefont {Schollw\"ock}(2005)}]{schollwoeck05}%
  \BibitemOpen
  \bibfield  {author} {\bibinfo {author} {\bibfnamefont {U.}~\bibnamefont
  {Schollw\"ock}},\ }\href@noop {} {\bibfield  {journal} {\bibinfo  {journal}
  {Rev. Mod. Phys.}\ }\textbf {\bibinfo {volume} {77}},\ \bibinfo {pages} {259}
  (\bibinfo {year} {2005})}\BibitemShut {NoStop}%
\bibitem [{\citenamefont {Schollw\"ock}(2011)}]{schollwoeck11}%
  \BibitemOpen
  \bibfield  {author} {\bibinfo {author} {\bibfnamefont {U.}~\bibnamefont
  {Schollw\"ock}},\ }\href@noop {} {\bibfield  {journal} {\bibinfo  {journal}
  {Ann. Phys. (NY)}\ }\textbf {\bibinfo {volume} {326}},\ \bibinfo {pages} {96}
  (\bibinfo {year} {2011})}\BibitemShut {NoStop}%
\bibitem [{\citenamefont {Rigol}\ and\ \citenamefont
  {Muramatsu}(2004{\natexlab{a}})}]{rigol04a}%
  \BibitemOpen
  \bibfield  {author} {\bibinfo {author} {\bibfnamefont {M.}~\bibnamefont
  {Rigol}}\ and\ \bibinfo {author} {\bibfnamefont {A.}~\bibnamefont
  {Muramatsu}},\ }\href@noop {} {\bibfield  {journal} {\bibinfo  {journal}
  {Phys. Rev. A}\ }\textbf {\bibinfo {volume} {70}},\ \bibinfo {pages}
  {031603(R)} (\bibinfo {year} {2004}{\natexlab{a}})}\BibitemShut {NoStop}%
\bibitem [{\citenamefont {Rigol}\ and\ \citenamefont
  {Muramatsu}(2004{\natexlab{b}})}]{rigol04}%
  \BibitemOpen
  \bibfield  {author} {\bibinfo {author} {\bibfnamefont {M.}~\bibnamefont
  {Rigol}}\ and\ \bibinfo {author} {\bibfnamefont {A.}~\bibnamefont
  {Muramatsu}},\ }\href@noop {} {\bibfield  {journal} {\bibinfo  {journal}
  {Phys. Rev. Lett.}\ }\textbf {\bibinfo {volume} {93}},\ \bibinfo {pages}
  {230404} (\bibinfo {year} {2004}{\natexlab{b}})}\BibitemShut {NoStop}%
\bibitem [{\citenamefont {Mierzejewski}\ and\ \citenamefont
  {Prelov\v{s}ek}(2010)}]{mierzejewski10}%
  \BibitemOpen
  \bibfield  {author} {\bibinfo {author} {\bibfnamefont {M.}~\bibnamefont
  {Mierzejewski}}\ and\ \bibinfo {author} {\bibfnamefont {P.}~\bibnamefont
  {Prelov\v{s}ek}},\ }\href@noop {} {\bibfield  {journal} {\bibinfo  {journal}
  {Phys. Rev. Lett.}\ }\textbf {\bibinfo {volume} {105}},\ \bibinfo {pages}
  {186405} (\bibinfo {year} {2010})}\BibitemShut {NoStop}%
\bibitem [{\citenamefont {Langer}\ \emph {et~al.}(2009)\citenamefont {Langer},
  \citenamefont {Heidrich-Meisner}, \citenamefont {Gemmer}, \citenamefont
  {McCulloch},\ and\ \citenamefont {Schollw\"ock}}]{langer09}%
  \BibitemOpen
  \bibfield  {author} {\bibinfo {author} {\bibfnamefont {S.}~\bibnamefont
  {Langer}}, \bibinfo {author} {\bibfnamefont {F.}~\bibnamefont
  {Heidrich-Meisner}}, \bibinfo {author} {\bibfnamefont {J.}~\bibnamefont
  {Gemmer}}, \bibinfo {author} {\bibfnamefont {I.}~\bibnamefont {McCulloch}}, \
  and\ \bibinfo {author} {\bibfnamefont {U.}~\bibnamefont {Schollw\"ock}},\
  }\href@noop {} {\bibfield  {journal} {\bibinfo  {journal} {Phys. Rev. B}\
  }\textbf {\bibinfo {volume} {79}},\ \bibinfo {pages} {214409} (\bibinfo
  {year} {2009})}\BibitemShut {NoStop}%
\bibitem [{\citenamefont {Langer}\ \emph {et~al.}(2011)\citenamefont {Langer},
  \citenamefont {Heyl}, \citenamefont {McCulloch},\ and\ \citenamefont
  {Heidrich-Meisner}}]{langer11}%
  \BibitemOpen
  \bibfield  {author} {\bibinfo {author} {\bibfnamefont {S.}~\bibnamefont
  {Langer}}, \bibinfo {author} {\bibfnamefont {M.}~\bibnamefont {Heyl}},
  \bibinfo {author} {\bibfnamefont {I.~P.}\ \bibnamefont {McCulloch}}, \ and\
  \bibinfo {author} {\bibfnamefont {F.}~\bibnamefont {Heidrich-Meisner}},\
  }\href@noop {} {\bibfield  {journal} {\bibinfo  {journal} {Phys. Rev. B}\
  }\textbf {\bibinfo {volume} {84}},\ \bibinfo {pages} {205115} (\bibinfo
  {year} {2011})}\BibitemShut {NoStop}%
\bibitem [{\citenamefont {Karrasch}\ \emph {et~al.}(shed)\citenamefont
  {Karrasch}, \citenamefont {Moore},\ and\ \citenamefont
  {Heidrich-Meisner}}]{karrasch-unpub}%
  \BibitemOpen
  \bibfield  {author} {\bibinfo {author} {\bibfnamefont {C.}~\bibnamefont
  {Karrasch}}, \bibinfo {author} {\bibfnamefont {J.~E.}\ \bibnamefont {Moore}},
  \ and\ \bibinfo {author} {\bibfnamefont {F.}~\bibnamefont
  {Heidrich-Meisner}},\ }\href@noop {} {\ ,\ \bibinfo {pages} {preprint
  arXiv:1312.2938} (\bibinfo {year} {unpublished})}\BibitemShut {NoStop}%
\bibitem [{\citenamefont {Jreissaty}\ \emph {et~al.}(2013)\citenamefont
  {Jreissaty}, \citenamefont {Carrasquilla},\ and\ \citenamefont
  {Rigol}}]{jreissaty13}%
  \BibitemOpen
  \bibfield  {author} {\bibinfo {author} {\bibfnamefont {A.}~\bibnamefont
  {Jreissaty}}, \bibinfo {author} {\bibfnamefont {J.}~\bibnamefont
  {Carrasquilla}}, \ and\ \bibinfo {author} {\bibfnamefont {M.}~\bibnamefont
  {Rigol}},\ }\href@noop {} {\bibfield  {journal} {\bibinfo  {journal} {Phys.
  Rev. A}\ }\textbf {\bibinfo {volume} {88}},\ \bibinfo {pages} {031606(R)}
  (\bibinfo {year} {2013})}\BibitemShut {NoStop}%
\bibitem [{\citenamefont {Eisler}\ and\ \citenamefont
  {R\'acz}(2013)}]{eisler13}%
  \BibitemOpen
  \bibfield  {author} {\bibinfo {author} {\bibfnamefont {V.}~\bibnamefont
  {Eisler}}\ and\ \bibinfo {author} {\bibfnamefont {Z.}~\bibnamefont
  {R\'acz}},\ }\href@noop {} {\bibfield  {journal} {\bibinfo  {journal} {Phys.
  Rev. Lett.}\ }\textbf {\bibinfo {volume} {110}},\ \bibinfo {pages} {060602}
  (\bibinfo {year} {2013})}\BibitemShut {NoStop}%
\bibitem [{\citenamefont {Polini}\ and\ \citenamefont
  {Vignale}(2007)}]{polini07}%
  \BibitemOpen
  \bibfield  {author} {\bibinfo {author} {\bibfnamefont {M.}~\bibnamefont
  {Polini}}\ and\ \bibinfo {author} {\bibfnamefont {G.}~\bibnamefont
  {Vignale}},\ }\href@noop {} {\bibfield  {journal} {\bibinfo  {journal} {Phys.
  Rev. Lett.}\ }\textbf {\bibinfo {volume} {98}},\ \bibinfo {pages} {266403}
  (\bibinfo {year} {2007})}\BibitemShut {NoStop}%
\bibitem [{\citenamefont {Micheli}\ \emph {et~al.}(2004)\citenamefont
  {Micheli}, \citenamefont {Daley}, \citenamefont {Jaksch},\ and\ \citenamefont
  {Zoller}}]{micheli04}%
  \BibitemOpen
  \bibfield  {author} {\bibinfo {author} {\bibfnamefont {A.}~\bibnamefont
  {Micheli}}, \bibinfo {author} {\bibfnamefont {A.~J.}\ \bibnamefont {Daley}},
  \bibinfo {author} {\bibfnamefont {D.}~\bibnamefont {Jaksch}}, \ and\ \bibinfo
  {author} {\bibfnamefont {P.}~\bibnamefont {Zoller}},\ }\href@noop {}
  {\bibfield  {journal} {\bibinfo  {journal} {Phys. Rev. Lett.}\ }\textbf
  {\bibinfo {volume} {93}},\ \bibinfo {pages} {140408} (\bibinfo {year}
  {2004})}\BibitemShut {NoStop}%
\bibitem [{\citenamefont {Daley}\ \emph {et~al.}(2005)\citenamefont {Daley},
  \citenamefont {Clark}, \citenamefont {Jaksch},\ and\ \citenamefont
  {Zoller}}]{daley05}%
  \BibitemOpen
  \bibfield  {author} {\bibinfo {author} {\bibfnamefont {A.~J.}\ \bibnamefont
  {Daley}}, \bibinfo {author} {\bibfnamefont {S.~R.}\ \bibnamefont {Clark}},
  \bibinfo {author} {\bibfnamefont {D.}~\bibnamefont {Jaksch}}, \ and\ \bibinfo
  {author} {\bibfnamefont {P.}~\bibnamefont {Zoller}},\ }\href@noop {}
  {\bibfield  {journal} {\bibinfo  {journal} {Phys. Rev. A}\ }\textbf {\bibinfo
  {volume} {72}},\ \bibinfo {pages} {043618} (\bibinfo {year}
  {2005})}\BibitemShut {NoStop}%
\bibitem [{\citenamefont {Lancaster}\ and\ \citenamefont
  {Mitra}(2010)}]{lancaster10}%
  \BibitemOpen
  \bibfield  {author} {\bibinfo {author} {\bibfnamefont {J.}~\bibnamefont
  {Lancaster}}\ and\ \bibinfo {author} {\bibfnamefont {A.}~\bibnamefont
  {Mitra}},\ }\href@noop {} {\bibfield  {journal} {\bibinfo  {journal} {Phys.
  Rev. E}\ }\textbf {\bibinfo {volume} {81}},\ \bibinfo {pages} {061134}
  (\bibinfo {year} {2010})}\BibitemShut {NoStop}%
\bibitem [{\citenamefont {Heidrich-Meisner}\ \emph {et~al.}(2008)\citenamefont
  {Heidrich-Meisner}, \citenamefont {Rigol}, \citenamefont {Muramatsu},
  \citenamefont {Feiguin},\ and\ \citenamefont {Dagotto}}]{hm08}%
  \BibitemOpen
  \bibfield  {author} {\bibinfo {author} {\bibfnamefont {F.}~\bibnamefont
  {Heidrich-Meisner}}, \bibinfo {author} {\bibfnamefont {M.}~\bibnamefont
  {Rigol}}, \bibinfo {author} {\bibfnamefont {A.}~\bibnamefont {Muramatsu}},
  \bibinfo {author} {\bibfnamefont {A.~E.}\ \bibnamefont {Feiguin}}, \ and\
  \bibinfo {author} {\bibfnamefont {E.}~\bibnamefont {Dagotto}},\ }\href@noop
  {} {\bibfield  {journal} {\bibinfo  {journal} {Phys. Rev. A}\ }\textbf
  {\bibinfo {volume} {78}},\ \bibinfo {pages} {013620} (\bibinfo {year}
  {2008})}\BibitemShut {NoStop}%
\bibitem [{\citenamefont {K\"{u}hner}\ \emph {et~al.}(2000)\citenamefont
  {K\"{u}hner}, \citenamefont {White},\ and\ \citenamefont
  {Monien}}]{kuhner00}%
  \BibitemOpen
  \bibfield  {author} {\bibinfo {author} {\bibfnamefont {T.~D.}\ \bibnamefont
  {K\"{u}hner}}, \bibinfo {author} {\bibfnamefont {S.~R.}\ \bibnamefont
  {White}}, \ and\ \bibinfo {author} {\bibfnamefont {H.}~\bibnamefont
  {Monien}},\ }\href@noop {} {\bibfield  {journal} {\bibinfo  {journal} {Phys.
  Rev. B}\ }\textbf {\bibinfo {volume} {61}},\ \bibinfo {pages} {12474}
  (\bibinfo {year} {2000})}\BibitemShut {NoStop}%
\bibitem [{\citenamefont {Rodriguez}\ \emph {et~al.}(2006)\citenamefont
  {Rodriguez}, \citenamefont {Manmana}, \citenamefont {Rigol}, \citenamefont
  {Noack},\ and\ \citenamefont {Muramatsu}}]{rodriguez06}%
  \BibitemOpen
  \bibfield  {author} {\bibinfo {author} {\bibfnamefont {K.}~\bibnamefont
  {Rodriguez}}, \bibinfo {author} {\bibfnamefont {S.}~\bibnamefont {Manmana}},
  \bibinfo {author} {\bibfnamefont {M.}~\bibnamefont {Rigol}}, \bibinfo
  {author} {\bibfnamefont {R.}~\bibnamefont {Noack}}, \ and\ \bibinfo {author}
  {\bibfnamefont {A.}~\bibnamefont {Muramatsu}},\ }\href@noop {} {\bibfield
  {journal} {\bibinfo  {journal} {New J. Phys.}\ }\textbf {\bibinfo {volume}
  {8}},\ \bibinfo {pages} {169} (\bibinfo {year} {2006})}\BibitemShut {NoStop}%
\bibitem [{\citenamefont {Heidrich-Meisner}\ \emph {et~al.}(2009)\citenamefont
  {Heidrich-Meisner}, \citenamefont {Manmana}, \citenamefont {Rigol},
  \citenamefont {Muramatsu}, \citenamefont {Feiguin},\ and\ \citenamefont
  {Dagotto}}]{hm09}%
  \BibitemOpen
  \bibfield  {author} {\bibinfo {author} {\bibfnamefont {F.}~\bibnamefont
  {Heidrich-Meisner}}, \bibinfo {author} {\bibfnamefont {S.~R.}\ \bibnamefont
  {Manmana}}, \bibinfo {author} {\bibfnamefont {M.}~\bibnamefont {Rigol}},
  \bibinfo {author} {\bibfnamefont {A.}~\bibnamefont {Muramatsu}}, \bibinfo
  {author} {\bibfnamefont {A.~E.}\ \bibnamefont {Feiguin}}, \ and\ \bibinfo
  {author} {\bibfnamefont {E.}~\bibnamefont {Dagotto}},\ }\href@noop {}
  {\bibfield  {journal} {\bibinfo  {journal} {Phys. Rev. A}\ }\textbf {\bibinfo
  {volume} {80}},\ \bibinfo {pages} {041603} (\bibinfo {year}
  {2009})}\BibitemShut {NoStop}%
\bibitem [{\citenamefont {Kajala}\ \emph {et~al.}(2011)\citenamefont {Kajala},
  \citenamefont {Massel},\ and\ \citenamefont {T\"orm\"a}}]{kajala11}%
  \BibitemOpen
  \bibfield  {author} {\bibinfo {author} {\bibfnamefont {J.}~\bibnamefont
  {Kajala}}, \bibinfo {author} {\bibfnamefont {F.}~\bibnamefont {Massel}}, \
  and\ \bibinfo {author} {\bibfnamefont {P.}~\bibnamefont {T\"orm\"a}},\
  }\href@noop {} {\bibfield  {journal} {\bibinfo  {journal} {Phys. Rev. Lett.}\
  }\textbf {\bibinfo {volume} {106}},\ \bibinfo {pages} {206401} (\bibinfo
  {year} {2011})}\BibitemShut {NoStop}%
\bibitem [{\citenamefont {Karlsson}\ \emph {et~al.}(2011)\citenamefont
  {Karlsson}, \citenamefont {Verdozzi}, \citenamefont {Odashima},\ and\
  \citenamefont {Capelle}}]{karlsson11}%
  \BibitemOpen
  \bibfield  {author} {\bibinfo {author} {\bibfnamefont {D.}~\bibnamefont
  {Karlsson}}, \bibinfo {author} {\bibfnamefont {C.}~\bibnamefont {Verdozzi}},
  \bibinfo {author} {\bibfnamefont {M.}~\bibnamefont {Odashima}}, \ and\
  \bibinfo {author} {\bibfnamefont {K.}~\bibnamefont {Capelle}},\ }\href@noop
  {} {\bibfield  {journal} {\bibinfo  {journal} {EPL}\ }\textbf {\bibinfo
  {volume} {93}},\ \bibinfo {pages} {23003} (\bibinfo {year}
  {2011})}\BibitemShut {NoStop}%
\bibitem [{\citenamefont {Ke\ss{}ler}\ \emph {et~al.}(2013)\citenamefont
  {Ke\ss{}ler}, \citenamefont {McCulloch},\ and\ \citenamefont
  {Marquardt}}]{kessler13}%
  \BibitemOpen
  \bibfield  {author} {\bibinfo {author} {\bibfnamefont {S.}~\bibnamefont
  {Ke\ss{}ler}}, \bibinfo {author} {\bibfnamefont {I.~P.}\ \bibnamefont
  {McCulloch}}, \ and\ \bibinfo {author} {\bibfnamefont {F.}~\bibnamefont
  {Marquardt}},\ }\href@noop {} {\bibfield  {journal} {\bibinfo  {journal} {New
  J. Phys.}\ }\textbf {\bibinfo {volume} {15}},\ \bibinfo {pages} {053043}
  (\bibinfo {year} {2013})}\BibitemShut {NoStop}%
\bibitem [{\citenamefont {\v{Z}nidari\v{c}}(2011)}]{znidaric11}%
  \BibitemOpen
  \bibfield  {author} {\bibinfo {author} {\bibfnamefont {M.}~\bibnamefont
  {\v{Z}nidari\v{c}}},\ }\href@noop {} {\bibfield  {journal} {\bibinfo
  {journal} {Phys. Rev. Lett.}\ }\textbf {\bibinfo {volume} {106}},\ \bibinfo
  {pages} {220601} (\bibinfo {year} {2011})}\BibitemShut {NoStop}%
\bibitem [{\citenamefont {Steinigeweg}\ and\ \citenamefont
  {Brenig}(2011)}]{steinigeweg11}%
  \BibitemOpen
  \bibfield  {author} {\bibinfo {author} {\bibfnamefont {R.}~\bibnamefont
  {Steinigeweg}}\ and\ \bibinfo {author} {\bibfnamefont {W.}~\bibnamefont
  {Brenig}},\ }\href@noop {} {\bibfield  {journal} {\bibinfo  {journal} {Phys.
  Rev. Lett.}\ }\textbf {\bibinfo {volume} {107}},\ \bibinfo {pages} {250602}
  (\bibinfo {year} {2011})}\BibitemShut {NoStop}%
\bibitem [{\citenamefont {Sch\"onmeier-Kromer}\ and\ \citenamefont
  {Pollet}(2013)}]{pollet13}%
  \BibitemOpen
  \bibfield  {author} {\bibinfo {author} {\bibfnamefont {J.}~\bibnamefont
  {Sch\"onmeier-Kromer}}\ and\ \bibinfo {author} {\bibfnamefont
  {L.}~\bibnamefont {Pollet}},\ }\href@noop {} {\bibfield  {journal} {\bibinfo
  {journal} {arXiv:1308.2229}\ } (\bibinfo {year} {2013})}\BibitemShut
  {NoStop}%
\bibitem [{\citenamefont {Hen}\ and\ \citenamefont {Rigol}(2010)}]{hen10}%
  \BibitemOpen
  \bibfield  {author} {\bibinfo {author} {\bibfnamefont {I.}~\bibnamefont
  {Hen}}\ and\ \bibinfo {author} {\bibfnamefont {M.}~\bibnamefont {Rigol}},\
  }\href@noop {} {\bibfield  {journal} {\bibinfo  {journal} {Phys. Rev. Lett.}\
  }\textbf {\bibinfo {volume} {105}},\ \bibinfo {pages} {180401} (\bibinfo
  {year} {2010})}\BibitemShut {NoStop}%
\bibitem [{\citenamefont {Jreissaty}\ \emph {et~al.}(2011)\citenamefont
  {Jreissaty}, \citenamefont {Carrasquilla}, \citenamefont {Wolf},\ and\
  \citenamefont {Rigol}}]{jreissaty11}%
  \BibitemOpen
  \bibfield  {author} {\bibinfo {author} {\bibfnamefont {M.}~\bibnamefont
  {Jreissaty}}, \bibinfo {author} {\bibfnamefont {J.}~\bibnamefont
  {Carrasquilla}}, \bibinfo {author} {\bibfnamefont {F.~A.}\ \bibnamefont
  {Wolf}}, \ and\ \bibinfo {author} {\bibfnamefont {M.}~\bibnamefont {Rigol}},\
  }\href@noop {} {\bibfield  {journal} {\bibinfo  {journal} {Phys. Rev. A}\
  }\textbf {\bibinfo {volume} {84}},\ \bibinfo {pages} {043610} (\bibinfo
  {year} {2011})}\BibitemShut {NoStop}%
\bibitem [{\citenamefont {Chen}\ \emph {et~al.}(2011)\citenamefont {Chen},
  \citenamefont {Huber}, \citenamefont {Trotzky}, \citenamefont {Bloch},\ and\
  \citenamefont {Altman}}]{chen11}%
  \BibitemOpen
  \bibfield  {author} {\bibinfo {author} {\bibfnamefont {Y.-A.}\ \bibnamefont
  {Chen}}, \bibinfo {author} {\bibfnamefont {S.~D.}\ \bibnamefont {Huber}},
  \bibinfo {author} {\bibfnamefont {S.}~\bibnamefont {Trotzky}}, \bibinfo
  {author} {\bibfnamefont {I.}~\bibnamefont {Bloch}}, \ and\ \bibinfo {author}
  {\bibfnamefont {E.}~\bibnamefont {Altman}},\ }\href@noop {} {\bibfield
  {journal} {\bibinfo  {journal} {Nature Phys.}\ }\textbf {\bibinfo {volume}
  {7}},\ \bibinfo {pages} {61} (\bibinfo {year} {2011})}\BibitemShut {NoStop}%
\bibitem [{\citenamefont {Kollath}\ \emph {et~al.}(2004)\citenamefont
  {Kollath}, \citenamefont {Schollw\"ock}, \citenamefont {von Delft},\ and\
  \citenamefont {Zwerger}}]{kollath04}%
  \BibitemOpen
  \bibfield  {author} {\bibinfo {author} {\bibfnamefont {C.}~\bibnamefont
  {Kollath}}, \bibinfo {author} {\bibfnamefont {U.}~\bibnamefont
  {Schollw\"ock}}, \bibinfo {author} {\bibfnamefont {J.}~\bibnamefont {von
  Delft}}, \ and\ \bibinfo {author} {\bibfnamefont {W.}~\bibnamefont
  {Zwerger}},\ }\href@noop {} {\bibfield  {journal} {\bibinfo  {journal} {Phys.
  Rev. A}\ }\textbf {\bibinfo {volume} {69}},\ \bibinfo {pages} {031601(R)}
  (\bibinfo {year} {2004})}\BibitemShut {NoStop}%
\bibitem [{\citenamefont {Rigol}\ \emph {et~al.}(2009)\citenamefont {Rigol},
  \citenamefont {Batrouni}, \citenamefont {Rousseau},\ and\ \citenamefont
  {Scalettar}}]{rigol09}%
  \BibitemOpen
  \bibfield  {author} {\bibinfo {author} {\bibfnamefont {M.}~\bibnamefont
  {Rigol}}, \bibinfo {author} {\bibfnamefont {G.~G.}\ \bibnamefont {Batrouni}},
  \bibinfo {author} {\bibfnamefont {V.~G.}\ \bibnamefont {Rousseau}}, \ and\
  \bibinfo {author} {\bibfnamefont {R.~T.}\ \bibnamefont {Scalettar}},\
  }\href@noop {} {\bibfield  {journal} {\bibinfo  {journal} {Phys. Rev. A}\
  }\textbf {\bibinfo {volume} {79}},\ \bibinfo {pages} {053605} (\bibinfo
  {year} {2009})}\BibitemShut {NoStop}%
\bibitem [{\citenamefont {Batrouni}\ \emph {et~al.}(2002)\citenamefont
  {Batrouni}, \citenamefont {Rousseau}, \citenamefont {Rigol}, \citenamefont
  {Muramatsu}, \citenamefont {Denteneer},\ and\ \citenamefont
  {Troyer}}]{batrouni02}%
  \BibitemOpen
  \bibfield  {author} {\bibinfo {author} {\bibfnamefont {G.~G.}\ \bibnamefont
  {Batrouni}}, \bibinfo {author} {\bibfnamefont {V.~G.}\ \bibnamefont
  {Rousseau}}, \bibinfo {author} {\bibfnamefont {M.}~\bibnamefont {Rigol}},
  \bibinfo {author} {\bibfnamefont {A.}~\bibnamefont {Muramatsu}}, \bibinfo
  {author} {\bibfnamefont {P.~J.~H.}\ \bibnamefont {Denteneer}}, \ and\
  \bibinfo {author} {\bibfnamefont {M.}~\bibnamefont {Troyer}},\ }\href@noop {}
  {\bibfield  {journal} {\bibinfo  {journal} {Phys. Rev. Lett.}\ }\textbf
  {\bibinfo {volume} {89}},\ \bibinfo {pages} {117203} (\bibinfo {year}
  {2002})}\BibitemShut {NoStop}%
\bibitem [{\citenamefont {Wessel}\ \emph {et~al.}(2004)\citenamefont {Wessel},
  \citenamefont {Alet}, \citenamefont {Troyer},\ and\ \citenamefont
  {Batrouni}}]{wessel04}%
  \BibitemOpen
  \bibfield  {author} {\bibinfo {author} {\bibfnamefont {S.}~\bibnamefont
  {Wessel}}, \bibinfo {author} {\bibfnamefont {F.}~\bibnamefont {Alet}},
  \bibinfo {author} {\bibfnamefont {M.}~\bibnamefont {Troyer}}, \ and\ \bibinfo
  {author} {\bibfnamefont {G.~G.}\ \bibnamefont {Batrouni}},\ }\href@noop {}
  {\bibfield  {journal} {\bibinfo  {journal} {Phys. Rev. A}\ }\textbf {\bibinfo
  {volume} {70}},\ \bibinfo {pages} {053615} (\bibinfo {year}
  {2004})}\BibitemShut {NoStop}%
\end{thebibliography}%

%\bibliographystyle{apsrev}
%\bibliography{expansion_mi.bbl}

\end{document}